\documentclass[
  twocolumn,aps,pra,showpacs,amsmath,
  amssymb,superscriptaddress,bibnotes,
  ]{revtex4-1} 

\usepackage{hyperref}
\usepackage{graphicx}
\usepackage{mathtools}
\usepackage{hyperref}
\usepackage{color}

\usepackage[export]{adjustbox}

\newcommand{\Ekin}{E_{\mathrm{kin}}}
\newcommand{\Etot}{E_{\mathrm{tot}}}

\newcommand{\ssa}{{n}}
\newcommand{\ssb}{{m}}

\newcommand{\mbf}[1]{\mathbf{#1}}

\newcommand{\ba}{b^{\phantom{\dagger}}}
\newcommand{\bc}{b^\dagger}

\newcommand{\LW}{\Phi_{\text{\tiny LW}}}
\newcommand{\BK}{\Gamma_{\text{\tiny BK}}}
\newcommand{\DMFT}{\Gamma_{\textrm{\tiny DMFT}}}

\newcommand{\SE}{\Gamma_{\text{\tiny SE}}}
\newcommand{\SED}{\tilde{\Gamma}_{\text{\tiny SE}}}
\newcommand{\SFT}{\Gamma_{\textrm{\tiny SFT}}}
\newcommand{\hF}{\mathcal{F}}
\newcommand{\hFA}{\tilde{\mathcal{F}}}

\newcommand{\n}{\hat{n}}
\newcommand{\C}{\mathcal{C}}

\newcommand{\Tr}{\textrm{Tr}}

\newcommand{\astcycl}{\mathrlap{\kern0.085em{\circlearrowright}}\ast}

\newcommand{\Z}{\mathcal{Z}}

\newcommand{\taucycl}{\mathrlap{\kern0.42em{\bullet}}\circlearrowright}

\newcommand{\bbb}{\mbf{b}}
\newcommand{\bba}{\mbf{b}^{\phantom{\dagger}}}
\newcommand{\bbc}{\mbf{b}^{\dagger}}

\newcommand{\bG}{\mbf{G}}

\newcommand{\bD}{\mbf{D}}
\newcommand{\bF}{\mbf{F}}
\newcommand{\bt}{\mbf{t}}
\newcommand{\bS}{\boldsymbol{\Sigma}}
\newcommand{\bSp}{\boldsymbol{\Sigma}_{1/2}}

\newcommand{\bPhi}{\boldsymbol\Phi}

\newcommand{\bDelta}{\boldsymbol{\Delta}}

\newcommand{\la}{\langle}
\newcommand{\ra}{\rangle}

\newcommand{\TC}{\mathcal{T}_\C}

\begin{document}
% --------------------------------------------------------------------

\title{Bosonic self-energy functional theory}

\author{Dario H\"ugel}
\affiliation{Department of Physics, Arnold Sommerfeld Center for Theoretical Physics and Center for NanoScience, Ludwig-Maximilians-Universit\"at M\"unchen, Theresienstrasse 37, 80333 Munich, Germany} 

\author{Philipp Werner}
\affiliation{Department of Physics, University of Fribourg, 1700 Fribourg, Switzerland} 

\author{Lode Pollet}
\affiliation{Department of Physics, Arnold Sommerfeld Center for Theoretical Physics and Center for NanoScience, Ludwig-Maximilians-Universit\"at M\"unchen, Theresienstrasse 37, 80333 Munich, Germany} 

\author{Hugo U.~R.~Strand}
\email{hugo.strand@unifr.ch}
\affiliation{Department of Physics, University of Fribourg, 1700 Fribourg, Switzerland} 

\date{\today} 

% Hubbard model electronic structure, 71.10.Fd
% Boson systems, 05.30.Jp
% Bose-Einstein statistics, 05.30.-d
% Ultracold gases, 67.85.-d


% --------------------------------------------------------------------
\begin{abstract}
We derive the self-energy functional theory for bosonic lattice systems with broken $U(1)$ symmetry by parametrizing the bosonic Baym-Kadanoff effective action in terms of one- and two-point self-energies.
The formalism goes beyond other approximate methods such as the pseudoparticle variational cluster approximation, the cluster composite boson mapping, and the Bogoliubov+U theory. It simplifies to bosonic dynamical-mean field theory when constraining to local fields, whereas when neglecting kinetic contributions of non-condensed bosons it reduces to the static mean-field approximation.
To benchmark the theory we study the Bose-Hubbard model on the two- and three-dimensional cubic lattice, comparing with exact results from path integral quantum Monte Carlo.
We also study the frustrated square lattice with next-nearest neighbor hopping, which is beyond the reach of Monte Carlo simulations.
A reference system comprising a single bosonic state, corresponding to three variational parameters, is sufficient to quantitatively describe phase-boundaries, and thermodynamical observables,
while qualitatively capturing the spectral functions, as well as the enhancement of kinetic fluctuations in the frustrated case.
On the basis of these findings we propose self-energy functional theory as the omnibus framework for treating bosonic lattice models, in particular, in cases where path integral quantum Monte Carlo methods suffer from severe sign problems (\textit{e.g.} in the presence of non-trivial gauge fields or frustration).
Self-energy functional theory enables the construction of diagrammatically sound approximations that are quantitatively precise and controlled in the number of optimization parameters, but nevertheless remain computable by modest means.
\end{abstract}
% --------------------------------------------------------------------

\maketitle
\makeatletter
\let\toc@pre\relax
\let\toc@post\relax
\makeatother

% --------------------------------------------------------------------
\section{Introduction}
% --------------------------------------------------------------------

During the last century interest in strongly correlated bosonic systems was driven by experimental work on superfluid Helium \cite{Wheatley:1975aa}, giving rise to a number of theoretical advances in the field of interacting symmetry broken bosonic systems \cite{ANDERSON:1966aa}.
Recent experimental advances in cold atom systems \cite{Morsch:2006vn, Bloch:2008uq} have revived this field, especially for strongly correlated bosonic systems described by the Bose-Hubbard model \cite{Fisher:1989kl, Jaksch:1998vn}.
Theoretically, simple lattice boson models with real Hamiltonians are solvable using numerically exact path integral quantum Monte Carlo (QMC) methods \cite{Pollet:2012ly}. However, one of the forefronts of cold atom research is the exploration of artificial gauge fields \cite{Struck:2012aa, Greschner:2014aa, Goldman:2014aa}, synthetic spin-orbit interactions \cite{Lin:2011aa, Struck:2014aa, Jimenez-Garcia:2015aa}, and non-local interactions \cite{Baier:2015aa, Zeiher:2016aa}.
Handling complex valued terms such as gauge fields is a challenge for QMC due to the resulting sign-problem \cite{Troyer:2005aa, Pollet:2012ly}, motivating the need for development of new theoretical methods for strongly correlated bosons.

One interesting approach is the self-energy functional theory (SFT) \cite{Potthoff:2003aa, Potthoff:2003ab, Potthoff:2006aa, Springer:2012}, originally developed for fermionic systems.
While the formalism contains the dynamical mean-field theory (DMFT) \cite{Georges:1996aa, Kotliar:2006aa}  in the limit of local fields (with retardation effects) \cite{Potthoff:2003aa} it has also been extended to non-local correlations \cite{Potthoff:2003ac} and disorder \cite{Potthoff:2007aa}. 
The bosonic version of SFT, initially formulated without symmetry breaking \cite{Koller:2006aa}, was recently extended to incorporate superfluidity \cite{Arrigoni:2011aa}.
However, in Ref.\ \onlinecite{Arrigoni:2011aa} no attempt was made to connect SFT to previous works on diagrammatic theory and the bosonic effective-action formalism \cite{De-Dominicis:1964aa, De-Dominicis:1964ab}.
In fact, we show that the ansatz for the one-point propagator's equation of motion used in Ref.\ \onlinecite{Arrigoni:2011aa} is in contradiction with standard literature \cite{De-Dominicis:1964aa, De-Dominicis:1964ab}.
To remedy this, we will put bosonic SFT on firm diagrammatic, functional, and variational grounds, paying special attention to the intricacies of bosonic $U(1)$ symmetry breaking. The result will be a functional which differs in a subtle, but significant way from the one proposed in Ref.~\onlinecite{Arrigoni:2011aa}. 

We derive a self-energy effective action $\SE$ for symmetry-broken interacting lattice bosons starting from De Dominicis and Martin's generalization \cite{De-Dominicis:1964aa, De-Dominicis:1964ab} of the Baym-Kadanoff effective action $\BK$ \cite{Baym:1961tw,Baym:1962qo}.
In analog to the fermionic formulation by Potthoff \cite{Potthoff:2003aa}, this involves a Legendre transform of the universal part of $\BK$, the two particle irreducible (2PI) Luttinger-Ward functional $\LW$ \cite{Luttinger:1960aa}. The transform changes the functional dependence from the one- and two-point response functions, $\bPhi$ and $\bG$, to their respective self-energies $\bSp$ and $\bS$, producing a universal self-energy functional $\hF \equiv \hF[\bSp, \bS]$.

Using the self-energy effective action $\SE$ we formulate the self-energy functional theory (SFT) approximation by exploiting the universality of $\hF$, which enables an exact evaluation of $\SE$ in the sub-space of self-energies of any reference system having the same interactions as the original lattice system \cite{Potthoff:2003aa}. By constraining the variational principle of $\SE$ to this subspace we arrive at the bosonic generalization of the SFT functional $\SFT$.
We show that for a local reference system with a completely general imaginary-time dependent hybridization function $\bDelta(\tau)$ the variations of $\SFT$ yield the self-consistency equations of bosonic dynamical mean-field theory (BDMFT) \cite{Byczuk:2008nx, Hubener:2009cr, Hu:2009qf, Anders:2010uq, Snoek:2010uq, Anders:2011uq}.
On the other hand, when omitting the hybridization function completely and neglecting the kinetic energy contributions of non-condensed bosons, static mean-field theory \cite{Fisher:1989kl, Sachdev:1999fk} is recovered.

As a proof of concept, we use SFT to study the Bose-Hubbard model \cite{Fisher:1989kl} at finite temperature on the two- and three-dimensional cubic lattice with nearest neighbor hopping.
For this purpose we make use of the simplest imaginable Hamiltonian reference system comprising a single bosonic state and three variational parameters:
a symmetry-breaking field $F'$ coupling to the particle-creation/annihilation operators ($b$ and $\bc$)
and the two fields $\Delta_{00}$ and $\Delta_{01}$ that are coupled with the density ($\bc b$) and pair-creation/annihilation operators ($bb$ and $\bc \bc$), respectively.
Hence, the fields $\Delta_{00}$ and $\Delta_{01}$ enter as an instantaneous imaginary-time Nambu hybridization function $\bDelta(\tau) = \delta(\tau) \bDelta$
in the reference system action.
This reference system Hamiltonian has also been used in the recently developed Bogoliubov+U theory (B+U) \cite{Hugel:2015ab}.

We compare our SFT results, employing the minimal reference system, to exact lattice quantum Monte Carlo (QMC) results \cite{Capogrosso-Sansone:2007lh} and find quantitative agreement on the location of phase-boundaries, energetics, and local observables throughout the normal and superfluid phases.
We also compare with BDMFT results \cite{Anders:2010uq, Anders:2011uq}, corresponding to the local SFT approximation with an infinite number of variational parameters.
The deviation of the three parameter SFT from QMC (and BDMFT) is surprisingly small and only noticeable close to the normal to superfluid phase transition, where kinetic quantum fluctuations become prominent.
The B+U calculations in Ref.\ \onlinecite{Hugel:2015ab} use the same reference system Hamiltonian and show excellent agreement with QMC at zero temperature.
The SFT method presented here, however, gives quantitative agreement with QMC also at finite temperature.
We also study the spectral function in both the normal and symmetry broken phase and provide a detailed analysis of the high-energy resonances.

While the calculations presented here employ a local self-energy approximation, SFT trivially extends to non-local self-energies and cluster reference systems \cite{Potthoff:2003ac}.
The great promise of the SFT formalism lies in its ability to treat systems with gauge fields \cite{Struck:2012aa, Greschner:2014aa, Goldman:2014aa} and other complex terms such as spin-orbit coupling \cite{Lin:2011aa, Struck:2014aa, Jimenez-Garcia:2015aa}, where lattice quantum Monte Carlo approaches suffer from a sign problem.
To explicitly show that SFT is sign-problem agnostic we study the frustrated Bose-Hubbard model on the square lattice with next-nearest neigbor hopping, and find a substantial shift of the phase-boundaries with respect to the Bose-Hubbard model without frustration, due to the enhancement of kinetic fluctuations in the frustrated regime.

The fermionic version of SFT has also been extended to systems out of equilibrium \cite{Hofmann:2013aa, Hofmann:2015ab}. This makes bosonic SFT an interesting alternative to the recently developed real-time dynamical mean-field theory \cite{Aoki:2014kx} and its bosonic generalization \cite{Strand:2015aa}, for studies of, e.g., the superfluid to normal phase transition in quenched or driven non-equilibrium systems.
 
This paper is organized as follows. In Sec.\ \ref{Sec:Theo} we present a general derivation of the SFT formalism for bosons. We discuss the Baym-Kadanoff effective-action $\BK$ derived by De Dominicis and Martin \cite{De-Dominicis:1964aa, De-Dominicis:1964ab} in Sec.\ \ref{Sec:Theo_BK} and show how the self-consistency conditions of BDMFT can be derived from it in Sec.\ \ref{sec:DMFT}.
In Sec.\ \ref{sec:SE} we derive the self-energy effective action $\SE$ as a Legendre transform of $\BK$.
The SFT formalism is then developed in Sec.\ \ref{Sec:Theo_SFT} for a general bosonic lattice system (Sec.\ \ref{Sec:Theo_SFT_General}), and a general reference system (Sec.\ \ref{Sec:Theo_SFT_RefSyst}). We further show how BDMFT (Sec.\ \ref{Sec:Theo_SFT_DMFT}) and the mean-field approximation (Sec.\ \ref{Sec:Theo_SFT_MeanField}) are obtained as limits of SFT.
In Sec.\ \ref{sec:BHModel} we introduce the Bose-Hubbard model, and the minimal reference system (Sec.\ \ref{sec:minref}). Sec.\  \ref{Sec:Res} is devoted to numerical results, in particular phase boundaries (Sec.\ \ref{Sec:Res_Phase}) and thermodynamical observables (Sec.\ \ref{Sec:Res_Obs}). We also discuss the superfluid phase transition in Sec. \ref{sec:res_free_energy} and the Hugenholtz-Pines relation in Sec.\ \ref{Sec:Res_HP}.
We then present the lattice spectral function in Sec.\ \ref{Sec:Res_SF} and study the effect of frustration due to next-nearest neigbor hopping in Sec.\ \ref{Sec:Res_NNN}.
Finally in Sec.\ \ref{Sec:Conc} we conclude and give an outlook on future applications.

% --------------------------------------------------------------------
\section{Theory}
\label{Sec:Theo}
% --------------------------------------------------------------------

Consider a general system of lattice bosons with local interactions having the Hamiltonian
\begin{equation}
  H =
  \sum_{i} (\bc_i F_i + F_i^\dagger \ba_i )
  + \sum_{ij} t_{ij} \bc_i \ba_j
  + \hat{V}_3 + \hat{V}_4
  \, ,
\end{equation}
where $\bc_i$ ($b_i$) creates (annihilates) a boson at site $i$, $t_{ij}$ is the single-particle hopping, $F$ is an external field, which couples linearly to the bosonic operators, and $\hat{V}_3$ and $\hat{V}_4$ are general interactions with three and four legs, respectively.
Using Einstein summation and the Nambu operators $\bbc_\alpha = \bbc_{i\nu} = ( \bc_i , \, \ba_i )_\nu$ with commutator $[\bbb^{\alpha}, \bbc_\beta] = (\mbf{1} \otimes \sigma_z)^\alpha_\beta$, where $\alpha$ is a composite index comprising the site and Nambu indices $i$ and $\nu$, the Hamiltonian can be written compactly as
\begin{align}
  H & =
  \bF^\dagger_\alpha \bbb^\alpha
  + \frac{1}{2} \bbc_\alpha \bt^\alpha_\beta \bbb^\beta
  + \hat{V}_3 + \hat{V}_4
  \, ,
  \\
  \hat{V}_3 & =
  V^{(3)}_{\alpha \beta \gamma} \,
  \bbb^{\alpha} \bbb^{\beta} \bbb^{\gamma}
  \, , \quad
  \hat{V}_4 =
  V^{(4)}_{\alpha \beta \gamma \delta} \,
  \bbb^{\alpha} \bbb^{\beta} \bbb^{\gamma} \bbb^{\delta}
  \, . \label{eq:V3V4}  
\end{align}
where $\bt^{\alpha}_\beta = \bt^{i\eta}_{j\nu} = t_{ij} \otimes \mbf{1}_{\eta \nu}$, up to an irrelevant constant.
For brevity in the following we will drop all tensor indices whenever contractions are well defined.

The partition function $\Z$ is given by the trace of the imaginary-time-ordered exponential $\Z = \Tr[ \TC e^{-S} ]$, where $S$ is the action
\begin{multline}
  S[\bbb] =
  \int_0^\beta d\tau (\hat{V}_3[\bbb(\tau)] + \hat{V}_4[\bbb(\tau)])
  + \int_0^\beta d\tau \mbf{F}^\dagger \bbb(\tau)
  \\
  + \frac{1}{2} \iint_0^\beta d\tau d\tau'
  \bbc(\tau) [ - \bG_0^{-1}(\tau, \tau') ] \bbb(\tau')
  \, ,
  \label{eq:GeneralActionS}
\end{multline}
$\beta$ is the inverse temperature, and the hopping $t_{ij}$ is absorbed in the non-interacting propagator
\begin{equation}
  \bG_0^{-1}(\tau, \tau') = \delta(\tau - \tau')
  ( - [\mbf{1} \otimes \sigma_z] \partial_{\tau'} - \mbf{t} )
  \, .
\end{equation}
The partition function $\Z$'s functional dependence on $\bF$ and $\bG_0^{-1}$, $\Z = \Z[\bF, \bG_0^{-1}]$, make the free energy $\Omega[\bF, \bG_0^{-1}] \equiv - \ln[\Z]/\beta$ a generating functional for the propagators
\begin{align}
  \beta \frac{\delta\Omega}{\delta \bF^\dagger}
  & = \la \mbf{b} \ra \equiv \bPhi
  \, , \label{eq:FreeEnergyFvar} \\
  2 \beta \frac{\delta\Omega}{\delta \bG_0^{-1}(\tau', \tau)}
  & = - \la \bbb (\tau) \bbc (\tau') \ra = \bG(\tau, \tau') - \bPhi \bPhi^\dagger
  \, , \label{eq:FreeEnergyG0var} 
\end{align}
where $\bPhi$ is the expectation value of the bosonic Nambu annihilation operator $\bbb$, $\bG$ is the connected single-particle Green's function, and the expectation value of an operator $\hat{O}(\tau)$ is defined as the time-ordered trace $\la \hat{O}(\tau) \ra = \Tr[\mathcal{T} e^{-S} \hat{O}(\tau)] / \Z$.

% --------------------------------------------------------------------
\subsection{Baym-Kadanoff effective action}
\label{Sec:Theo_BK}
% --------------------------------------------------------------------

The effective action formulation is a useful starting point for approximations to the many-body system. It is based on a Legendre transform of the free energy functional $\Omega$ in both $\bF$ and $\bG_0^{-1}$ to the interacting system propagators $\bPhi$ and $\bG$, see Refs.\ \cite{Kleinert:1982aa, Berges:2005aa} for an overview.

The resulting functional $\BK = \BK[\bPhi, \bG]$ was derived by Baym and Kadanoff \cite{Baym:1961tw,Baym:1962qo} for fermions and generalized to bosons by De Dominicis and Martin \cite{De-Dominicis:1964aa, De-Dominicis:1964ab} and later to relativistic systems \cite{Cornwall:1974aa}. The functional has the form
\begin{multline}
  \BK[ \bPhi, \bG ] =
  S_0[ \bPhi ]
  + \frac{1}{2} \Tr [ \bG_0^{-1} \bG ]
\\
  + \frac{1}{2} \Tr\ln [-\bG^{-1}]
  + \LW[ \bPhi, \bG ],
  \label{eq:BK}
\end{multline}
where $S_0$ is the non-interacting part of the system action,
$S_0[\bPhi] = \bF^\dagger \bPhi - \frac{1}{2}\bPhi^\dagger \bG_0^{-1} \bPhi$.
For explicit definitions of the products and traces, see Appendix \ref{app:TraceProduct}. The Baym-Kadanoff functional $\BK$ is stationary in $\bPhi$ and $\bG$ at the physical solution
\begin{equation}
  \frac{\delta \BK}{\delta \bPhi^\dagger} = 0
  \, , \quad
  \frac{\delta \BK}{\delta \bG} = 0
  \, .
  \label{eq:BKVarPrinc}
\end{equation}

In Eq.\ \ref{eq:BK}, the whole complexity of the many-body system is contained in the Luttinger-Ward functional $\LW[\bPhi, \bG] \equiv \LW[\bPhi, \bG, \hat{\nu}_3, \hat{V}_4]$ \cite{Luttinger:1960aa} which contains all two-particle irreducible diagrams (2PI) in $\bG$ with the three- and four-point vertices $\hat{\nu}_3 = \hat{V}_3 + \hat{V}_4 \bPhi$ and $\hat{V}_4$, respectively \footnote{Note that the Luttinger-Ward functional $\LW$ for symmetry broken bosons with only a four-particle interaction vertex ($\hat{V}_3 =0$, $\hat{V}_4 \ne 0$) still acquires an effective three-particle vertex \cite{Kleinert:1982aa} ($\hat{\nu}_3 = \hat{V}_4 \bPhi$).}.
Note that the Luttinger-Ward functional $\LW$ is a universal functional, in that it depends only on the interacting one- and two-point propagators ($\bPhi$ and $\bG$ respectively) and the three- and four-point interaction vertices ($\hat{V}_3$ and $\hat{V}_4$). In particular, $\LW$ does not depend on the free propagator $\bG_0$ of the system.
Using the diagrammatic notation
\begin{multline}
  \includegraphics[valign=c]{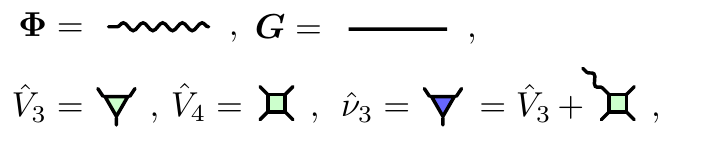}
  \label{eq:DiagrammaticNotation}
\end{multline}
the lowest order diagrams in $\LW$ can be written as
\begin{multline}
  \ \\[-10mm]
  \LW
  = \!\!\!
  \includegraphics[valign=c]{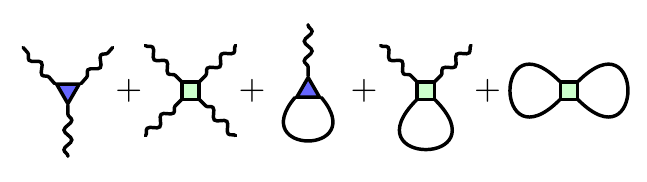}
  \\
  \includegraphics[valign=c]{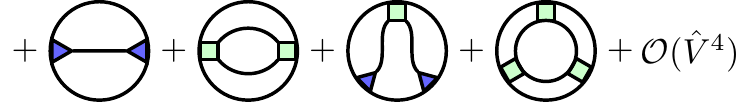}
  , \!\!\!\!
  \label{eq:LW}
\end{multline}
when omitting constant prefactors \cite{Kleinert:1982aa}.
The functional derivative $\delta_{\bPhi^\dagger} \LW$ amounts to removing one $\bPhi^\dagger$ term in the first order terms and in the effective three point vertex $\hat{\nu}_3 = \hat{V}_3 + \hat{V}_4 \bPhi$, which yields a one-point 2PI vertex \\[-8mm]
\begin{equation}
  \frac{\delta \LW}{\delta \bPhi^\dagger}
  = \! \! \! \!
  \includegraphics[valign=c]{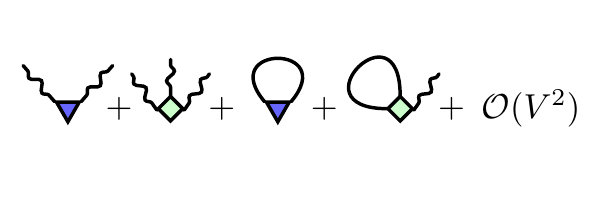}
  \, ,
\end{equation}\\[-8mm]
while $\delta_{\bG} \LW$ corresponds to cutting one propagator line $\bG$, which yields a two-point 2PI vertex \\[-8mm]
\begin{equation}
  \frac{\delta \LW}{\delta \bG}
  = \, \, \, 
  \includegraphics[valign=c]{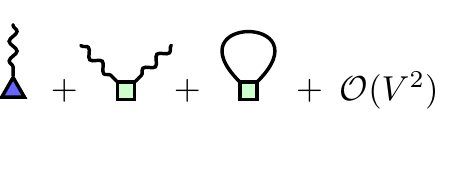}
  \, .
\end{equation}\\[-8mm]
The explicit form of the stationary condition [Eq.\ (\ref{eq:BKVarPrinc})] gives the equations of motion for the propagators
\begin{align}
  \frac{\delta \BK}{ \delta \bPhi^\dagger } & =
  \bF
  - \bG_0^{-1} \bPhi
  + \frac{\delta \LW}{\delta \bPhi^\dagger}
  = 0 \, ,
  \\
  \frac{\delta \BK}{ \delta \bG } & =
  \frac{1}{2} \bG_0^{-1}
  - \frac{1}{2} \bG^{-1}
  + \frac{\delta \LW}{\delta \bG}
  = 0 \, .
\end{align}
In the last equation we directly identify the two-point 2PI vertex as the self-energy $\delta_{\bG} \LW = - \bS / 2 $. The one-point vertex is less renowned, and will be denoted here as $\delta_{\bPhi^\dagger} \LW = - \bSp$. Hence, the stationary condition yields the Dyson  equations
\begin{align}
  \bG_0^{-1} \bPhi & = \bF - \bSp \, ,
  \label{eq:DysonS12}
  \\
  \bG^{-1} & = \bG_0^{-1} - \bS
  \label{eq:DysonS}
  \, .
\end{align}

The power of the effective action formalism is that approximations of the Luttinger-Ward functional $\LW$ produce non-perturbative approximations, i.e., sums to all orders in the interactions $\hat{V}_3$ and $\hat{V}_4$ and the non-interacting propagator $\bG_0$, that still obey the symmetries of the original system. In particular the approximations conserve total energy, density, and momentum \cite{Baym:1961tw,Baym:1962qo}. In (relativistic) quantum field theory it is common to make consistent approximations in $\LW$ to a given ``loop-order'' in the 2PI diagrams \cite{Kleinert:2001fk}.

Recently, interesting issues regarding the single-valuedness of the Luttinger-Ward functional $\LW$ have been raised within the framework of truncated (but high-order) expansions using diagrammatic Monte Carlo, dynamical mean-field theory, and the GW-approximation \cite{Kozik:2015aa, 1367-2630-17-9-093045}.
The findings show that particular self-consistent schemes to sum the boldified diagrams to infinite order can produce non-physical solutions, where a solution is given by the resulting propagator(s) $\bG$ (and $\bPhi$) and self-energy(s) $\bS$ (and $\bSp$).
This should come as no surprise as the construction of the Baym-Kadanoff functional $\BK$ is a Legendre-transform of the free energy $\Omega$. Thus, while $\BK$ and $\Omega$ have the same stationary points, there is no guarantee that maximas and inflection points of $\Omega$ do not become minimas of $\BK$. In such a case a stationary point of $\Omega$, which is not a local minimum, can very well become a local minimum of $\BK$ and hence an attractive fix-point for self-consistent calculations of $\BK$ through the evaluation of $\LW$.

% --------------------------------------------------------------------
\subsection{Dynamical mean-field theory}
\label{sec:DMFT}
% --------------------------------------------------------------------

An interesting class of approximations to the effective action amounts to evaluate the exact Luttinger-Ward functional, but only for a selected subset of propagators. One example is the local real-space approximation
\begin{equation}
  \LW[\bPhi, \bG] \approx \LW[\bPhi, \bG_{ii}] \, ,
  \label{eq:LocalLW}
\end{equation}
which accounts for all diagrams with site-local propagators $\bG_{ii}$ of the lattice. For number conserving systems ($\bPhi=\bF=0$), Eq.\ (\ref{eq:LocalLW}) becomes an equality in the limit of infinite dimensions \cite{Metzner:1989aa, Georges:1996aa} yielding the exact solution. Only accounting for local diagrams in $\LW$ trivially generates a site-local self-energy
\begin{equation}
  -2 \delta_{\bG_{ij}} \LW[\bPhi, \bG_{ll}] = \delta_{ij} \bS_{ii} \, .
\end{equation}
This approximation is not interesting \textit{per se} as the calculation of $\LW[\bPhi, \bG_{ii}]$ remains a formidable problem. The ingenuity of dynamical mean-field theory (DMFT), however, is the observation that there exists a simpler and exactly solvable many-body system with the same local Luttinger-Ward functional. In fact, there is a reference system (here denoted with primed quantities) with the same propagators $\bPhi$ and $\bG_{ii}$, and thus the same Luttinger-Ward functional $\LW[\bPhi, \bG_{ii}]$, but with \textit{a priori} unknown local sources $\bF'_i$ and $\bG_{0,ij}' = \delta_{ij} \bG_0'$. The corresponding reference-system effective action, 
\begin{multline}
  \BK' =
  S'_0[\bPhi]
  + \frac{1}{2} \Tr[ \bG_{0}'^{-1} \bG_{ii} ]
\\
  + \frac{1}{2}\Tr\ln[-\bG_{ii}^{-1}]
  + \LW[\bPhi, \bG_{ii}] \, ,
  \label{eq:DMFTRefSystBK}
\end{multline}
is also stationary at $\bPhi$ and $\bG_{ii}$, $\delta_{\bPhi^\dagger} \BK' = \delta_{\bG_{ii}} \BK' = 0$.

The DMFT effective action can be constructed as the difference, $\DMFT = \BK - \BK'$, which remains stationary, $\delta_{\bPhi} \DMFT = \delta_{\bG_{ii}} \DMFT = 0$, and whose variations give
\begin{align}
  \frac{\delta \DMFT}{\delta \bPhi_{i}^\dagger} & =
  \sum_{j} \bG_{0, ij}^{-1} \bPhi_j - \bG_{0}'^{-1} \bPhi_i - \bF_i + \bF'_i= 0
  \, , \label{eq:DMFTF} \\  
  \frac{\delta \DMFT}{\delta \bG_{ii}} & =
  [\bG_{0}^{-1}]_{ii} + [-\bG^{-1}]_{ii}
  - \bG_{0}'^{-1} - [-\bG_{ii}]^{-1}
  \nonumber \\ & =
  [\bG_{ii}]^{-1} + \bS_{ii} - \bG_{0}'^{-1}
  = 0
  \, . \label{eq:DMFTWeiss}
\end{align}
These stationarity conditions are equivalent to the DMFT self-consistency equations \cite{Georges:1996aa} which are used to determine the reference system's source fields $\bF'$ and $\bG_0'^{-1}$;
Eq.~(\ref{eq:DMFTWeiss}) fixes the reference system's Weiss field $\bG_0'$ \cite{Georges:1996aa} and Eq.~(\ref{eq:DMFTF}) determines the effective symmetry breaking field $\bF'$ of the reference system in the bosonic generalization of DMFT \cite{Byczuk:2008nx, Hubener:2009cr, Hu:2009qf, Anders:2010uq, Snoek:2010uq, Anders:2011uq}.

Solving the reference system while imposing these relations yields a non-trivial and non-perturbative solution of the original lattice system, including all local diagrams in $\LW$.
Note that the reference system, commonly called ``the impurity problem'' in DMFT, has a general (retarded) non-interacting propagator $\bG_0'^{-1}$, and exact solutions can only be obtained by infinite summations of diagrams using, e.g., continuous-time quantum Monte Carlo \cite{Werner:2006rt, Gull:2011lr}.

% --------------------------------------------------------------------
\subsection{Self-energy effective action}
\label{sec:SE}
% --------------------------------------------------------------------

An interesting reformulation of the Baym-Kadanoff functional $\BK$ has been devised by Potthoff \cite{Potthoff:2003aa} for fermions. The starting point is a Legendre transform of the Luttinger-Ward functional $\LW$, changing the functional dependence from the dressed propagators $\bPhi$ and $\bG$ to the one- and two-point vertices $\bS_{1/2}$ and $\bS$. Here we generalize this procedure for the bosonic action. Using the Dyson equations [Eqs.\ (\ref{eq:DysonS12}) and (\ref{eq:DysonS})] we can write $\BK$ [Eq.\ (\ref{eq:BK})] as
\begin{multline}
  \BK[\bPhi, \bG] =
  \frac{1}{2} \bPhi^\dagger \bG_0^{-1} \bPhi
  + \frac{1}{2} \Tr \ln [-\bG^{-1}]
  \\
  + \LW[\bPhi, \bG]
  + \bS_{1/2}^\dagger \bPhi
  + \frac{1}{2} \Tr [\bS \bG],
\end{multline}
where the last line can be viewed as a Legendre transform of $\LW$ \cite{Boyd:2004fu}. This is possible because the two last terms are in fact derivatives of $\LW$, {\it i.e.,} the last line can be replaced by the universal functional
\begin{multline}
  \hF [\bS_{1/2}, \bS]
  =
  \LW[\bPhi, \bG]
  - (\delta_{\bPhi} \LW) \bPhi
  - \Tr [ ( \delta_{\bG} \LW ) \bG ]
  \\ =
  \LW[\bPhi, \bG]
  + \bS_{1/2}^\dagger \bPhi
  + \frac{1}{2} \Tr [\bS \bG]
  \, , \label{eq:UniversalF}
\end{multline}
that depends only on the one- and two-point self-energies $\bSp$ and $\bS$, having (by construction) the variations
\begin{equation}
  \delta_{\bS_{1/2}^\dagger} \hF = \bPhi
  \, , \quad
  \delta_{\bS} \hF = \bG/2
  \, .
\end{equation}
In terms of $\hF$ the Baym-Kadanoff functional $\BK[\bPhi, \bG]$ can be rewritten as a self-energy effective action $\SE$ parametrized by the self-energies of $\bS_{1/2}$ and $\bS$
\begin{multline}
  \SE[\bS_{1/2}, \bS] =
  \frac{1}{2} (\bF - \bS_{1/2})^\dagger \bG_0 (\bF - \bS_{1/2})
  \\ + \frac{1}{2} \Tr \ln [-(\bG_0^{-1} - \bS)]
  + \hF[\bS_{1/2}, \bS],
  \label{eq:SESigma}
\end{multline}
which remains stationary at the physical solution, as the variations with respect to $\bSp$ and $\bS$ still yield the Dyson equations [Eqs.\ (\ref{eq:DysonS12}) and (\ref{eq:DysonS})]
\begin{align}
  \frac{\delta \SE}{\delta \bSp^\dagger} & =
  - \bG_0 (\bF - \bSp) + \bPhi = 0
  \, ,   \label{eq:SEVarPrinc_1} \\
  2\frac{\delta \SE}{\delta \bS} & =
  -(\bG_0^{-1} - \bS)^{-1} + \bG = 0
  \, .
  \label{eq:SEVarPrinc_2}
\end{align}

This self-energy effective action $\SE = \SE[\bS_{1/2}, \bS]$ can be used to construct generalized approximations in the spirit of dynamical mean-field theory. The resulting class of approximations is commonly denoted as \emph{self-energy functional theory} (SFT) approximations \cite{Potthoff:2003aa}.

We note that the self-energy effective action $\SE$ derived here in Eq.\ (\ref{eq:SESigma}) differs from the one previously derived in Ref.\ \onlinecite{Arrigoni:2011aa}.
The difference lies in the one-point Dyson equation [Eq.\ (\ref{eq:SEVarPrinc_1})] obtained at stationarity of the self-energy functional $\SE$. The result we arrive at in Eq.\ (\ref{eq:SEVarPrinc_1}) is a direct consequence of the bosonic Baym-Kadanoff effective action $\BK$ [Eq.\ (\ref{eq:BK})] and its one-point Dyson equation [Eq.\ (\ref{eq:DysonS12})], while Ref.\ \onlinecite{Arrigoni:2011aa} uses an ansatz for the one-point Dyson equation [Eq.\ (\ref{eq:DysonD})] that is inconsistent with $\BK$ and standard literature \cite{De-Dominicis:1964aa, De-Dominicis:1964ab}, see Appendix \ref{app:Arrigoni} for a detailed discussion.

% --------------------------------------------------------------------
\subsection{Self-energy functional theory}
\label{Sec:Theo_SFT}
% --------------------------------------------------------------------

As pointed out in the seminal work of Potthoff \cite{Potthoff:2003aa}, the universality of the self-energy functional $\hF = \hF[\bSp, \bS]$ can be used to construct a generalized class of approximations to interacting many-body systems.
It is instructive to recall the main steps in the construction of the dynamical mean-field theory approximation in Section \ref{sec:DMFT}. It was based on (i) an initial approximation of the universal part of the effective action [Eq.\ (\ref{eq:LocalLW})] (the Luttinger-Ward functional $\LW$), (ii) the introduction of an exactly solvable reference system with the same universal functional [Eq.\ (\ref{eq:DMFTRefSystBK})], and (iii) the use of the variational principle of the effective action to obtain self-consistent equations for the reference system [Eqs.\ (\ref{eq:DMFTF}) and (\ref{eq:DMFTWeiss})].
In the construction of self-energy functional theory the approximation is moved from the functional to the variational principle.

% --------------------------------------------------------------------
\subsubsection{Functional formulation}
\label{Sec:Theo_SFT_General}
% --------------------------------------------------------------------

Let us again introduce an (analytically or numerically) exactly solvable reference system, with linear field $\bF'$ and free propagator $\bG'_0$. The self-energy effective action $\SE'$ of the reference system is then given by
\begin{multline}
  \SE'[\bSp, \bS] =
  \frac{1}{2} ( \bF' - \bSp)^\dagger \bG'_0 ( \bF' - \bSp )
  \\
  + \frac{1}{2} \Tr \ln [ - (\bG'^{-1}_0 - \bS) ] + \hF[\bSp, \bS]
  \, ,
  \label{eq:BKSigmaRefSyst}
\end{multline}
which, at the physical solution $\bSp = \bSp'$ and $\bS = \bS'$, is stationary, $\delta_{\bSp'} \SE'[\bSp', \bS'] = \delta_{\bS'} \SE'[\bSp', \bS'] = 0$, and equal to the reference system's free energy
\begin{equation}
  \SE'[\bSp', \bS'] = \beta \Omega'[\bF', \bG_0']
  \, . \label{eq:SERefSysFreeEnergy}
\end{equation}

We can now use the universality of $\hF$ to evaluate the self-energy effective action $\SE$ of the original lattice system at the physical solution ($\bSp'$ and $\bS'$) of the reference system. The $\SE$ functional evaluated at $\bSp = \bSp'$ and $\bS = \bS'$ is given by
\begin{multline}
  \SE[\bS', \bSp'] =
  \beta \Omega'
  + \frac{1}{2} ( \bF \! - \bSp')^\dagger \bG_0 ( \bF - \bSp' )
  \\
  - \frac{1}{2} ( \bF' - \bSp')^\dagger \bG'_0 ( \bF' - \bSp' )
  + \frac{1}{2} \Tr \ln
    \left[ \frac{\bG^{-1}_0 - \bS'}{\bG'^{-1}_0 - \bS'} \right]
  \, , \label{eq:SFAFunc}
\end{multline}
where we have replaced $\hF$ in Eq.\ (\ref{eq:SESigma}) using the equations of the reference system [Eqs.\ (\ref{eq:BKSigmaRefSyst}) and (\ref{eq:SERefSysFreeEnergy})].

In solving the reference system exactly, the self-energies $\bSp'$ and $\bS'$ are parametrized by $\bF'$ and $\bG'_0$, i.e., $\bSp' = \bSp'[\bF', \bG_0']$ and $\bS' = \bS'[\bF', \bG_0']$ and we can formally construct the self-energy functional theory approximation $\SFT$ to the self-energy effective action $\SE$ according to
\begin{equation}
  \SFT[\bF', \bG_0'] =
  \SE[
    \bSp'[\bF', \bG_0'], 
    \bS[\bF', \bG_0']
  ]
  \label{eq:SFAFuncDef}
  \, .
\end{equation}
In terms of $\SFT$ we can now approximate the self-energy effective action variational principle $\delta_{\bSp} \SE = \delta_{\bS} \SE = 0$ [Eqs.\ (\ref{eq:SEVarPrinc_1}) and (\ref{eq:SEVarPrinc_2})] by constraining the variations to the subspace of self-energies spanned by the reference system, giving the Euler equations
\begin{equation}
  \frac{\delta \SFT}{\delta \bF'^\dagger} = 0
  \, , \quad
  \frac{\delta \SFT}{\delta \bG_0'^{-1}} = 0
  \, .
  \label{eq:SFAVarPrinc}
\end{equation}
If we explicitly perform the variations, using the variational relations of the free energy [Eq.\ (\ref{eq:FreeEnergyFvar}) and (\ref{eq:FreeEnergyG0var})], only the self-energy dependent variations are nonzero, and the Euler equations take the form
\begin{multline}
  0 = \frac{\delta \SFT}{\delta \bF'^\dagger}
  =   
  \frac{\delta \SE}{\delta \bSp'}
  \frac{\delta \bSp'}{\delta \bF'^\dagger}
  +
  \frac{\delta \SE}{\delta \bS'}
  \frac{\delta \bS'}{\delta \bF'^\dagger}
  \\ =
  (\bPhi' - \bPhi)
  \frac{\delta \bSp'}{\delta \bF'^\dagger}
  +
  \frac{1}{2}(\bG' - \bG)
  \frac{\delta \bS'}{\delta \bF'^\dagger}
  \, , \label{eq:SFTVarPrincF}
\end{multline}
\begin{multline}
  0 =
  \frac{\delta \SFT}{\delta \bG_0'^{-1}}
  =
  \frac{\delta \SE}{\delta \bSp'}
  \frac{\delta \bSp'}{\delta \bG_0'^{-1}}
  +
  \frac{\delta \SE}{\delta \bS'}
  \frac{\delta \bS'}{\delta \bG_0'^{-1}}
  \\ =
  (\bPhi' - \bPhi)
  \frac{\delta \bSp'}{\delta \bG_0'^{-1}}
  +
  \frac{1}{2}(\bG' - \bG)
  \frac{\delta \bS'}{\delta \bG_0'^{-1}}
  \, , \label{eq:SFTVarPrincG0}
\end{multline}
where the self-energy variations of $\SE$ are obtained using Eq.\ (\ref{eq:SFAFunc}).
From the form of these equations one can see that the approximate variational principle for $\SFT$ [Eq.\ (\ref{eq:SFAVarPrinc})] corresponds to finding the stationary point of $\SE$ with respect to $\bSp$ and $\bS$ projected onto the plane of reference-system representable self-energies $\bSp = \bSp'$ and $\bS = \bS'$.

The self-energy functional theory approximation $\SFT$ of the self-energy effective action [Eqs.\ (\ref{eq:SFAFunc}) and (\ref{eq:SFAFuncDef})] and its corresponding variational principle [Eqs.\ (\ref{eq:SFTVarPrincF}) and (\ref{eq:SFTVarPrincG0})] are the two main results of this paper.

% --------------------------------------------------------------------
\subsubsection{Reference system}
\label{Sec:Theo_SFT_RefSyst}
% --------------------------------------------------------------------

% - - - - - - - - - - - - - - - - - - - - - - - - - - - - - - - - - - 
\begin{figure}
  \includegraphics[scale=1.0]{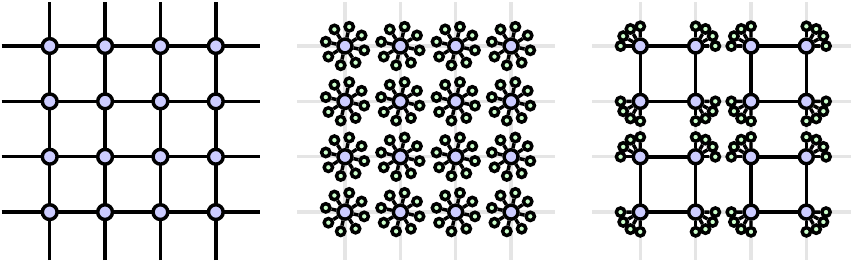}
  \caption{\label{fig:RefSyst}(Color online) Schematic examples of a physical system and two types of reference-system constructions. Left: A two-dimensional square lattice with correlated sites (big blue circles). Center: Reference systems with local non-interacting Green's functions $\bG_{0,ij}' = \delta_{ij} \bG_{0,i}'$ and additional non-interacting bath sites (small green circles). Right: Two-by-two plaquette reference system, with a non-local free-propagator $\bG_0'$ and non-local self-energy $\bS'$.}
\end{figure}
% - - - - - - - - - - - - - - - - - - - - - - - - - - - - - - - - - - 

The versatility of the self-energy functional theory approach lies in the freedom of constructing the reference system. While keeping a subset of lattice sites $i$ with the same interaction vertices as the physical system [$\hat{V}_3$ and $\hat{V}_4$ in Eq.\ (\ref{eq:V3V4})] the reference system's free propagator $\bG_0'$ can be parametrized by hybridizing the interacting lattice sites with non-interacting ``bath sites''. In the case of a two-dimensional square lattice, two such choices of reference systems are shown schematically in Fig.\ \ref{fig:RefSyst}.
In general $\bG_{0,ij}'$ can be written as
\begin{equation}
  \bG_{0,ij}'^{-1}(i\omega_n) =
  \sigma_z i \omega_n - \mbf{1} (\mu  - \bt'_{ij}) - \bDelta_{ij}(i\omega_n)
  \, , 
\end{equation}
where $\bDelta(i\omega_n) = \bDelta_{ij}(i \omega_n)$ is the reference system hybridization function, parametrized by the non-interacting bath sites. Labeling the bath sites with Greek indices and denoting the reference-system hopping with $\bt'$ the hybridization function can be expressed as
\begin{equation}
  \bDelta_{ij}(i\omega_n) =
  \frac{1}{2} \sum_{\alpha \beta} \bt'_{i\alpha} \tilde{\bG}_{0, \alpha \beta}(i\omega_n) \bt'_{\beta j}
  \,, 
\end{equation}
where $\tilde{\bG}_{0, \alpha \beta}^{-1}(i \omega_n) = \sigma_z i \omega_n - \mbf{1} \bt'_{\alpha \beta}$ is the free propagator restricted to the bath sites.
Under these assumptions the reference system can be written in Hamiltonian form 
\begin{equation}
  H'[\bF', \bt'] =
  \hat{V}_3 + \hat{V}_4
  + \sum_i ( \bc_i F'_i + F_i'^* \ba_i )
  + \bbc \bt' \bbb
  \, , 
\end{equation}
where $\bbc$ is a Nambu vector in both correlated and bath sites. As $H'$ comprises a finite number of bosonic states the reference system free energy $\Omega'$ and self-energies $\bSp'$ and $\bS'$ can be calculated using exact diagonalization, see Appendix \ref{app:ReferenceSystem}. With these results the SFT functional $\SFT$ [Eq.\ (\ref{eq:SFAFunc})] can be evaluated and its stationary points with respect to variations in $\bF'$ and $\bt'$, 
\begin{equation}
  \frac{\delta \SFT}{\delta \bF'^\dagger} = 0
  \, , \quad
  \frac{\delta \SFT}{\delta \bt'} = 0
  \, ,
\end{equation}
can be located using, e.g., a multi-dimensional root solver. This general formalism will be applied to the canonical model for interacting lattice bosons, the Bose-Hubbard model, in Sec.\ \ref{sec:BHModel}.

% --------------------------------------------------------------------
\subsubsection{Dynamical mean-field theory limit}
\label{Sec:Theo_SFT_DMFT}
% --------------------------------------------------------------------

The self-energy functional theory approximation contains the bosonic version
\cite{Byczuk:2008nx, Hubener:2009cr, Hu:2009qf, Anders:2010uq, Snoek:2010uq, Anders:2011uq} of dynamical mean-field theory \cite{Georges:1996aa} as a special limit, in direct analogy to the fermionic case \cite{Potthoff:2003aa}.
When allowing the reference system to have a completely general (retarded) but local free propagator $\bG'_{0,ij} = \delta_{ij} \bG'_{0,ii}$, the Euler equations of SFT [Eqs.\ (\ref{eq:SFTVarPrincF}) and (\ref{eq:SFTVarPrincG0})] simplify to the DMFT self-consistency equations [Eqs. (\ref{eq:DMFTF}) and (\ref{eq:DMFTWeiss})]. In terms of the reference system parametrization of the previous section, this amounts to taking the limit of an infinite number of bath sites.

With $\bG_0'$ being local, also the reference system's self-energy is local, $\bS'_{ij} = \delta_{ij} \bS'_{ii}$, and the $\bS'$-variation in the SFT Euler equation [Eq.\ (\ref{eq:SFTVarPrincG0})] reduces to
$2\delta_{\bS'_{ii}} \SE = \bG'_{ii} - \bG_{ii}$.
Furthermore, the retardedness of $\bG_0'$ provides sufficient freedom to fulfill the SFT Euler equations [Eqs.\ (\ref{eq:SFTVarPrincF}) and (\ref{eq:SFTVarPrincG0})] by enforcing that the local Green's functions and the symmetry breaking order parameters of the physical and reference systems are identical
\begin{equation}
  \bPhi' - \bPhi = 0
  \, , \quad
  \bG'_{ii} - \bG_{ii} = 0
  \, .
  \label{eq:DMFTlimitSFT}
\end{equation}
Using the Dyson equation [Eq.\ (\ref{eq:DysonS})] on the last relation directly gives the DMFT self-consistency equation for the reference system's Weiss field $\bG_0'^{-1} = [\bG_{ii}]^{-1} + \bS'_{ii}$ [Eq.\ (\ref{eq:DMFTWeiss})]. The analogous relation for the symmetry breaking order parameters requires the insertion of the Dyson equation for $\bSp'$  [Eq. (\ref{eq:DysonS12})] twice.
In terms of the imagnary-time products defined in Appendix \ref{app:TraceProduct} this reads
\begin{equation}
  0 = \bPhi' - \bPhi
  =
  \bG_0 \left[
    \bG_0^{-1} \bPhi' - \bG_0'^{-1} \bPhi' - \bF + \bF'
    \right],
  \label{eq:DMFTFv2}
\end{equation}
where the relation in brackets is equal to zero, in direct agreement with the DMFT self-consistency relation for the symmetry breaking field $\bF'$ [Eq.\ (\ref{eq:DMFTF})].

% --------------------------------------------------------------------
\subsubsection{Static mean-field theory limit}
\label{Sec:Theo_SFT_MeanField}
% --------------------------------------------------------------------

While dynamical mean-field theory is a specific limit of self-energy functional theory, the static mean-field theory (MFT) approximation \cite{Fisher:1989kl, Sachdev:1999fk} can only be obtained by making one further approximation. 
Contrary to SFT, the static mean-field theory only accounts for the kinetic energy of the bosonic condensate, and neglects all kinetic energy contributions from non-condensed bosons.
Hence, to arrive at MFT from SFT one has to drop the trace log term in the SFT functional [Eq.\ (\ref{eq:SFAFunc})], which accounts for the kinetic energy contributions from non-condensed bosons.
Upon dropping the trace log terms the variations of $\SFT$ [Eqs.\ (\ref{eq:SFTVarPrincF}) and (\ref{eq:SFTVarPrincG0})] reduce to
\begin{equation}
 \left(\bPhi-\bPhi'\right) \frac{\delta \bSp'}{\delta \bF'^\dagger} = 0
  \, , \quad
  \left(\bPhi-\bPhi'\right) \frac{\delta \bSp'}{\delta \bG_0'^{-1}} = 0
  \, ,
  \label{eq:MFVarPrinc}
\end{equation}
which are trivially fulfilled if the lattice and reference systems' one-point propagators are equal, $\bPhi = \bPhi'$.
As the only variational parameter in mean-field theory is the symmetry breaking field $\bF'$ the reference-system free propagator is fixed to  $\bG_0'^{-1} = \sigma_z i\omega_n + \mu \mbf{1}$. Thus, stationarity $\bPhi = \bPhi'$ amounts to inserting $\bG_0'$ in Eq.\ (\ref{eq:DMFTFv2}), which for a homogeneous lattice system with nearest-neighbor hopping $J$ and coordination number $z$, reduces to
\begin{equation}
  \bF' = \bF - (\bG_0^{-1} - \bG_0'^{-1}) \bPhi'
  = \bF - zJ \bPhi'
  \, .
  \label{eq:MF_Selfcon}
\end{equation}
The resulting equation for the reference-system linear symmetry breaking field $\bF'$ is identical to the self-consistency relation of the static mean-field approximation \cite{Fisher:1989kl, Sachdev:1999fk}.

% --------------------------------------------------------------------
\section{The Bose-Hubbard model}
\label{sec:BHModel}
% --------------------------------------------------------------------

To test our generalization of self-energy functional theory to bosons we apply it to the canonical model for interacting lattice bosons, the Bose-Hubbard model \cite{Fisher:1989kl}, which is described by the Hamiltonian
\begin{equation}
  H =
  - J \sum_{\la i,j \ra} ( \bc_i \ba_j + \bc_j \ba_i )
  + \frac{U}{2} \sum_i \bc_i \bc_i \ba_i \ba_i
  - \mu \sum_i \n_i
  \, , \label{eq:BHHamiltonian}
\end{equation}
with nearest neighbor hopping $J$, local pair interaction $U$, and chemical potential $\mu$, where $\bc_i$ ($b_i$) creates (annihilates) a boson at site $i$ and $\n_i = \bc_i b_i$ is the density operator. We will consider the model on the two- and three-dimensional square lattice and study its phase boundaries, observables, and energetics at finite temperature.

% --------------------------------------------------------------------
\subsection{Minimal reference system}
\label{sec:minref}
% --------------------------------------------------------------------

For the reference system we focus on the simplest possible construction, and use a single bosonic state with the Hamiltonian
\begin{equation}
  H'[\bF', \bDelta] =
  \frac{U}{2} \bc \bc b \, b
  - \mu \n
  + \bF'^\dagger \bbb
  + \frac{1}{2} \bbc \bDelta \bba
  \, ,
  \label{eq:SFA3Hamiltonian}
\end{equation}
where $\bbc = (\bc \,\, b)$ is a Nambu operator, and $\bF'$ and $\bDelta$ are defined as
\begin{equation}
  \bF' =
  \left( \begin{matrix}
    F' & F'^* 
  \end{matrix} \right)
  \, , \quad
  \bDelta =
  \left( \begin{matrix}
    \Delta_{00} & \Delta_{01} \\
    \Delta_{01}^* & \Delta_{00}
  \end{matrix} \right)
  \, .
  \label{eq:SFA3Parms}
\end{equation}
Hence, the reference system is parametrized by the three parameters $F'$, $\Delta_{00}$, and $\Delta_{01}$. The linear symmetry breaking field $F'$ is the conjugate variable to the anomalous expectation value $\la b \ra$ while $\Delta_{00}$ and $\Delta_{01}$ are conjugate to the density $\la \bc b \ra$ and the anomalous density $\la b b \ra$, respectively.
In the normal phase the number of variational parameters reduces to only $\Delta_{00}$ as the absence of symmetry breaking requires $F' = \Delta_{01} = 0$.
Henceforth, we will denote this three parameter self-energy functional theory approximation as SFA3.

Clearly, the restriction of the reference system to a single bosonic state is a drastic approximation.
Temporal retardation effects can be treated by adding additional non-interacting bath sites to the reference system, producing additional variational parameters, where in the limit of infinite number of bath-sites the BDMFT solution \cite{Anders:2010uq, Anders:2011uq} is obtained.
However, as we will show, already SFA3 quantitatively describes the Bose-Hubbard model, both deep in the superfluid and the Mott/normal phase.

Note that the SFA3 minimal reference system Hamiltonian in Eq.\ (\ref{eq:SFA3Hamiltonian}) has the same variational degrees of freedom as the reference system employed in Bogoliubov+U theory (B+U) \cite{Hugel:2015ab}.

% --------------------------------------------------------------------
\subsection{Numerical implementation}
\label{sec:numimpl}
% --------------------------------------------------------------------

\begin{figure}
  \includegraphics[scale=1]{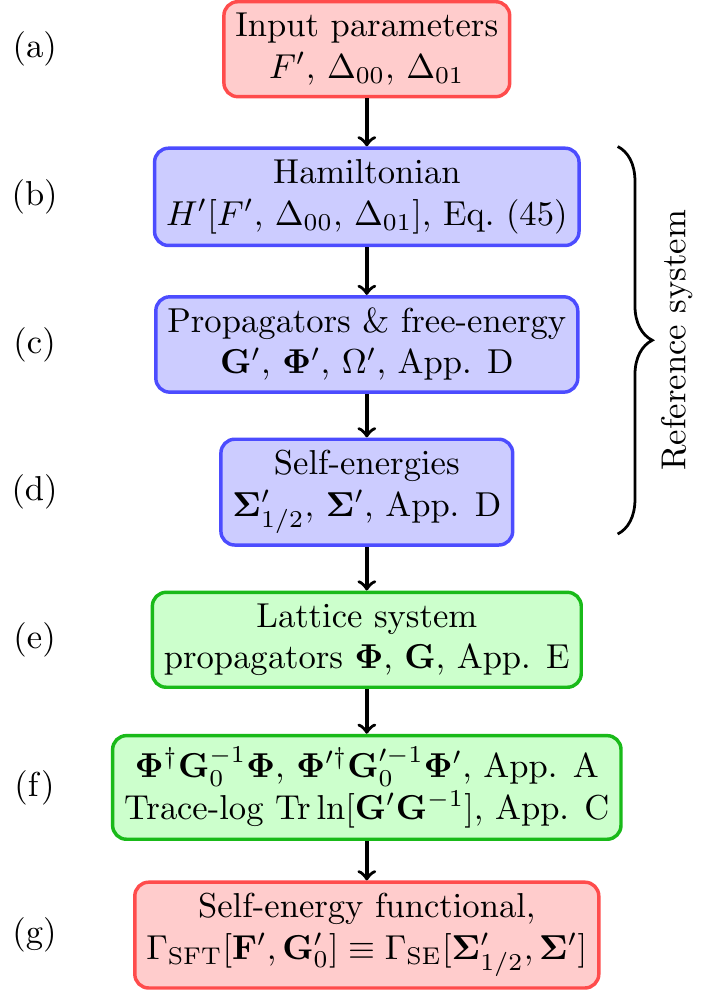}
  \caption{\label{fig:FlowChart}(Color online) Flow chart for the evaluation of the SFT functional $\SFT \equiv \SFT[F', \Delta_{00}, \Delta_{01}]$ for given values of $F'$, $\Delta_{00}$ and $\Delta_{01}$. The functional $\SFT$ is used to numerically locate stationary points $\nabla \SFT = \mbf{0}$.
}
\end{figure}

To find stationary solutions of the SFT functional $\SFT$ [Eq.\ (\ref{eq:SFAFunc})] for the Bose-Hubbard model [Eq.\ (\ref{eq:BHHamiltonian})] and the reference system [Eq.\ (\ref{eq:SFA3Hamiltonian})]
we implement a direct evaluation of $\SFT$ and use a root finder \footnote{The hybrd and hybrj methods of MINPACK as wrapped in SciPy \cite{Jones:2001aa}.} in combination with numerical evaluation of the gradient $\nabla \SFT$ to locate stationary solutions $\nabla \SFT=\mbf{0}$.
The procedure for evaluating $\SFT$ is shown schematically in Fig.\ \ref{fig:FlowChart} and consists of the steps:
(a) starting from given values of $F'$, $\Delta_{00}$ and $\Delta_{01}$,
(b) construct the reference systems Hamiltonian $H'$ using Eq.\ (\ref{eq:SFA3Hamiltonian}),
(c) compute the free-energy $\Omega'$ and the one- and two-point propagators $\bPhi'$ and $\bG'$ of the reference system using Eqs.\ (\ref{eq:ApRefSys1}) to (\ref{eq:ApRef7}) in Appendix \ref{app:ReferenceSystem},
(d) compute the reference system self-energies $\bSp'$ and $\bS'$ using the one- and two-point Dyson Equations (\ref{eq:ApRef9}) and (\ref{eq:ApRef10}) in Appendix \ref{app:ReferenceSystem},
(e) compute the lattice system one- and two-point propagators $\bPhi$ and $\bG$ using the relations in Appendix \ref{app:LatticeSystem},
(f) calculate the products $\bPhi^\dagger \bG_0^{-1} \bPhi$ and $\bPhi'^\dagger \bG_{0}'^{-1} \bPhi'$ using the algebraic rules in Appendix \ref{app:TraceProduct} and the trace log $\Tr \ln [ \bG' \bG^{-1} ]$ using Eq.\ (\ref{eq:TrLnSecondOrderCorrected}) in Appendix \ref{app:TrLn},
and finally (g) evaluate the self-energy functional $\SFT$ using Eq.\ (\ref{eq:SFAFunc}).

In order to achieve high accuracy in the evaluation of $\SFT$, the trace log term in Eq.\ (\ref{eq:SFAFunc}) is evaluated using Eq.\ (\ref{eq:TrLnSecondOrderCorrected}) and third-order high-frequency tail coefficients. Calculations at temperatures $T/J \sim 1$ -- $10$ then require $10^3$ -- $10^4$ Matsubara frequencies in order to reach a relative accuracy of $10^{-9}$, for details see Appendix~\ref{app:TrLn}.
Once as stationary point of $\SFT$ is located in terms of the reference system parameters $F'$, $\Delta_{00}$ and $\Delta_{01}$ (i.e.\ $\nabla \SFT[F', \Delta_{00}, \Delta_{01}] = \mbf{0}$), lattice system observables can be computed as described in Appendix \ref{app:LatticeSystem}.

% --------------------------------------------------------------------
\section{Results}
\label{Sec:Res}
% --------------------------------------------------------------------

% - - - - - - - - - - - - - - - - - - - - - - - - - - - - - - - - - - 
\begin{figure*}
  \includegraphics[scale=1]{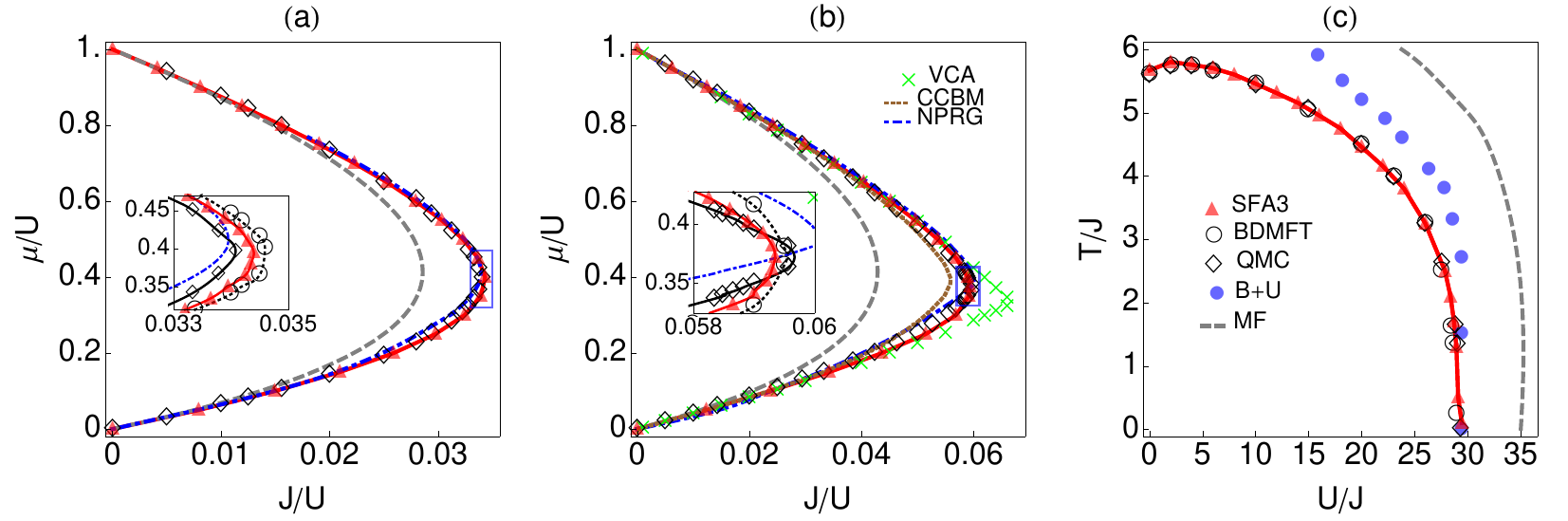}
  \caption{\label{fig:PhaseDiags}(Color online) Phase boundaries for the Bose-Hubbard model, at temperature zero on the three-dimensional [panel (a)] and two-dimensional [panel (b)] cubic lattices, and for unit-filling ($n = 1$) in three dimensions at finite temperature [panel (c)]. The SFA3 results (red triangles) are compared with MFT (dashed gray line), B+U \cite{Hugel:2015ab} (blue dots), QMC \cite{Capogrosso-Sansone:2007lh} (diamonds), BDMFT \cite{Anders:2010uq, Anders:2011uq} (circles), VCA \cite{Knap:2010aa, Knap:2011aa} (green crosses), CCBM \cite{Huerga:2013aa} (dashed brown line) and NPRG  \cite{Ranifmmode-celse-cfion:2011aa, Ranifmmode-celse-cfion:2011ab} (dashed blue line).
The B+U results are suppressed in panel (a) and (b), since they overlap with the QMC data within $1 \%$, for the same reason the BDMFT results are only partially shown [insets in panel (a) and (b)].
}    
\end{figure*}
% - - - - - - - - - - - - - - - - - - - - - - - - - - - - - - - - - - 

The Bose-Hubbard model is an ideal model for benchmarking SFA3 as ample numerical results are already available on the two- and three-dimensional cubic lattices.
In particular, since the model is free of sign problems, quantum Monte Carlo (QMC) \cite{Capogrosso-Sansone:2007lh} provides numerically exact results (after finite size scaling).
However, since SFT is inherently an approximate method we also compare with the other approximate schemes: static mean-field theory (MFT) \cite{Fisher:1989kl}, Bogoliubov+U theory (B+U) \cite{Hugel:2015ab}, bosonic dynamical mean-field theory (BDMFT) \cite{Anders:2010uq, Anders:2011uq}, the pseudo-particle based variational cluster approximation (VCA) \cite{Knap:2010aa, Knap:2011aa}, the cluster composite boson mapping method (CCBM) \cite{Huerga:2013aa}, and the nonperturbative renormalization group (NPRG) \cite{Ranifmmode-celse-cfion:2011aa, Ranifmmode-celse-cfion:2011ab}.

% --------------------------------------------------------------------
\subsection{Superfluid phase boundaries}
\label{Sec:Res_Phase}
% --------------------------------------------------------------------

The zero-temperature SFA3 results for the phase boundary between the superfluid and the Mott-insulator at unit-filling on the three- and two-dimensional lattice are shown in Figs.~\ref{fig:PhaseDiags}a and \ref{fig:PhaseDiags}b, respectively. 

At zero temperature mean-field is already expected to give qualitatively correct results for the three-dimensional lattice \cite{Sachdev:1999fk}.
Quantitatively, however, kinetic fluctuation corrections beyond mean-field stabilize the Mott phase and strongly shift the tip of the unit-filling Mott-lobe to larger $J/U$, see Fig.\ \ref{fig:PhaseDiags}a.
However, as shown in previous BDMFT studies \cite{Anders:2010uq, Anders:2011uq}, local self-energy approximations are sufficient to quantitatively capture these kinetic fluctuations.
Surprisingly our SFA3 results, where kinetic effects are tuneable by only two variational parameters ($\Delta_{00}$ and $\Delta_{01}$), yield the same level of accuracy as BDMFT.
We expect the SFA3 phase-boundary, see inset in Fig.\ \ref{fig:PhaseDiags}a, to move towards the BDMFT result when extending the reference system with additional bath sites.
While SFA3 and BDMFT slightly but systematically over-estimate the critical value of $J/U$, see inset in Fig.\ \ref{fig:PhaseDiags}a, we expect this behavior to diminish when accounting for short-range non-local fluctuations by extending to multi-site (cluster) reference systems.
This is in contrast to methods where both local and non-local fluctuations are treated approximately, such as NPRG, which both over- and under-estimates the cricital $J/U$ depending on $\mu / U$, see inset in Fig.\ \ref{fig:PhaseDiags}a.

The Bose-Hubbard model on the two-dimensional lattice is an even greater challenge for local approximations such as SFA3, as non-local correlations grow in importance with reduced dimension. 
For this model interesting results are available from the two semi-local schemes VCA \cite{Knap:2010aa, Knap:2011aa} and CCBM \cite{Huerga:2013aa}. The VCA results employ an eight-site cluster comprising three edge sharing two-by-two plaquettes and determine the phase-boundary from the closing of the Mott gap \cite{Knap:2010aa}, while the CCBM calculations are performed using a single two-by-two plaquette cluster. 
Hence, both methods require the solution of much more complex effective models than the single-site SFA3 reference system.
However, while SFA3 yields quantitatively correct results, see Fig.\ \ref{fig:PhaseDiags}b, apart from a narrow region at the tip of the Mott-lobe (see inset), VCA and CCBM show large deviations in this region, even though both methods are semi-local and incorporate short-ranged non-local correlations.
This behavior indicates that for the phase transition at the tip of the Mott-lobe treating all kinetic fluctuations with an approximate local self-energy (as in SFA3 and BDMFT) is more important than treating short ranged non-local fluctuations exactly (as in VCA and CCBM).
We also note that while NPRG \cite{Ranifmmode-celse-cfion:2011aa, Ranifmmode-celse-cfion:2011ab} excel over both VCA and CCBM in two dimensions it can not compete with SFA3 and BDMFT. Seemingly the upwards shift in $\mu/U$ of the NPRG phase boundary in the vicinity of the tip of the Mott-lobe becomes more severe with reduced dimension.

On the three-dimensional lattice we further present results on the temperature driven normal to superfluid phase transition at unit-filling ($\la \n \ra = 1$), see Fig.\ \ref{fig:PhaseDiags}c.
Also in this case the phase boundary of SFA3 lies on top of both the BDMFT and QMC results, while MFT and B+U deviates substantially.
SFA3 also captures the weakly interacting Bose gas (WIBG) limit, indicated by a downturn in the critical temperature at low $U/J$. For a detailed discussion in the context of BDMFT see Ref.\ \onlinecite{Anders:2011uq}.
%
% - - - - - - - - - - - - - - - - - - - - - - - - - - - - - - - - - - 
\begin{figure}
  \includegraphics[scale=1]{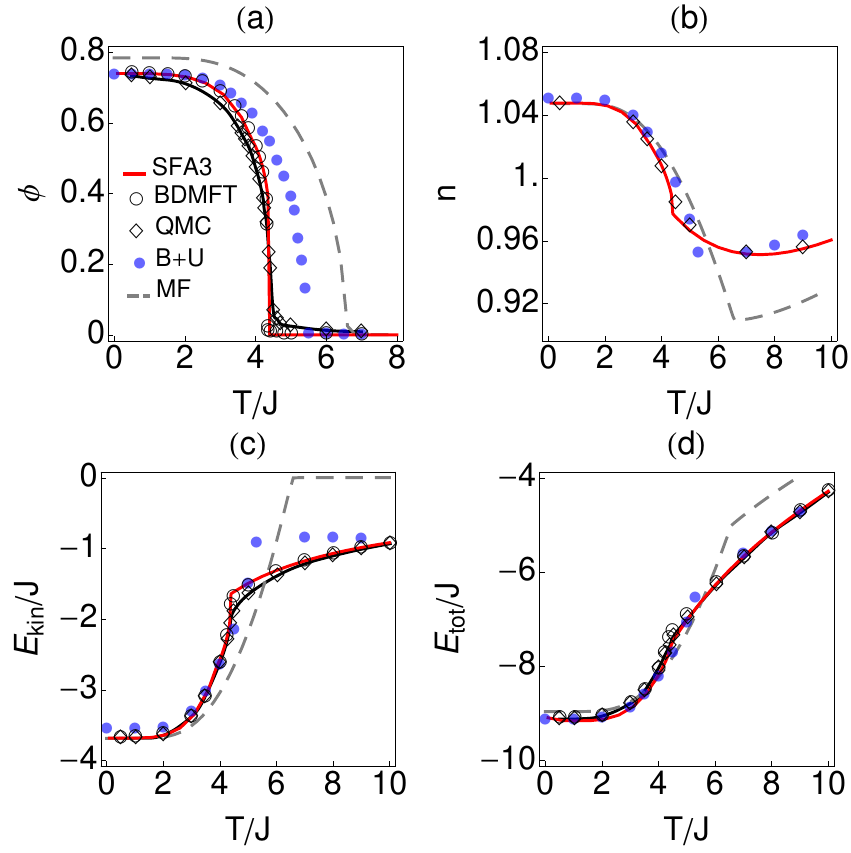}
  \caption{\label{fig:Thermo}(Color online)
    Local observables and energies vs.\ temperature $T$ for the Bose-Hubbard model on the three dimensional cubic lattice with $U/J=20$ $\mu/U=0.4$ ($n \approx 1$). Panel (a): condensate order parameter $\phi$, panel (b): local density $n$, panel (c): kinetic energy $\Ekin$, and panel (d): total energy $\Etot$. Results for SFA3 (red line), MFT (dashed gray line), B+U \cite{Hugel:2015ab} (blue dots), QMC \cite{Capogrosso-Sansone:2007lh} (diamonds), and BDMFT \cite{Anders:2010uq, Anders:2011uq} (circles) are shown. The systematic errors are smaller than the marker size.
  }
\end{figure}
% - - - - - - - - - - - - - - - - - - - - - - - - - - - - - - - - - - 

% - - - - - - - - - - - - - - - - - - - - - - - - - - - - - - - - - - 
\begin{figure*}
  \includegraphics[scale=1]{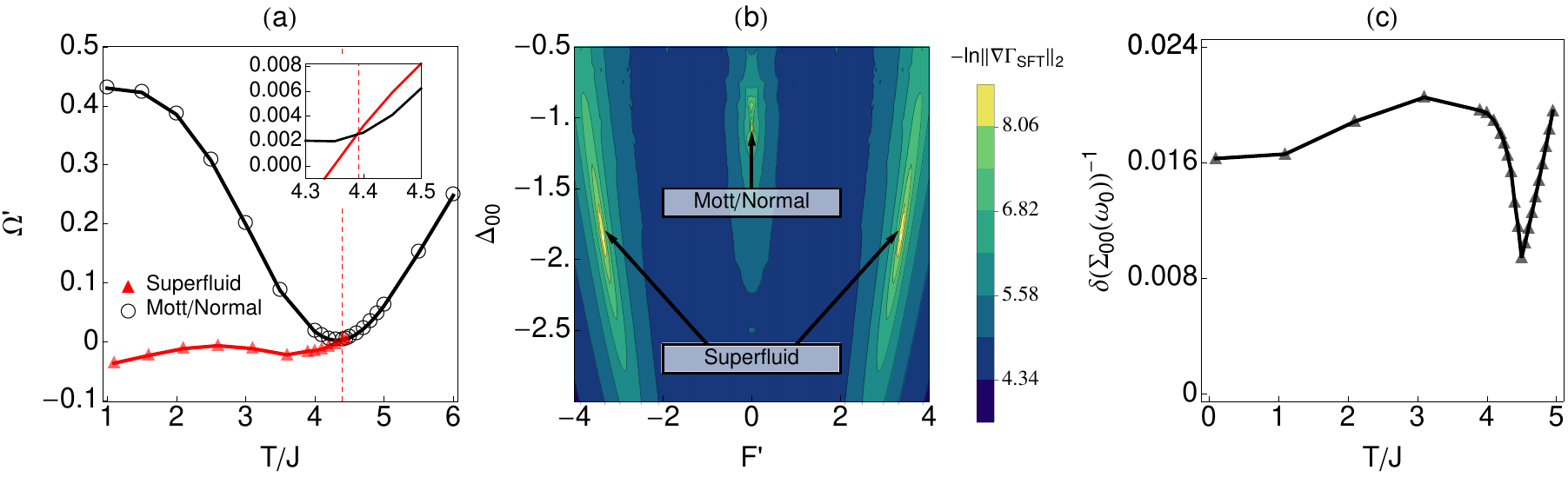}
  \caption{\label{fig:Omega}(Color online)
    Panel~(a): Free energy $\Omega$ vs. temperature $T$ of the Bose-Hubbard model on the three-dimensional cubic lattice at $U/J=20$ and $\mu/U=0.4$ ($n  \approx 1$), showing the SFA3 normal-phase (circles) and superfluid (red triangles) stationary points. To discern the two solutions we subtract a fixed asymptotic model $f(T) = -T/2 -\sqrt{ 0.3 + (T-4.3)^{2}/4} - 10.85$ from $\Omega$ and show $\tilde{\Omega} = \Omega - f(T)$. A detailed view of the crossing (dashed red line) of the free energies is shown in the inset.
    Panel~(b): Gradient map of $\SFT$ as a function of $\Delta_{00}$ and $F'$ with $\Delta_{01}=0$ and $T/J=1$ (deep in the superfluid phase).
    Panel~(c): Relative breaking of the Hugenholtz-Pines relation $\delta/\Sigma_{00}(\mbf{k}=0, i\omega_0)$ vs. temperature $T$. The systematic error is smaller than the marker size.
  }
\end{figure*}
% - - - - - - - - - - - - - - - - - - - - - - - - - - - - - - - - - - 

% --------------------------------------------------------------------
\subsection{Energetics and observables}
\label{Sec:Res_Obs}
% --------------------------------------------------------------------

To further characterize SFA3 we study local observables and energy components of the Bose-Hubbard model on the three-dimensional lattice as a function of temperature at fixed interaction $U/J=20$ and chemical potential $\mu/U=0.4$, see Fig.\ \ref{fig:Thermo}.

The SFA3 superfluid order parameter $\phi = \la b \ra$ reproduces the BDMFT results quantitatively, see Fig.\ \ref{fig:Thermo}a. The phase transition occurs at the SFA3 critical temperature $T_c/J \approx 4.39778$, to be compared to BDMFT ($T_c/J \approx 4.365(3)$) and QMC ($T_c/J \approx 4.43(3)$) \cite{Anders:2011uq}. Note that the QMC results for $\phi$ in Fig.\ \ref{fig:Thermo} are computed for a finite system with $40^3$ sites, yielding a crossover rather than the (thermodynamical limit) phase transition. The QMC critical temperature $T_c$, however, is extrapolated to the thermodynamical limit using finite size scaling \cite{Anders:2011uq}.
We further note that MFT and finite-temperature B+U are not precise in locating the phase transition, as they both over-estimate $T_c$ by more than $20\%$, see Fig.\ \ref{fig:Thermo}.

For the average local density $n = \la \n \ra$, shown in Fig.\ \ref{fig:Thermo}b, we find that SFA3 agrees quantitatively with QMC in both phases, with slight deviations only in the immediate proximity of the phase transition, improving significantly on the MFT and B+U results. 

The kinetic energy $\Ekin$ and total energy $\Etot$ are shown in Fig.\ \ref{fig:Thermo}c and \ref{fig:Thermo}d respectively. The SFA3 result for $\Etot$ is again in quantitative agreement with QMC (and BDMFT). For the kinetic energy $\Ekin$ on the other hand we find a small but discernible deviation of SFA3 from QMC (and BDMFT) close to the phase transition in the normal phase.
This deviation directly shows the difference between accounting for kinetic fluctuations in the normal phase  (where $\Delta_{01} = F' = 0$) using a completely general imaginary-time dependent hybridization function $\Delta(\tau)$ (as in BDMFT) and using a single variational parameter $\Delta_{00}$ (as in SFA3).
However, from Fig.\ \ref{fig:Thermo}c it is evident that the major contribution to the kinetic energy in the normal phase is accounted for by the instantaneous SFA3 variational parameter $\Delta_{00}$. This can be understood from the tremendous difference between SFA3 and the MFT result, where the latter contains zero variational parameters in the normal phase, and thereby produces the atomic limit with zero kinetic energy.

While we find the SFA3 approximation to be quantitatively predictive, the extremely limited variational space is not sufficient to adiabatically connect the weak and deep superfluid, for an indepth discussion see Appendix \ref{App:Sta_Points}. However, as this phenomenon has not been observed for BDMFT we expect it to diminish when adding bath-sites to the reference system.

% --------------------------------------------------------------------
\subsection{Superfluid Phase Transition}
\label{sec:res_free_energy}
% --------------------------------------------------------------------

At the stationary points of the SFT functional $\SFT$ [Eq.\ (\ref{eq:SFAFunc})], the free energy $\Omega$ is directly given by the value of the functional $\SFT$ itself, i.e., $\Omega = \SFT : \delta \SFT = 0$.
The SFA3 free energy $\Omega$ as a function of temperature $T$ for the same parameters as in Fig.\ \ref{fig:Thermo} is shown in Fig.\ \ref{fig:Omega}a.
At low temperatures, SFA3 displays both superfluid and normal-phase stationary points, with the superfluid solution yielding the lowest free energy $\Omega$. At the phase transition ($T_c \approx 4.39778$) the free energies of the two solutions cross, with the superfluid solution vanishing at slightly higher temperatures, whence the transition is weakly first order, for a detailed discussion see Appendix \ref{App:Sta_Points}.
While the phase transition in the Bose-Hubbard model is expected to be second order, the description of the symmetry breaking using a classical field $F'$ is known to change the phase-transition order, see Ref.\ \onlinecite{Pankov:2002aa} for a discussion of the issue in the context of EDMFT.

A map of the stationary points of the SFT functional $\SFT$ as a function of $F'$ and $\Delta_{00}$ can be obtained from the gradient-two-norm-logarithm $-\log || \nabla \SFT ||_2$ which diverges at the stationary points where $\nabla \SFT = \mbf{0}$. As seen in Fig.~\ref{fig:Omega}b, deep in the superfluid phase ($T/J=1$) $\SFT$ shows both a normal-phase stationary point (with $F'=0$) and two symmetry-breaking superfluid stationary points.
The symmetry breaking solutions are both part of the same class of $U(1)$ symmetry breaking solutions with $F'= |F'| e^{i\theta}$, where only $\theta = 0, \pi$ are seen in Fig.\ \ref{fig:Omega}c, as $F'$ is restricted to be real. Furthermore, the mirror symmetry $F' \rightarrow -F'$ in Fig.\ \ref{fig:Omega}c is a direct result of the global $U(1)$ symmetry of $\SFT$, $\SFT[F'] = \SFT[F' e^{i\theta}]$, $\forall \theta \in \Re$.

% --------------------------------------------------------------------
\subsection{Hugenholtz-Pines relation}
\label{Sec:Res_HP}
% --------------------------------------------------------------------

In the superfluid phase of the Bose-Hubbard model the broken $U(1)$ symmetry imposes a constraint on the zero-frequency single-particle Green's function, or equivalently the self-energy. In the continuum this constraint is the well known Hugenholtz-Pines relation \cite{Hugenholtz:1959aa, Rickayzen:1980aa},
\begin{equation*}
  \mu = \Sigma_{00}(\mbf{k}=0, i\omega_0)
  - \Sigma_{01}(\mbf{k}=0, i\omega_0)
  \, ,
\end{equation*}
valid in the symmetry-broken superfluid phase.
On the lattice the Hugenholtz-Pines relation is shifted to
$\mu = \Sigma_{00}(\mbf{k}=0, i\omega_0) - \Sigma_{01}(\mbf{k}=0, i\omega_0) - zJ$ \footnote{Here specialized for lattices with cosine-dispersion, e.g., hypercubic lattices with nearest-neighbor hopping.}, and is known to be weakly broken by local approximations such as MFT, SFT, and BDMFT \cite{Anders:2011uq}.
The relative deviation $\delta / \Sigma_{00}(\mbf{k}=0, i \omega_0)$ of SFA3 from the shifted Hugenholtz-Pines relation is shown in Fig.\ \ref{fig:Omega}c, where $\delta = \mu - \Sigma_{00}(\mbf{k}=0, i\omega_0) + \Sigma_{01}(\mbf{k}=0, i\omega_0) + zJ$.
The deviation is small and comparable to BDMFT \cite{Anders:2011uq}; starting from the critical temperature $T_c$ and going into the superfluid phase it shows an initial increase and then starts to decrease with temperature.

We note in passing that there are methods that obey the Hugenholtz-Pines relation exactly. One example is the nonperturbative renormalization group (NPRG) \cite{Ranifmmode-celse-cfion:2011aa, Ranifmmode-celse-cfion:2011ab}, which approximately treats both local and critical fluctuations.

% --------------------------------------------------------------------
\subsection{Spectral function}
\label{Sec:Res_SF}
% --------------------------------------------------------------------

% - - - - - - - - - - - - - - - - - - - - - - - - - - - - - - - - - - 
\begin{figure}
  \includegraphics[scale=1]{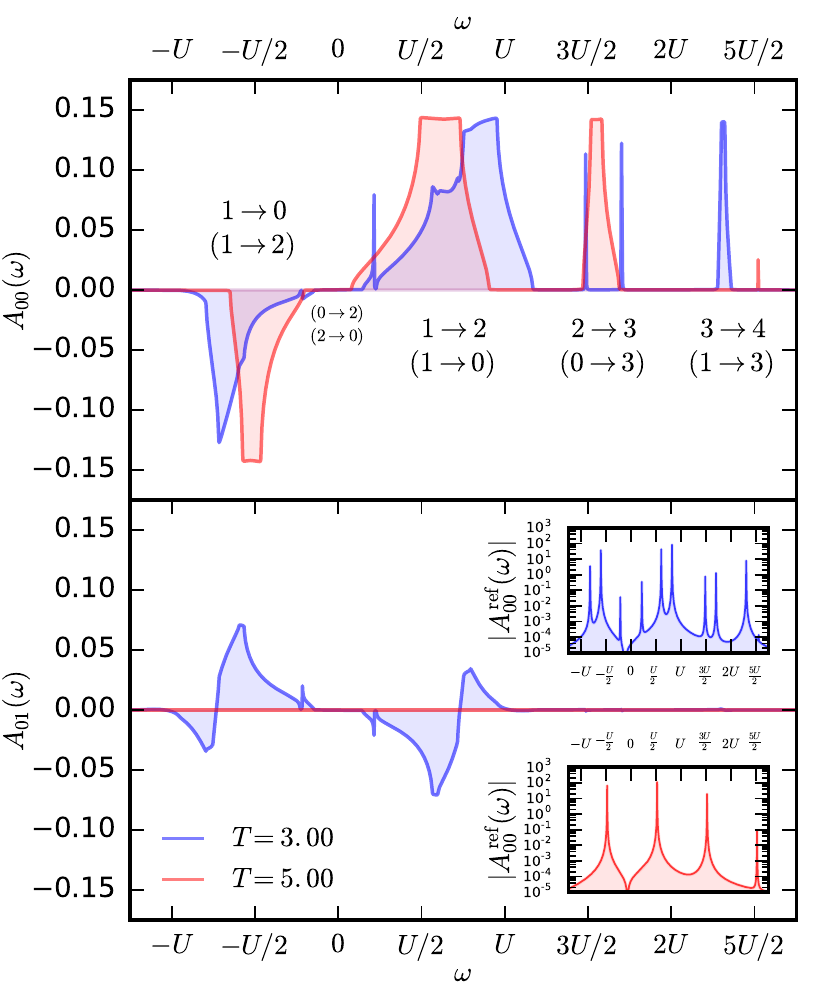}
  \caption{\label{fig:SpecFunc}(Color online)
    Local spectral functions of the Bose-Hubbard model on the three dimensional cubic lattice on both sides of the temperature driven superfluid to normal-phase transition at $U/J=20$, $\mu=0.4 U$, and $T/J=3$ (blue), and $T/J=5$ (red). Both the normal spectral functions $A_{00}(\omega)$ (upper panel) and the anomalous spectral functions $A_{01}(\omega)$ (lower panel) are shown as well as the SFA3 reference system spectral functions $A_{00}^{\rm ref}(\omega)$ (insets). The local transitions corresponding to each spectral feature are indicated, with the symmetry broken allowed transitions in parentheses (upper panel).
  }
\end{figure}
% - - - - - - - - - - - - - - - - - - - - - - - - - - - - - - - - - - 

While lattice quantum Monte Carlo (QMC) provides numerically exact results for sign-problem-free interacting bosonic systems such as the Bose-Hubbard model with nearest-neighbor hopping [Eq.\ (\ref{eq:BHHamiltonian})], this is only true when it comes to thermodynamical expectation values. Dynamical properties, such as the single-particle spectral function, can only be obtained through numerical analytic continuation of imaginary-time results to the real-frequency axis \cite{Jarrell:1996fj, Pippan:2009aa}. This is also an issue in local self-energy approximations such as BDMFT when using a Monte Carlo based reference-system solver, e.g. the continuous time quantum Monte Carlo (CT-QMC) method \cite{Anders:2010uq, Anders:2011uq, Anders:2012kx, Panas:2015ab}. Analytic continuation can resolve the low-energy spectral function but is limited when it comes to resolving high-energy features beyond the first Hubbard bands \cite{Pippan:2009aa, Panas:2015ab}.
Self energy functional theory combined with an exactly solvable reference system such as SFA3, on the other hand, gives direct access to the real-frequency spectral function without any restriction in frequency.

Here we report on the spectral function in the normal-phase using SFA3 and compare with previous results in the low-energy range from BDMFT and CT-QMC \cite{Panas:2015ab} and in the high-energy range from BDMFT and the non-crossing approximation (NCA) \cite{Strand:2015ac}. We also study the finite temperature superfluid spectral function, previously studied in the low-energy range \cite{Pippan:2009aa, Knap:2010aa}, and make an extended analysis of its high-energy resonances.
Entering from the normal phase into the superfluid we find that the high-energy resonances change character and fundamental behavior.

We choose to study the local spectral function in both phases of the temperature driven superfluid to normal-phase transition at $U/J=20$ and $\mu=0.4U$ ($\langle \n \rangle \approx 1$) previously discussed in Sec.\ \ref{Sec:Res_Obs} and Fig.\ \ref{fig:Thermo}.
The normal and anomalous lattice spectral functions, $A_{00}$ and $A_{01}$ respectively, are shown in Fig.\ \ref{fig:SpecFunc}. The normal phase has been studied in detail elsewhere \cite{Strand:2015ac} using BDMFT+NCA and agrees qualitatively with the result we obtain using SFA3.

The features of the lattice spectral function can be understood by studying the corresponding SFA3 reference-system spectral function where the resonances can be understood in terms of transitions between local occupation number states.
In the low energy range the normal phase exhibits a lower and upper Hubbard band, corresponding to singlon-holon ($1 \rightarrow 0$) and singlon-doublon ($1 \rightarrow 2$) transitions, respectively.
The lower band has roughly half the spectral weight of the upper band due to boson prefactors, see Appendix A in Ref.\ \onlinecite{Strand:2015ac}. Beyond the Hubbard bands resonances only occur at positive frequencies as the (local) bosonic states are not bound with respect to the addition of particles. The resonance at $\omega = 3U/2$ only occurs at elevated temperatures and corresponds to a thermally activated doublon-triplon transition ($2 \rightarrow 3$).
Similarly, at $\omega = 5U/2$ we observe a much weaker resonance that, within the SFA3 reference system, is a thermally activated triplon-quadruplon transition ($3 \rightarrow 4$). However, when going beyond SFA3 and adding additional bath sites to the reference system (improving the description of kinetic fluctuations), we expect this resonance to persist down to zero temperature, where it turns into a pure lattice fluctuation of a singlon-triplon with a dispersing holon ($1 \rightarrow 3 \otimes h$), as shown in Ref.\ \onlinecite{Strand:2015ac}.

% - - - - - - - - - - - - - - - - - - - - - - - - - - - - - - - - - - 
\begin{figure}
  \includegraphics[scale=1]{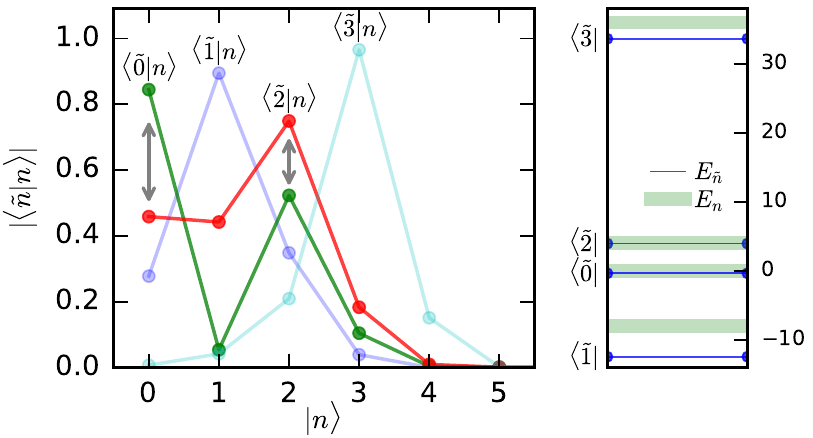}\\[-3mm]
  \caption{\label{fig:SFA3_SF_eigenstates}(Color online) SFA3 reference system eigenstate $| \tilde{n} \rangle$ overlap with the occupation number states $| n \rangle$ (left) in the superfluid phase with parameters identical to Fig.\ \ref{fig:SpecFunc}. The eigenstate energies $E_{\tilde{n}}$ and the zero-hopping limit energies $E_n$ are also shown (right). The symmetry breaking and proximity in energy between $|\tilde{0}\rangle$ and $| \tilde{2} \rangle$ causes both eigenstates to have appreciable weights for both the $|0\rangle$ and $|2\rangle$ occupation number states (arrows in left panel).
  }
\end{figure}
% - - - - - - - - - - - - - - - - - - - - - - - - - - - - - - - - - - 

This picture is heavily modified when entering the superfluid phase.
For $U/J=20$ the superfluid is strongly correlated (i.e.\ $U \gg J, F, \Delta_{01}$) and on the SFA3 reference system level the eigenstates $| \tilde{n} \rangle$ maintain their main occupation number character $| \tilde{n} \rangle \approx |n \rangle$,
whoose eigen energies $E_{\tilde{n}}$ are to first order given by the zero-hopping limit energies $E_{\tilde{n}} \approx E_n \equiv Un(n-1)/2 - \mu n$.
However, the symmetry breaking terms $F$ and $\Delta_{01}$ in the reference-system Hamiltonian [Eq.\ (\ref{eq:SFA3Hamiltonian})] cause the eigenstates to have small but finite admixtures of all other occupation-number states, see Fig.\ \ref{fig:SFA3_SF_eigenstates}.
This turns many more overlaps for one-particle addition $\langle \tilde{n} | b | \tilde{n}' \rangle$ (and one-particle removal $\langle \tilde{n} | \bc | \tilde{n}' \rangle$) non-zero in the symmetry broken phase, as compared to the normal phase where only $\langle n | b | n+1 \rangle$ and $\langle n | \bc | n - 1 \rangle$ contribute in Eq.\ (\ref{eq:LehmannGF}).

On the reference system level, this causes the Green's function [Eq.\ (\ref{eq:LehmannGF})] and thus the spectral function, to exhibit an increased number of resonances in the superfluid phase, see insets in Fig.\ \ref{fig:SpecFunc}.
One example are the two additional low energy resonances at $\pm U/4$, which correspond to the holon-doublon transition ($0 \rightarrow 2$) upon adding \emph{only one particle} and the doublon-holon transition ($2 \rightarrow 0$) removing \emph{only one particle}, respectively. The extra allowed transitions also split the reference system resonances at $\pm U/2$, causing a broadening of the upper and lower Hubbard bands of the lattice system.
The thermally activated doublon-triplon transition at $\omega = 3U/2$ is also split into two separate resonances, that both diminish as temperature is lowered (not shown).
Studying the SFA3 eigenstate overlap with the occupation number states $\langle \tilde{n} | n \rangle$ in Fig.\ \ref{fig:SFA3_SF_eigenstates} we observe a substantial admixture of empty $|0\rangle$ and doubly occupied $|2\rangle$ states in the SFA3 eigenstates $|\tilde{0}\rangle$ and $|\tilde{2}\rangle$. This mixing causes the observed doubling of all resonances in the superfluid phase. In the normal phase on the other hand these resonances either start or end in a simple empty or doubly occupied state.

The high energy resonance at $\omega \approx 5U/2$ also undergoes drastic change, but is not split since no doubly occupied or empty states are involved in the transition. Instead of the thermally activated form found in the normal phase, the resonance carries much more spectral weight and persists down to low temperatures in the superfluid phase. At low temperatures this implies that we have non-zero overlaps $\langle \tilde{n} | \bc | {\rm GS} \rangle$, where $| {\rm GS} \rangle = | \tilde{1} \rangle$ is the many-body groundstate of the SFA3 reference system.
Thus, comparing eigenstate energies we can attribute this resonance to the direct singlon-triplon transition ($1 \rightarrow 3$), which is clearly a forbidden one-particle transition of the SFA3 reference system in the normal phase.

Based on this result and the normal-phase results of Ref.\ \onlinecite{Strand:2015ac} we predict that the singlon-triplon resonance is a fundamental quantum fluctuation of the Bose-Hubbard model, both in the normal and superfluid phase.

% --------------------------------------------------------------------
\subsection{Frustration and next-nearest neighbor hopping}
\label{Sec:Res_NNN}
% --------------------------------------------------------------------

As an example of the broad applicability of SFT we go beyond previous works and investigate the effect of kinetic frustration on the superfluid to normal-phase transition, looking explicitly at effects that are beyond single-site mean-field and out-of-reach for QMC.

To this end we study the Bose-Hubbard model with additional diagonal next-nearest neighbor hopping $J'$ on the two dimensional square lattice, i.e.\
\begin{equation}
H=H_{\rm BH}-J'\sum_{\langle\langle i,j\rangle\rangle}\left(b^{\dagger}_i b^{}_j + b^{\dagger}_j b^{}_i\right)
  \, , \label{eq:NNNBH_JJp}
\end{equation}
where $H_{\rm BH}$ is the Bose-Hubbard Hamiltonian [Eq.\ (\ref{eq:BHHamiltonian})] and $\langle\langle i,j\rangle\rangle$ indicates summing over all next-nearest neighbors. For $J'/J<0$ Eq.\ (\ref{eq:NNNBH_JJp}) exhibits frustration making it inaccessible for QMC, due to a strong sign-problem \cite{Pollet:2012ly}.
The standard mean-field approximation \cite{Fisher:1989kl, Sachdev:1999fk} can be applied,
but the symmetry-breaking mean-field $F$ only depends on the total bandwidth $F = z\left(J+J'\right) \langle b \rangle$, and does not account for the spectral weight distribution within the band.
Hence, for fixed $\mu/U$ and $(J+J')/U$ mean-field yields constant results independent of the $J'/J$ ratio. In other words for finite next-nearest-neighbor hopping $J'$ mean-field is no longer expected to qualitatively describe the superfluid to normal phase transition, as was the case for only nearest-neighbor hopping $J$ in Sec.\ \ref{Sec:Res_Phase}.
%
% - - - - - - - - - - - - - - - - - - - - - - - - - - - - - - - - - - 
\begin{figure}
  \includegraphics[scale=1]{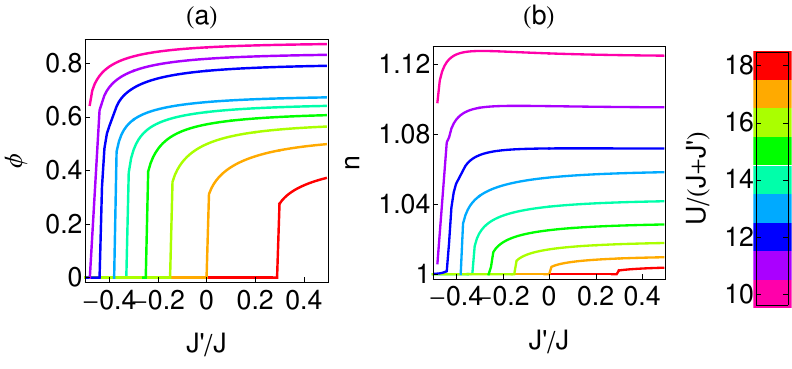}\\[-3mm]
  \caption{\label{fig:T_prime}(Color online)
    Condensate $\phi$ (panel a) and density $n$ (panel b) of the Bose-Hubbard model with next-nearest neighbor hopping $J'$ for $J=1$, $\mu/U=0.4$, $U/(J+J')=10, 11, 12,...18$, and $T=0.01$. The coloring indicates the respective $U/(J+J')$ values (see legend).
  }
\end{figure}
% - - - - - - - - - - - - - - - - - - - - - - - - - - - - - - - - - - 

% - - - - - - - - - - - - - - - - - - - - - - - - - - - - - - - - - - 
\begin{figure}
\includegraphics[scale=1.0]{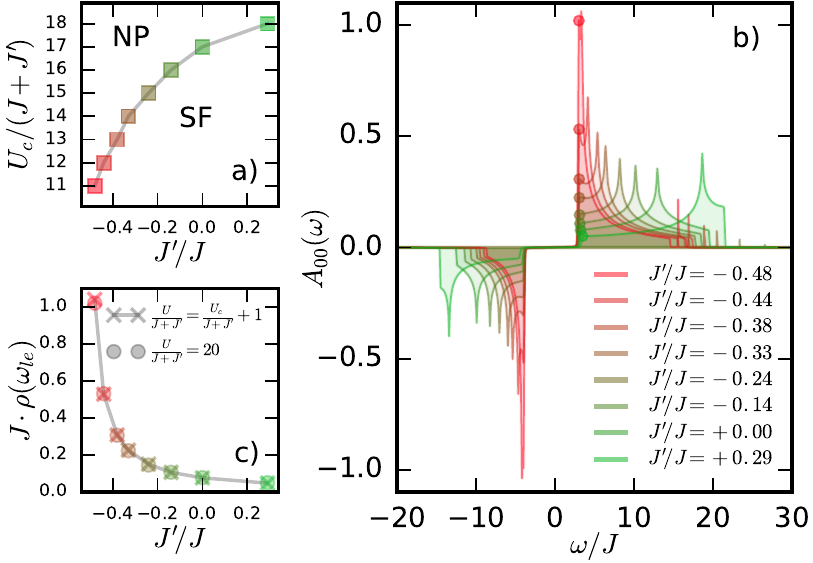}
  \\[-3mm]
  \caption{\label{fig:Jp_sweep_sf}(Color online)
    Critical interaction $U_c/(J+J')$ for the superfluid (SF) to normal phase (NP) transition on the two dimensional lattice with nearest and next-nearest neighbor hopping ($J$ and $J'$) as a function of $J'/J$ (panel a).
    The redistribution of spectral weight driven by $J'/J$ in the spectral function $A_{00}(\omega)$ is shown in panel b for $U/(J+J') = 20$, $\mu/U=0.4$, $T/J=1$, and $J=1$.
   The spectral weight $A_{00}(\omega)$ at the lower Hubbard-band edge anti-correlates with $U_c/(J+J')$ both for $U/(J+J')=20$ (circles) and in the vicinity of the phase transtition at $U/(J+J') = U_c/(J+J') + 1$ (crosses) (panel~c).
  }
\end{figure}
% - - - - - - - - - - - - - - - - - - - - - - - - - - - - - - - - - - 
%
Self-energy functional theory using the single-site reference system (SFA3) incorporates the spectral function and directly depends on the spectral distribution.
In Fig.\ \ref{fig:T_prime} we present SFA3 results for the condensate $\phi$ and local density $n$ for fixed $\mu/U=0.4$ and  $U/(J+J')=10, ..., 18$ as a function of $J'/J$.
We find that both $\phi$ and $n$ vary with $J'/J$ and that the transition between the superfluid and the normal phase strongly depends on $J'/J$.
The critical values of the interaction $U_c/(J + J')$ from the sweeps in Fig.\ \ref{fig:T_prime} are shown in Fig.\ \ref{fig:Jp_sweep_sf}a revealing a drastic reduction of the critical coupling when moving into the frustrated regime $J'/J < 0$.

To understand this trend we study the local spectral function $A_{00}(\omega)$ in the normal phase at $U/(J+J')=20$ and $\mu/U=0.4$, see Fig.\ \ref{fig:Jp_sweep_sf}b.
We see that fixing the bandwidth $z(J+J')/U$ and the chemical potential $\mu/U$ relative to the interaction $U$ corresponds to fixing the gap in $A_{00}(\omega)$.
However, tuning $J'/J$ produces substantial spectral weight redistribution within the Hubbard bands, see Fig.\ \ref{fig:Jp_sweep_sf}b. In particular, the spectral density at the low frequency edges of the Hubbard bands increases as $J'/J$ is lowered. As can be seen in Fig.\ \ref{fig:Jp_sweep_sf}c, this effect prevails also when approaching the phase transition from the normal phase. Thus, we attribute the change in the critical coupling $U_c/(J+J')$ as a function of $J'/J$ to this spectral weight redistribution.
To conclude we find that the enhancement of low energy kinetic fluctuations in the frustrated regime $J'/J < 0$ is detrimental for the superfluid and enhances the extent of the normal-phase.

% --------------------------------------------------------------------
\section{Conclusion and outlook}
\label{Sec:Conc}
% --------------------------------------------------------------------

We derived the self-energy effective action $\SE$ [Eq.\ (\ref{eq:SESigma})] for lattice bosons with $U(1)$ symmetry breaking by Legendre-transforming the bosonic generalization of the Baym-Kadanoff functional $\BK$ [Eq.\ (\ref{eq:BK})] of De Dominicis and Martin \cite{De-Dominicis:1964aa, De-Dominicis:1964ab}.
Our resulting functional $\SE$ differs from the previously derived self-energy effective action in Ref.\ \onlinecite{Arrigoni:2011aa} on the level of the one-point Dyson equation, which in Ref.\ \onlinecite{Arrigoni:2011aa} is inconsistent with the bosonic Baym-Kadanoff functional [Eq.\ (\ref{eq:BK})] and Refs.\ \cite{De-Dominicis:1964aa, De-Dominicis:1964ab}. 

Using the self-energy effective action $\SE$ we derived the generalization of self-energy functional theory (SFT) \cite{Potthoff:2003aa, Potthoff:2003ab, Potthoff:2006aa, Springer:2012} to bosonic systems.
As for fermions, SFT enables to construct non-trivial variational approximations of the self-energy functional using a exactly solvable auxiliary reference system.
We showed that SFT simplifies to bosonic dynamical mean-field theory (BDMFT) \cite{Byczuk:2008nx, Hubener:2009cr, Hu:2009qf, Anders:2010uq, Anders:2011uq} in the limit of a local reference system with an infinite number of local degrees of freedom, while the static mean-field approximation can be obtained by neglecting kinetic contributions of non-condensed bosons.

We apply SFT to the Bose-Hubbard model \cite{Fisher:1989kl, Jaksch:1998vn}, with a minimal reference system comprising a single site with three variational parameters, denoted as ``SFA3'', which was also used in Bogoliubov+U theory \cite{Hugel:2015ab}.

The SFA3 groundstate phase diagrams in two and three dimensions show quantitative agreement with numerically exact quantum Monte Carlo (QMC) \cite{Capogrosso-Sansone:2007lh} and BDMFT results  \cite{Byczuk:2008nx, Hubener:2009cr, Hu:2009qf, Anders:2010uq, Anders:2011uq}.
The accuracy of SFA3 is remarkable, considering the radically restricted variational space.
In fact, while the SFA3 phase-boundaries are in quantitative agreement with QMC, computationally much more demanding cluster based schemes, such as the pseudoparticle variational cluster approximation (VCA) \cite{Knap:2010aa, Knap:2011aa} and the cluster composite boson mapping (CCBM) \cite{Huerga:2013aa}, show substantial deviations at the tip of the Mott-lobe in two dimensions. While to a lesser degree, this is also the case for the nonperturbative renormalization group (NPRG) approach \cite{Ranifmmode-celse-cfion:2011aa, Ranifmmode-celse-cfion:2011ab}.
Within SFA3 we further study local observables and energetics throughout the temperature driven normal phase to superfluid transition in three dimensions, finding excellent agreement with QMC.
Dynamical properties are investigated in terms of the real frequency spectral function in both the normal and superfluid phase near unity filling $\mu/U=0.4$ ($\langle \n \rangle \approx 1$), where we find that the high energy singlon-triplon fundamental quantum fluctuation previously observed in the normal phase \cite{Strand:2015ac} persist in the strongly correlated superfluid phase.

While the benchmark results are very encouraging, the promise of SFT lies in its application to systems that cannot be treated with QMC.
Prime examples of such systems are experimentally realized models with artificial gauge fields \cite{Struck:2012aa, Greschner:2014aa, Goldman:2014aa} and spin-orbit interaction \cite{Lin:2011aa, Struck:2014aa, Jimenez-Garcia:2015aa}, where QMC suffers from a severe sign problem.
As a proof of concept we investigated the frustrated Bose-Hubbard model on the square lattice with next-nearest neighbor hopping using SFA3, finding a drastic suppression of the superfluid phase through enhanced kinetic fluctuations. This particular effect is out-of-reach for both QMC and single-site mean field.

Another interesting venue are bosonic non-equilibrium problems in real time. While SFT already has been extended to real time for fermionic models \cite{Hofmann:2013aa, Hofmann:2015ab}, the bosonic generalization to non-equilibrium would provide an interesting alternative to the recently developed non-equilibrium extension of BDMFT \cite{Strand:2015aa, Strand:2015ac}.

In a broader context the SFT formalism is based on the Baym-Kadanoff effective-action, constructed through two successive Legendre transforms of the free energy \cite{De-Dominicis:1964aa, De-Dominicis:1964ab}. The effect of the transformation is to change the functional dependence from bare to bold one- and two-point propagators. However, as already shown by De Dominicis and Martin \cite{De-Dominicis:1964aa, De-Dominicis:1964ab}, this procedure can be continued to higher orders, by boldifying also the three-point and four-point propagators.
A very interesting venue for future research is to extend SFT to the three-point irreducible vertex. Such a theory should contain the recently proposed local approximation of the dynamical three-leg interaction vertex \cite{Ayral:2015ab} as a special limit.

% --------------------------------------------------------------------
\begin{acknowledgments}
The authors would like to acknowledge fruitful discussions with 
E. Arrigoni,
T. Ayral,
L. Boehnke,
D. Golez,
A. Herrmann,
F. Hofmann,
M. Knap,
J. Panas,
and
M. Potthoff.
Part of the calculations have been performed on the UniFr cluster. HS and PW are supported by FP7/ERC starting grant No.\ 278023. DH and LP by FP7/ERC starting grant No.\ 306897 and FP7/Marie-Curie CIG grant  No.\ 321918.
\end{acknowledgments}
% --------------------------------------------------------------------

\appendix

% --------------------------------------------------------------------
\section{Imaginary time tensor products}
\label{app:TraceProduct}
% --------------------------------------------------------------------

The effective action formalism involves imaginary-time first and second order tensors, such as $\bF$ and $\bG_0$. The product $\mbf{C}^\alpha_\beta(\tau, \tau')$ of two second-order tensors $\mbf{A}_\alpha^\beta(\tau, \tau')$ and $\mbf{B}_\alpha^\beta(\tau, \tau')$, $\mbf{C} = \mbf{A}\mbf{B}$, is defined as the sum over one super-index and an integration in imaginary time
\begin{equation}
  \mbf{C}^\alpha_\beta(\tau, \tau')
  =
  \sum_{\gamma} \int_0^\beta d\bar{\tau} \, 
  \mbf{A}^{\alpha}_\gamma (\tau, \bar{\tau})
  \mbf{B}^\gamma_\beta(\bar{\tau}, \tau')
  \, , \label{eq:CAB}
\end{equation}
while the product $\mbf{R}^\alpha(\tau)$ of a first-order tensor $\mbf{F}^\alpha(\tau)$ and a second order tensor $\mbf{A}_\alpha^\beta(\tau, \tau')$, $\mbf{R} = \mbf{A} \mbf{F}$, is defined as
\begin{equation}
  \mbf{R}^\alpha(\tau) =
  \sum_\gamma
  \int_0^\beta d \bar{\tau} \,
  \mbf{A}^\alpha_\gamma(\tau, \bar{\tau})
  \mbf{F}^\gamma(\bar{\tau})
  \, . \label{eq:RAF}
\end{equation}
Hence, the scalar $S$, given by the sandwiched product $S = \mbf{R}^\dagger \mbf{A} \mbf{F}$ of two first-order tensors $\mbf{R}^\dagger_\alpha(\tau)$ and $\mbf{F}^\alpha(\tau)$ with a second order tensor $\mbf{A}^\alpha_\beta(\tau, \tau')$, becomes
\begin{equation}
  S = \mbf{R}^\dagger \mbf{A} \mbf{F}
  =
  \sum_{\alpha \beta}
  \iint_0^\beta d\tau d\tau' \,
  \mbf{R}^\dagger_\alpha(\tau)
  \mbf{A}^\alpha_\beta(\tau, \tau')
  \mbf{F}^{\beta}(\tau')
  \, . \label{eq:SRAF}
\end{equation}

In equilibrium first-order tensors are time independent, $\mbf{F}^\alpha(\tau) = \mbf{F}^\alpha$, while second-order tensors are time-translation invariant, $\mbf{A}(\tau, \tau') = \mbf{A}(\tau - \tau')$.
Thus, second-order tensors can be transformed to Matsubara frequency space using the relations \cite{Fetter:2003aa, Negele:1998aa, Mahan:2000tg}
\begin{align}
  \mbf{A}^\alpha_\beta (i\omega_n) & =
  \int_0^\beta d\tau
  e^{i\omega_n \tau} \mbf{A}^\alpha_\beta(\tau)
  \, ,
  \label{eq:TauToOmegan}
  \\
  \mbf{A}^\alpha_\beta(\tau) & =
  \frac{1}{\beta} \sum_{n=-\infty}^\infty e^{-i\omega_n \tau}
  \mbf{A}^\alpha_\beta(i\omega_n)
  \, ,
  \label{eq:OmeganToTau}
\end{align}
where $\omega_n = \frac{\pi}{\beta} (2n + \vartheta)$ with $\vartheta = (1 - \xi)/2$ and $\xi = \pm 1$ for bosons and fermions, respectively \cite{Matsubara:1955aa}.
Correspondingly first-order tensors transform like second-quantization operators \cite{Atland:2006nx}
\begin{align}
  \mbf{F}^\alpha(i\omega_n) & =
  \frac{1}{\sqrt{\beta}}
  \int_0^\beta d\tau e^{i\omega_n \tau} \mbf{F}^{\alpha}(\tau)
  \, ,
\\
  \mbf{F}^\alpha(\tau) & =
  \frac{1}{\sqrt{\beta}} \sum_{n=-\infty}^{\infty}
  e^{-i \omega_n \tau} \mbf{F}^\alpha(i \omega_n)
  \, ,
\end{align}
which simplifies to $\mbf{F}^\alpha(\tau) = \mbf{F}^\alpha$ and $\mbf{F}^\alpha(i\omega_n) = \sqrt{\beta} \delta_{n,0} \mbf{F}^\alpha$.

For spatially translation-invariant systems, first-order tensors are position independent, $\mbf{F}^\alpha(\tau) = \mbf{F}^{\mbf{r}_i, \eta}(\tau) \equiv \mbf{F}^\eta(\tau)$, while second-order tensors are invariant under simultaneous translations of the lattice vectors $\mbf{r}_i$ and $\mbf{r}_j$
\begin{equation}
  \mbf{A}^\alpha_\beta(\tau - \tau') =
  \mbf{A}^{\mbf{r}_i, \eta}_{\mbf{r}_j, \nu}(\tau - \tau')
  =
  \mbf{A}_{\nu}^{\eta}(\mbf{r}_i - \mbf{r}_j, \tau - \tau') 
  \, .
\end{equation}
Hence, in momentum space, $\mbf{A}$ is diagonal and given by the transforms
\begin{align}
  \mbf{A}_\nu^\eta(\mbf{k}, \tau) & =
  \sum_{\mbf{r}}
  e^{- i \mbf{k} \cdot \mbf{r}}
  \mbf{A}_{\nu}^{\eta}( \mbf{r}, \tau )
\, , \\
  \mbf{A}_{\nu}^{\eta}(\mbf{r}, \tau) & =
  \frac{1}{N}
  \sum_{\mbf{k}}
  e^{ i \mbf{k} \cdot \mbf{r}}
  \mbf{A}_\nu^\eta(\mbf{k}, \tau)
  \, ,
\end{align}
where $N$ is the number of lattice sites. Accordingly, first-order tensors only contribute at zero momentum
\begin{equation}
  \mbf{F}^\eta(\mbf{k}, \tau) =
  \delta_{\mbf{k}, \mbf{0}} \sqrt{N} \mbf{F}^\eta(\tau)
  \, ,
\end{equation}
as they again transform as \cite{Atland:2006nx}
\begin{align}
  \mbf{F}^\eta(\mbf{k}, \tau) & =
  \frac{1}{\sqrt{N}} \sum_{\mbf{r}} e^{- i \mbf{k} \cdot \mbf{r}} \mbf{F}^\eta(\mbf{r}, \tau)
\, , \\
  \mbf{F}^\eta(\mbf{r}, \tau) & =
  \frac{1}{\sqrt{N}} \sum_{\mbf{k}} e^{i \mbf{k} \cdot \mbf{r}} \mbf{F}^\eta(\mbf{k}, \tau)
  \, .
\end{align}
Thus, in momentum- and Matsubara frequency-space the product relations [Eqs.\ (\ref{eq:CAB}), (\ref{eq:RAF}) and (\ref{eq:SRAF})] simplify to
\begin{align}
  \mbf{C}_\nu^\eta(\mbf{k}, i\omega_n) & =
  \sum_{\mu}
  \mbf{A}^\eta_\mu(\mbf{k}, i\omega_n)
  \mbf{B}^\mu_\nu(\mbf{k}, i\omega_n)
  \, , \\
  \mbf{R}^\eta & =
  \sum_\mu
  \mbf{A}^\eta_\mu(\mbf{k}=\mbf{0}, i\omega_0) \mbf{F}^\mu
  \, , \\
  S & =
  \beta N
  \sum_{\eta\nu}
  \mbf{R}^\dagger_\eta
  \mbf{A}^\eta_\nu(\mbf{k} = \mbf{0}, i \omega_0)
  \mbf{F}^\nu
  \, .
\end{align}

% --------------------------------------------------------------------
\section{Imaginary time tensor traces}
\label{app:NumericalMatsubaraTrace}
% --------------------------------------------------------------------

Also the trace of a second-order tensor appears in the Baym-Kadanoff effective action and can be defined as the trace over super-indices and a double integral in imaginary time 
\begin{equation}
  \Tr [ \mbf{A} ]
  =
  \sum_{\gamma}
  \iint_0^\beta d\tau d\tau' \,
  \delta_\gamma(\tau - \tau')
  \mbf{A}^\gamma_\gamma(\tau, \tau')
  \, , \label{eq:TrA}
\end{equation}
where $\delta_\gamma(\tau)$ is the Nambu kernel $\delta_\alpha(\tau) = \delta_{\mbf{r}_i, \eta}(\tau) = \delta(\tau - (-1)^\eta 0^+)$ enforcing normal ordering (with Nambu index $\eta = 0$, $1$).
Imposing time- and spatial-translational invariance in Eq.\ (\ref{eq:TrA}) yields
\begin{multline}
  \Tr[ \mbf{A} ] =
  \beta N \sum_\mu \mbf{A}^\mu_\mu(\mbf{r} = \mbf{0}, \tau=(-1)^\mu 0^-)
  \\ =
  \sum_{\mu \mbf{k} n}
  e^{i\omega_n (-1)^\mu 0^+}
  \mbf{A}^\mu_\mu(\mbf{k}, i\omega_n)
  \, . \label{eq:TrAMatsubara}
\end{multline}
Note that the trace definition obeys the cyclicity conditions $\Tr[\mbf{A} \mbf{B}] = \Tr[ \mbf{B} \mbf{A}]$ and $S = \mbf{R}^\dagger \mbf{A} \mbf{F} = \Tr[ \mbf{A} \mbf{F} \mbf{R}^\dagger]$.

The fact that the Nambu-kernel $\delta(\tau - (-1)^\eta 0^+)$ is necessary to yield normal ordering can be understood by taking the trace of a Nambu Green's function $\bG$, given by the time-ordered expectation value $\bG^\eta_\nu(\tau) = - \la \bbb^\eta (\tau) \bbc_\nu \ra$,
\begin{equation}
  \Tr[ \bG ] =
  - \beta \sum_\mu \la \bbb^\mu( (-1)^\mu 0^- ) \bbc_\mu \ra 
  = - 2 \beta \la \bc b \ra
  \, .
%  \, ,
\end{equation}
Here the factor $\beta$ comes from the integrals over imaginary time, see Eq.\ (\ref{eq:TrA}), and the factor of two comes from the sum over Nambu indices.
Thus, the Nambu kernel $\delta_{\gamma}(\tau - \tau')$ in the imagnary-time trace $\Tr[ \cdot ]$ is normal-ordering both diagonal Nambu components of $\bG$, producing the second quantization normal ordered result.
This property is central for obtaining the correct free energy contribution from the $\frac{1}{2} \Tr\ln[-\bG^{-1}]$ terms in the Baym-Kadanoff functional [Eq.\ (\ref{eq:BK})], as will be shown in Appendix \ref{app:TrLn}.

% --------------------------------------------------------------------
\subsection{Reformulation using Matsubara asymptotic form}
% --------------------------------------------------------------------

One possible route for the numerical evaluation of the trace is to compute the sum over Matsubara frequencies in Eq.\ (\ref{eq:TrAMatsubara})
\begin{equation}
  \Tr[ \mbf{A} ] =
  \sum_{\mu \mbf{k}, n} e^{i \omega_n (-1)^\mu 0^+}
  \mbf{A}^\mu_\mu(\mbf{k}, i\omega_n)
  \, .
\end{equation}
However, in general, second-order tensors decay slowly with respect to $|\omega_n|$, with the asymptotic behavior $\mbf{A}^\mu_\mu \sim (i\omega_n)^{-1}$ whenever $\mbf{A}$ has a discontinuity at $\tau = 0$.
For a Green's function $\bG = - \la \bbb(\tau) \bbc \ra$ this is generated by the time-ordering operator $\mathcal{T}$ in the expectation value $\la \cdot \ra = \Z^{-1} \Tr[ \mathcal{T} e^{-S} \cdot ]$ and the (equal time) commutation relation $[\bbb, \bbc] = \sigma_z$.

To improve the convergence properties of the Matsubara frequency sum we introduce $\mathcal{A}$, the $N$th order high-frequency expansion of $\mbf{A}(i \omega_n) = \mathcal{A}(i\omega_n) + \mathcal{O}( [i \omega_n ]^{-(N+1)} )   $ given by
\begin{equation}
  \mathcal{A}(i\omega_n) = \sum_{p=1}^N \mbf{a}_p Q_p(i\omega_n)
  \, , \label{eq:HiFreqExpA}
\end{equation}
where $\mbf{a}_p$ are the high-frequency expansion coefficients of $\mbf{A}$ and $Q_p(i\omega_n)$ are the high frequency basis functions
\begin{equation}
  Q_p(i \omega_n ) =
  \left\{ \begin{array}{ll}
    (i\omega_n)^{-p} , & \omega_n \ne 0 \\
    0, & \omega_n = 0
  \end{array} \right.
  \, ,
  \label{eq:QpBasisFunctionsMatsubara}
\end{equation}
with the zeroth frequency mode removed.

Given $\mathcal{A}$ the trace of $\mbf{A}$ can be written as
\begin{multline}
  \Tr[\mbf{A}] =
  \sum_{\mu \mbf{k}} \Big(
  \sum_{n}
  [ \mbf{A}(\mbf{k}, i \omega_n) - \mathcal{A}(\mbf{k}, i\omega_n) ]^\mu_\mu
  \\ +
  \beta \mathcal{A}^\mu_\mu(\mbf{k}, \tau = (-1)^\mu 0^-)
  \Big)
  \, , \label{eq:TraceAsymptoticGeneral}
\end{multline}
where the summand in Matsubara frequency sum on the first row now decays as $[\mbf{A} - \mathcal{A}]^\mu_\mu \sim (i \omega_n)^{-N}$. Hence, the regularizing exponent factor $\exp(i\omega_n (-1)^\mu 0^+)$ is no longer needed.
The improved decay in the sum comes at the price of having to evaluate $\mathcal{A}$ in imaginary time
\begin{equation}
  \mathcal{A}^\mu_\mu(\mbf{k}, \tau = (-1)^\mu 0^-)
  =
  \sum_{p=1}^N [\mbf{a}_p(\mbf{k})]^\mu_\mu Q_p(\tau = (-1)^\mu 0^-)
  \, , \label{eq:AAsymptotic}
\end{equation}
which can be done analytically using the imaginary time form of $Q_p$ derived in Appendix \ref{sec:Qp}, see Eq.\ (\ref{eq:QpZeroTimeLimit}).

The asymptotic form can be used for constructing numerical approximants and to derive alternate analytic formulas for the trace.
For the latter, only the first-order term of the expansion is needed. Given $\mbf{A}(\mbf{k}, i \omega_n)$ and the first expansion coefficient $\mbf{a}_1(\mbf{k})$, Eq.\ (\ref{eq:TraceAsymptoticGeneral}) gives
\begin{equation}
  \Tr[\mbf{A}]
  =
  \sum_{\mu \mbf{k}} \Big(
  \sum_{n}
  \mbf{A}^\mu_\mu(\mbf{k}, i \omega_n)
  +
  \frac{\beta}{2} [\mbf{a}_1(\mbf{k})\sigma_z]^\mu_\mu
  \Big)
  \, , \label{eq:TraceAsymptoticFirstOrder}
\end{equation}
where we have used that the Matsubara sum over the asymptotic form including only the first order term is zero,
$\sum_n \mathcal{A}(\mbf{k}, i\omega_n) = \mbf{a}_1(\mbf{k}) \sum_n Q_1(i\omega_n) = \mbf{0}$, 
and that $\mbf{a}_1(\mbf{k})$ is diagonal, which gives $\left[\mbf{a}_1(\mbf{k})\right]^\mu_\mu Q_1(\tau = (-1)^\mu 0^-) =\left[\mbf{a}_1(\mbf{k})\sigma_z  / 2\right]^\mu_\mu$, see Eq.\ (\ref{eq:QpZeroTimeLimit}).

To numerically calculate tensor traces given $\mbf{A}(i\omega_n)$ at a finite number $N_\omega$ of Matsubara frequencies and $N$ high-frequency expansion coefficients $\mbf{a}_p$, the trace is readily approximated as the finite frequency sum
\begin{multline}
  \Tr[\mbf{A}] \approx
  \sum_{\mu \mbf{k}} \Bigg(
  \mbf{A}^\mu_\mu(\mbf{k}, i \omega_0)
  \\ +
  \sideset{}{'}\sum_{n=-N_\omega}^{N_\omega}  
  \bigg[
  \mbf{A}(\mbf{k}, i \omega_n)
    - \sum_{p=1}^N  \frac{\mbf{a}_p(\mbf{k})}{(i\omega_n)^p}
  \bigg]^\mu_\mu
  \\ +
  \beta \sum_{p=1}^N  \big[\mbf{a}_p(\mbf{k}) \big]^\mu_\mu Q_p(\tau = (-1)^\mu 0^-)
  \Bigg)
  \, ,
  \label{eq:TraceNumeric}
\end{multline}
which converges asymptotically as $\sim 1/N_\omega^{N+1}$ and where the primed sum excludes the zeroth term ($n = 0$). For the SFA3 calculations presented here we use second order tail corrections ($N=2$) see Appendix \ref{app:TrLn}.

% --------------------------------------------------------------------
\subsection{High-frequency basis functions in imaginary time}
\label{sec:Qp}
% --------------------------------------------------------------------

\begin{figure}
  \includegraphics[scale=1.0]{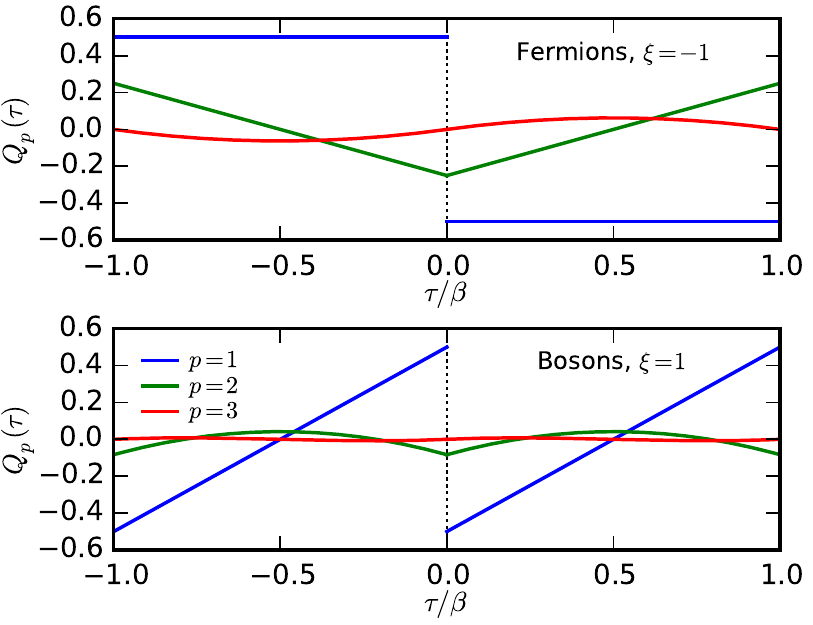}
  \caption{\label{fig:Qp}(Color online) High-frequency basis functions $Q_p(\tau)$ for Fermions (upper panel) and Bosons (lower panel).}
\end{figure}

The imaginary time form of the high-frequency basis functions $Q_p$ in Eq.\ (\ref{eq:QpBasisFunctionsMatsubara}) is given by the Fourier transform [Eq.\ (\ref{eq:OmeganToTau})]
\begin{equation*}
  Q_p(\tau) =
  \frac{1}{\beta} 
  \sideset{}{'}\sum_{n=-\infty}^\infty
  \frac{e^{- i\omega_n \tau}}{(i\omega_n)^p}
  =
  \sideset{}{'}\sum_{n = -\infty}^\infty
  \!\! \textrm{Res} \left[ \frac{ e^{(\beta - \tau)z}}{z^p} f(z), i\omega_n \right]
  \, ,
\end{equation*}
where the summand has been rewritten as a residue and $f(z)$ is the distribution function $f(z) = (e^{\beta z} - \xi)^{-1}$. The sum of residues is related to the contour integral
\begin{equation*}
 0 = \oint_C \frac{dz}{2\pi i} \frac{\xi e^{(\beta - \tau)z}}{z^p} f(z)
   =
  \! \! \! \sum_{n = -\infty}^\infty
  \! \! \! \textrm{Res} \left[ \frac{\xi e^{(\beta - \tau)z}}{z^p} f(z), i\omega_n \right]
  \, ,
\end{equation*}
whence $Q_p$ is given by the $n=0$ term
\begin{equation}
  Q_p(\tau) = 
  - \textrm{Res} \left[ \frac{e^{(\beta - \tau)z}}{z^p} f(z), 0 \right]
  \, . \label{eq:QpIntermediate}
\end{equation}
To evaluate the residue, bosons and fermions must be separated, as $f(z)$ contains a simple pole in the Bosonic case. For Fermions (with $\xi = -1$) Eq.\ (\ref{eq:QpIntermediate}) becomes
\begin{equation}
  Q_p(\tau) = 
  - \frac{1}{(p-1)!} \left(\frac{d}{dz} \right)^{p-1} 
  e^{(\beta - \tau)z} f(z)
  \bigg|_{z \rightarrow 0}
  \, ,
\end{equation}
whose first orders are
$Q_1(\tau) = - 1/2$, $Q_2(\tau) = ( 2\tau - \beta ) / 4$, and $Q_3(\tau) = \tau ( \beta - \tau ) / 4$ for $\tau \in (0, \beta]$. While on $\tau \in [-\beta ,\beta]$ the functions are anti-periodic, and the first order is a step function, see the upper panel in Fig.\ \ref{fig:Qp}.
For bosons one obtains
\begin{equation}
  Q_p(\tau) = 
  - \frac{1}{p!} \left(\frac{d}{dz} \right)^{p} 
  z e^{(\beta - \tau)z} f(z) 
  \bigg|_{z \rightarrow 0}
  \, ,
\end{equation}
yielding the first order terms
\begin{align}
Q_1(\tau) & = (-\beta + 2 \tau)/(2 \beta) \, , \\
Q_2(\tau) & = ( -\beta^2 + 6 \beta \tau - 6\tau^{2} ) /(12\beta) \, , \\
Q_3(\tau) & = \tau ( \beta^{2} - 3 \beta \tau + 2 \tau^{2} )/(12 \beta) \, .
\end{align}
The bosonic functions are periodic on $\tau \in [-\beta, \beta]$ and the first order term is a saw-tooth function, see lower panel in Fig.\ \ref{fig:Qp}.
From the bosonic basis functions $Q_p(\tau)$ we readily obtain the zero-time limits 
\begin{align}
  Q_1(\tau = (-1)^\eta 0^-) & = (-1)^\eta/2 \, , \nonumber \\ 
  Q_2(\tau = (-1)^\eta 0^-) & = - \beta/12 \, , \nonumber \\
  Q_3(\tau = (-1)^\eta 0^-) & = 0 \, ,
  \label{eq:QpZeroTimeLimit}
\end{align}
which are used in the Matsubara sum asymptotic expansion in Eqs.\ (\ref{eq:TraceAsymptoticGeneral}), (\ref{eq:AAsymptotic}), and (\ref{eq:TraceNumeric}).

% - - - - - - - - - - - - - - - - - - - - - - - - - - - - - - - - - - 
% - - - - - - - - - - - - - - - - - - - - - - - - - - - - - - - - - - 

% --------------------------------------------------------------------
\section{Matsubara trace logarithm}
\label{app:TrLn}
% --------------------------------------------------------------------

Apart from direct traces of second order tensors the Baym-Kadanoff functional in Eq.\ (\ref{eq:BK}) also contains the term $\Tr\ln[-\bG^{-1}]$, i.e., the trace of the functional logarithm of the interacting Green's function. In the non-interacting limit only this term remains and yields the free energy up to an infinite regularization factor.

To derive a closed formula for the trace log and the regularization factor reproducing the non-interacting limit we introduce the trace log functional $\Lambda[\bG]$,
\begin{equation}
  \beta \Lambda[\bG]
  = \frac{1}{2}\Tr\ln[-\bG^{-1}] -C_{\infty}
  \, ,
  \label{eq:TraceLogFunctional}
\end{equation}
where $C_\infty$ is an infinite constant, $C_{\infty} = \frac{1}{2} \Tr \ln [-\mbf{R}^{-1} ]$, defined in terms of the regularizing second-order tensor
\begin{equation}
  \mbf{R}(i\omega_n )
  =
  \left\{ \begin{array}{cl}
    -\beta \mbf{1}, & \omega_n = 0 \\[2mm]
    \displaystyle \frac{\sigma_z}{i\omega_n}, & \omega_n \ne 0 
  \end{array} \right.
  \, .
  \label{eq:TrLnReguariationFactor}
\end{equation}
As we will see (Eq.\ (\ref{eq:Zero_int_Free})), this definition of $C_{\infty}$ imposes that the trace log functional $ \Lambda[\bG]$ correctly yields the free energy in the non-interacting case. Since $C_{\infty}$ is constant it will have no effect on the variations of the SFT functional.

Using Eq.\ (\ref{eq:TrAMatsubara}) to write the logarithm trace in momentum- and Matsubara-frequency-space gives
\begin{multline}
  \beta \Lambda[\bG] = 
  - \frac{1}{2}
  \sum_{\mu \mbf{k} n}
  e^{i \omega_n (-1)^\mu 0^+}
  \Big(
    \ln [ \mbf{R}^{-1}(i\omega_n) \bG(\mbf{k}, i\omega_n)  ]
  \Big)^\mu_\mu
  \\
  =
  - \frac{1}{2}
  \sum_{\mu \mbf{k}} \Big(
  \ln [ - \bG(\mbf{k}, i\omega_0) / \beta ]
  +
  \sideset{}{'}\sum_{n}
  e^{i \omega_n (-1)^\mu 0^+}
  \\ \times
  \ln [ (\sigma_z i\omega_n) \bG(\mbf{k}, i\omega_n)  ]
  \Big)^\mu_\mu
  \, .
\end{multline}
To get rid of the exponential convergence factor we use the high-frequency expansion of the logarithm
\begin{multline}
  \ln [ (\sigma_z i\omega_n) \bG(\mbf{k}, i\omega_n)  ]
  =
  \frac{\sigma_z \mbf{c}_2(\mbf{k})}{ i\omega_n }
  \\ + \frac{ \sigma_z \mbf{c}_3 - (\sigma_z \mbf{c}_2)^2 / 2} { (i\omega_n)^2 }
  + \mathcal{O}( [i\omega_n]^{-3} )
  \, ,
  \label{eq:LnGTailExpansion}
\end{multline}
where $\mbf{c}_1 = \sigma_z$, $\mbf{c}_2$ and $\mbf{c}_3$ are the three first coefficients in the high-frequency expansion of $\bG$ [Eq.\ (\ref{eq:HiFreqExpA})].

Using the first order correction and Eq.\ (\ref{eq:TraceAsymptoticFirstOrder}) therefore yields $\Lambda[\bG]$ as
\begin{multline}
  \beta \Lambda [\bG]
  =
  - \frac{1}{2}
  \sum_{\mu \mbf{k}} \Big(
  \ln [ - \bG(\mbf{k}, i\omega_0) / \beta ]
  \\ +
  \sideset{}{'}\sum_{n}
  \ln [ (\sigma_z i\omega_n) \bG(\mbf{k}, i\omega_n)  ]
  +
  \mbf{c}_2(\mbf{k}) \frac{\beta}{2}
  \Big)^\mu_\mu
  \, .
  \label{eq:TrLnFirstOrderCorrected}
\end{multline}

To show that the trace log functional $\beta \Lambda[\bG]$ is correctly regularized as to reproduce the non-interacting limit we consider the free Green's function
$\bG_0(i\omega_n) = [ \sigma_z i\omega_n - \mbf{1} \epsilon_{\mbf{k}} ]^{-1}$,
and its high-frequency expansion
\begin{equation}
  \bG_0(i\omega_n) = \frac{\sigma_z}{i\omega_n} + \frac{\mbf{1} \epsilon_{\mbf{k}}}{(i\omega_n)^2}
  + \mathcal{O}( [i\omega_n]^{-3})
  \, .
\end{equation}
Hence, $\bG_0$ yields $\mbf{c_2}(\mbf{k}) = \mbf{1} \epsilon_{\mbf{k}}$ and $\bG_0(\mbf{k}, i\omega_0) = -\mbf{1} / \epsilon_{\mbf{k}}$ which inserted in Eq.\ (\ref{eq:TrLnFirstOrderCorrected}) gives
\begin{multline}
  \beta \Lambda[\bG] =
  - \sum_{\mbf{k}} \Big(
  - \ln [ \beta \epsilon_{\mbf{k}} ]
  +
  \sideset{}{'}\sum_{n}
  \ln \left[ \frac{i\omega_n}{i\omega_n - \epsilon_{\mbf{k}}} \right]
  +
  \frac{\beta \epsilon_{\mbf{k}}}{2}
  \Big)
  \\ =
  - \sum_{\mbf{k}}
  \ln \left[
    \frac{e^{\frac{\beta \epsilon_{\mbf{k}}}{2} }}{ \beta \epsilon_{\mbf{k}}}
    \sideset{}{'}\prod_n
    \frac{i\omega_n}{i\omega_n - \epsilon_{\mbf{k}}}
    \right]
  =
  \sum_{\mbf{k}} \ln ( 1 - e^{-\beta \epsilon_{\mbf{k}}} )
  \, ,
\end{multline}
where in the last step we have used the relation $ (\beta \epsilon)^{-1} \prod'_{n} \frac{i\omega_n}{i\omega_n - \epsilon} = [2 \sinh (\beta \epsilon /2 )]^{-1}$, see e.g.\ Ref.\ \cite{Kleinert:2009aa}. Thus, for a free Green's function $\bG_0$ the regularized trace log functional $\Lambda[\bG_0]$ is equal to the non-interacting bosonic free energy $\Omega_0$ \cite{Negele:1998aa}
\begin{equation}
  \beta \Lambda[\bG_0] =
  \sum_{\mbf{k}} \ln ( 1 - e^{-\beta \epsilon_{\mbf{k}}} )
  = \beta \Omega_0
  \, ,
  \label{eq:Zero_int_Free}
\end{equation}
confirming the ansatz in Eq.\ (\ref{eq:TrLnReguariationFactor}) for the regularizing tensor $\mbf{R}$.
An intuitive understanding of $\mbf{R}$ can be obtained by rewriting $\beta \Lambda[\bG]$ in Eq.\ (\ref{eq:TraceLogFunctional}) using functional determinants
\begin{equation*}
  \beta \Lambda[\bG] =
  \ln \left[
    \frac{
      \det \sqrt{ - \bG^{-1} }
    }{
      \det \sqrt{ - \mbf{R}^{-1} } 
    }
    \right]  
  =
  \ln \left[
    \frac{
      \det \sqrt{ - \bG^{-1} }
    }{
      \beta^{-1} \det' \sqrt{ \sigma_z \partial_\tau } 
    }
    \right]
  \, .
\end{equation*}
I.e., the regularization $\det \sqrt{ - \mbf{R}^{-1} }$ corresponds to the functional determinant of the free inverse propagator, $\det \sqrt{ - \mbf{R}^{-1} } = \beta^{-1} \det' \sqrt{ \sigma_z \partial_\tau }$, where the primed determinant indicates removal of all nullspace-eigenmodes of the argument.

\begin{figure}
  \includegraphics[scale=1.0]{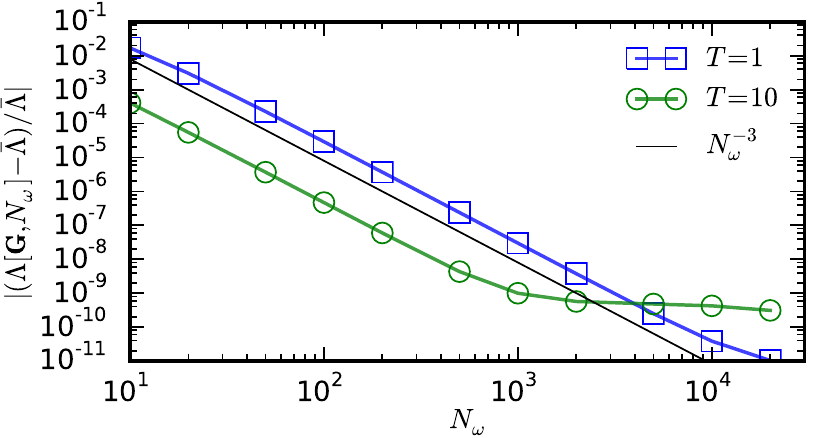}
  \caption{\label{fig:TrLnConvergence}(Color online) Convergence of the lattice Green's function trace log $\Lambda[\bG, N_\omega]$ in $N_\omega$ at $U=20$ and $\mu/U = 0.4$ (with baseline $\bar{\Lambda} = \Lambda[\bG, N_\omega]$ with $N_\omega = 5 \cdot 10^4$) for both the superfluid at $T/J=1$ (blue squares) and the normal phase at $T/J=10$ (green circles), for reference $N_\omega^{-3}$ is also shown (black line).}
\end{figure}

In the SFT calculations presented here we use a second order high-frequency expansion and a finite number $N_\omega$ of Matsubara frequencies [Eq.\ (\ref{eq:TraceNumeric})] which from Eq.\ (\ref{eq:TrLnFirstOrderCorrected}) gives the trace log functional $\Lambda[\bG]$ as
\begin{multline}
  \beta \Lambda [\bG]
  \approx
  - \frac{1}{2}
  \sum_{\mu \mbf{k}} \Big(
  \ln [ - \bG(\mbf{k}, i\omega_0) / \beta ]
  + \frac{\beta}{2}  \mbf{c}_2(\mbf{k}) 
  -  \frac{\beta^2}{12} \mbf{q}_2(\mbf{k})
  \\ +
  \sideset{}{'}\sum_{n=-N_\omega}^{N_\omega} \big[
    \ln [ (\sigma_z i\omega_n) \bG(\mbf{k}, i\omega_n)  ]
    -
    \frac{\mbf{q}_2(\mbf{k})}{ (i\omega_n)^2 }
  \big]
  \Big)^\mu_\mu
  \, ,
  \label{eq:TrLnSecondOrderCorrected}  
\end{multline}
where  $\mbf{q}_2$ is the second-order coefficient in Eq.\ (\ref{eq:LnGTailExpansion}), $\mbf{q}_2 = \sigma_z \mbf{c}_3 - (\sigma_z \mbf{c}_2)^2/2$.
Equation (\ref{eq:TrLnSecondOrderCorrected}) converges cubically with $N_\omega^{-3}$, such that for the lattice Green's function trace log in both the normal and super-fluid phase at the parameters used in Fig.\ \ref{fig:Thermo} we reach a precision of $10^{-9}$ with $N_\omega = 10^3$ and $10^4$ respectively, see Fig.\ \ref{fig:TrLnConvergence}.

% - - - - - - - - - - - - - - - - - - - - - - - - - - - - - - - - - - 
% - - - - - - - - - - - - - - - - - - - - - - - - - - - - - - - - - - 

% --------------------------------------------------------------------
\section{Reference system}
\label{app:ReferenceSystem}
% --------------------------------------------------------------------

To compute properties of the reference system with the Hamiltonian $H'$ in Eq.\ (\ref{eq:SFA3Hamiltonian}) we use the occupation number states $|\psi_n\ra$ of the single bosonic state, where $n \ge 0$. Annihilating and creating a boson yields $ b |\psi_n\ra = \sqrt{n} | \psi_{n-1} \ra$ and $ \bc | \psi_n \ra = \sqrt{n+1} |\psi_{n+1} \ra$, respectively. In this basis we generate matrix representations of $H'$ and the bosonic second quantization operators $\bbb$ and $\bbc$.
However, as the occupation number $n$ of a bosonic state is not bound from above we introduce an occupation number cut-off $N_\text{max}$, as to obtain a finite matrix representation, and disregard all occupation number states $|\psi_n\ra$ with $n > N_\text{max}$.
All reference system calculations thus have to be converged in $N_\text{max}$. For the calculations presented here we find that 10 - 20 states suffice.

To calculate static observables and dynamic response functions we first diagonalize $H'$ to determine its eigenvalues $E_n$ and eigenstates $| n \ra$, where $H' |n \ra = E_n |n \ra$. Repeatedly using the closure relation $1 = \sum_n |n \ra \la n |$ one can then determine
the partition function $\Z$,
\begin{equation}
  \Z = \Tr[ e^{-\beta H'} ] = \sum_n e^{-\beta E_n}
  \, ,
  \label{eq:ApRefSys1}
\end{equation}
the reference system free energy $\Omega'= -\ln[ \Z ] / \beta$,
static expectation values such as
\begin{equation}
  \bPhi' =
  \la \bbb \ra =
  \frac{1}{\Z} \Tr[ e^{-\beta H'} \bbb ]
  =
  \frac{1}{\Z} \sum_n e^{-\beta E_n} \la n | \bbb | n \ra
  \, ,
  \label{eq:ApRefSysPhiPrime}
\end{equation}
and the full single particle Green's function $\tilde{\bG}'$, 
\begin{multline}
  \tilde{\bG}'^\eta_{\nu}(\tau) =
  - \langle \bbb^\eta(\tau) \bbc_\nu \rangle
  %  \\ =
  =
  -\frac{1}{\Z} \Tr [ e^{-\beta H'} e^{\tau H'}
    \bbb^\eta e^{-\tau H'} \bbc_\nu ]
  \\ = 
  -\frac{1}{\Z} \sum_{\ssa \ssb}
  e^{-\beta E_\ssa + \tau (E_\ssa -  E_\ssb)}
  \langle \ssa | \bbb^\eta | \ssb \rangle
  \langle \ssb | \bbc_\nu | \ssa \rangle
  \, . \label{eq:ImaginaryTimeGreensFunction}
\end{multline}
In the last equation, the time dependent operators are defined in the Heisenberg representation $\bbb(\tau) = e^{\tau H'} \bbb e^{-\tau H'}$.
Given $\tilde{\bG}'$ and $\bPhi'$ the connected Green's function $\bG'$, defined in Eq.\ (\ref{eq:FreeEnergyG0var}), is obtained as
\begin{equation}
  \bG'(\tau) = \tilde{\bG}'(\tau) + \bPhi' \bPhi'^\dagger
  \, . \label{eq:AppConnGf}
\end{equation}

To solve the Dyson equations [Eqs.\ (\ref{eq:DysonS12}) and (\ref{eq:DysonS})] we use the Matsubara frequency representation of the Green's functions.
Transforming the $\tau$ dependence in Eq.\ (\ref{eq:ImaginaryTimeGreensFunction}) then gives
\begin{multline}
  e^{-\beta E_\ssa} \int_0^\beta d\tau \,
  e^{\tau ( i\omega_n + E_\ssa - E_\ssb) }
  \\ =
  \left\{\begin{array}{cl}
  - \frac{e^{-\beta E_\ssa} - \xi e^{-\beta E_\ssb}}{i\omega_n + E_\ssa - E_\ssb}
  \, , &
  i \omega_n + E_\ssa - E_\ssb \ne 0 \\[2mm]
  \beta e^{-\beta E_\ssa}
  \, , &
  i \omega_n + E_\ssa - E_\ssb = 0
  \end{array}\right.
  \, ,
\end{multline}
and the full Matsubara frequency Green's function can be expressed by the generalized Lehmann \cite{Negele:1998aa} expression
\begin{multline}
  \tilde{\bG}'^\eta_{\nu}(i\omega_n)
  =
  \frac{1}{\Z} \sum_{\ssa \ssb}
  \frac{
  \langle \ssa | \bbb^\eta | \ssb \rangle
  \langle \ssb | \bbc_\nu | \ssa \rangle
  }{i\omega_n + E_\ssa - E_\ssb} ( e^{-\beta E_\ssa} - \xi e^{-\beta E_\ssb} )
  \\
  - \beta \delta_{\omega_n, 0}
  \frac{1}{\Z} \sum_{\ssa}
  e^{-\beta E_\ssa}
  \langle \ssa | \bbb^\eta | \ssa \rangle
  \langle \ssa | \bbc_\nu | \ssa \rangle
  \, , \label{eq:LehmannGF}
\end{multline}
where we have assumed no accidental degeneracies $E_n \ne E_m$, $\forall$ $m \ne n$, in the last zero-frequency term.
Note that the connected Green's function $\bG'(i\omega_n)$ has an additional zero-frequency contribution, as seen in Eq.\ (\ref{eq:AppConnGf}), which becomes
\begin{equation}
  \bG'(i\omega_n) = \tilde{\bG}'(i\omega_n)
  + \beta \delta_{\omega_n, 0}
  \bPhi' \bPhi'^\dagger
  \, 
  \label{eq:ApRef7}
\end{equation}
in Matsubara frequency space.

Apart from $\bPhi'$ and $\bG'$ the evaluation of the SFT functional [Eq.\ (\ref{eq:SFAFunc})] also requires the non-interacting Green's function $\bG_0'$ and self-energies $\bSp'$ and $\bS$ of the reference system. Setting the interaction $U$ to zero in the reference system Hamiltonian in Eq.\ (\ref{eq:SFA3Hamiltonian}), $\bG_0'$ is obtained as
\begin{equation}
  \bG_0'^{-1}(i\omega_n) =
  \sigma_z i \omega_n + \mbf{1} \mu - \bDelta
  \, ,
    \label{eq:ApRef8}
\end{equation}
and the self-energies are determined by Dyson's equations [Eq.\ (\ref{eq:DysonS12}) and (\ref{eq:DysonS})]
\begin{align}
  \bSp' & = \bF' - \bG_0'^{-1}(i \omega_0) \bPhi'
  \, , \label{eq:ApRef9} \\
  \bS'(i\omega_n) & =
  \bG_0'^{-1}(i\omega_n) - \bG^{-1}(i\omega_n)
  \, .
    \label{eq:ApRef10}
\end{align}

% --------------------------------------------------------------------
\section{Lattice system}
\label{app:LatticeSystem}
% --------------------------------------------------------------------

To compute the response functions of the Bose-Hubbard model defined in Eq.\ (\ref{eq:BHHamiltonian}) at the self-energies $\bSp'$ and $\bS'$ of the reference system we transform to momentum space. The nearest neighbor single particle hopping in Eq.\ (\ref{eq:BHHamiltonian}) gives the dispersion
\begin{equation}
  \epsilon_{\mbf{k}} = -2J \sum_{i=1}^d \cos( k_{i} )
  \, ,
\end{equation}
where $d$ is the dimension of the hypercubic lattice ($d = 2$, $3$). Thus the free lattice Green's function can be written in Nambu form as
\begin{equation}
  \bG_{0}^{-1}(\mbf{k}, i\omega_n) =
  \sigma_z i \omega_n + \mbf{1} ( \mu - \epsilon_{\mbf{k}} )
  \, ,
\end{equation}
and using the Dyson equation [Eq. (\ref{eq:DysonS})] the interacting lattice Green's function evaluated at the reference system self-energy $\bS'$ is given by
\begin{equation}
  \bG^{-1}(\mbf{k}, i\omega_n) =
  \bG_{0}^{-1}(\mbf{k}, i\omega_n) - \bS'(i\omega_n)
  \, .
\end{equation}
Further, by using $\bSp'$ we can determine the condensate of the lattice system by
\begin{equation}
  \bPhi=-\bG_0(\mbf{k}= \mbf{0}, i\omega_0)\bSp'
  \, ,
\end{equation}
where we have used the fact that there is no symmetry-breaking field on the lattice system, $\bF=0$.
From the connected lattice Green's function $\bG_{\mbf{k}}$ and the one-point propagator $\bPhi$ we can compute a number of observables for the lattice system. The momentum space single-particle density matrix $\rho_{\mbf{k}}$ of non-condensed bosons is given by the trace of $\bG$ at fixed momentum $\mbf{k}$
\begin{multline}
  \rho_{\mbf{k}} =
  \la \n_{\mbf{k}} \ra =
  \frac{1}{2\beta} \Tr [ - \bG(\mbf{k}) ]
  \\ =
  - \frac{1}{2\beta} \sum_{\mu n} e^{i\omega_n (-1)^\mu 0^+} \bG^\mu_\mu(\mbf{k}, i \omega_n)
  \, ,
\end{multline}
while the condensate density $\rho_c$ is given by the one-point propagator $\rho_c = \frac{1}{2} \bPhi^\dagger \bPhi$ and the total density $n$ and the kinetic energy $E_{\textrm{kin}}$ is obtained by integrating $\mbf{k}$ over the Brillouin zone, 
\begin{align}
  n & =
  \Tr[ \rho_{\mbf{k}} ]
  + \rho_c
  \, , \\
  E_{\textrm{kin}} & =
  \Tr[ \epsilon_{\mbf{k}} \rho_{\mbf{k}} ] + \epsilon_{\mbf{k}=0} \rho_c
  \, .
\end{align}
Finally the interaction energy can be obtained from the trace of the Green's function and self-energy \cite{Negele:1998aa}
\begin{equation}
  E_{\textrm{int}} =
  - \frac{1}{4\beta} \Tr [ \bS' \bG ]
  \, .
\end{equation}

% --------------------------------------------------------------------
\section{High frequency tail expansions}
\label{sec:highfreq}
% --------------------------------------------------------------------

To evaluate the self-energy functional to high precision we use tail-corrected Matsubara traces as described in Appendix \ref{app:NumericalMatsubaraTrace}. This procedure requires the high-frequency tail coefficients of all Green's functions and self-energies. The details on how to obtain these coefficients for the reference system and the lattice system are detailed in this Appendix.

% --------------------------------------------------------------------
\subsection{Hamiltonian reference system}
% --------------------------------------------------------------------

For a reference system which is given by a finite-dimensional Hamiltonian the expansion can be calculated exactly.
Formally, the high frequency tail of a Matsubara Green's function, i.e. the $1/(i\omega_n)^k$ expansion, can be obtained from the imaginary-time Green's function by partial integration of the Fourier transform expression
\begin{multline}
  \bG(i\omega_n) = \int_0^\beta d\tau \,
  \bG(\tau) e^{i\omega_n \tau}
  \\ =
  \sum_{k=0}^\infty
  (-1)^k
  \frac{\xi \partial^k_\tau \bG(\beta^-) - \partial^k_\tau \bG(0^+)}
       {(i\omega_n)^{k+1}}
  =
  \sum_{k=0}^\infty
  \frac{\mbf{c}_{k+1}}{(i\omega_n)^{k+1}}
  \, ,
\end{multline}
where the $c_k$ are the high-frequency tail expansion coefficients.
Derivatives of the Green's function can be obtained directly from the imaginary-time expression
\begin{equation}
  \bG(\tau) = - \la \bbb(\tau) \bbc(0) \ra =
  -\frac{1}{\Z} \Tr [ e^{-\beta H} e^{\tau H} \bbb e^{-\tau H} \bbc ]
  \, ,
\end{equation}
or, equivalently, from the equation of motion of the operator $\bbb(\tau)$, $\partial_\tau \bbb(\tau) = [H, \bbb(\tau)]$. The first and $k$th order derivatives take the form
\begin{align}
  \partial_\tau \bG(\tau) &
  = - \la [H, \bbb(\tau)], \bbc \ra
  \, , \nonumber \\
  \partial^k_\tau \bG(\tau)
  & = - \la [[ H, \bbb(\tau) ]]^{(k)}, \bbc \ra
  \, ,
\end{align}
where $[[H, \bbb(\tau)]]^{(k)} = [H, ... [H, [H, \bbb(\tau)] ] $ is the $k$th order left side commutator of $H$ with $\bbb(\tau)$. For the specific imaginary times $\tau = 0^+$ and $\beta^-$ the time ordering can be made explicit yielding the static expectation values
\begin{align}
  \partial^k_\tau \bG(0^+) & = - \la [[ H, \bbb ]]^{(k)} \bbc \ra
  \, , \nonumber \\
  \partial^k_\tau \bG(\beta^-) & = - \la \bbc [[ H, \bbb ]]^{(k)} \ra
  \, .
\end{align}
Combining these two relations the coefficients $c_k$ of the high-frequency tail expansion can be written in terms of the static expectation value
\begin{multline}
  \mbf{c}_{k+1} =
  (-1)^k [ \xi \partial^k_\tau \bG(\beta^-) - \partial^k_\tau \bG(0^+) ]
  \\ =
  (-1)^{k+1} \la [ \, [[ H , \bbb ]]^{(k)}, \bbc ]_{-\xi} \ra
  \, .
\end{multline}

% --------------------------------------------------------------------
\subsection{Lattice system}
\label{App:LattSys}
% --------------------------------------------------------------------

Consider the free Nambu Green's function $\bG_0^{-1}(z) = \sigma_z z - \mbf{1} h$ (with $z = i \omega_n$, for brevity).
Inversion and Taylor expansion gives
\begin{multline}
  \bG_0(z)
  =
  \Bigg[ \mathbf{1} - \frac{\sigma_z h}{z} \Bigg]^{-1} \frac{\sigma_z}{z}
  =
  \Bigg[ \sum_{p=0}^\infty \left( \frac{ \sigma_z h }{ z } \right)^p \Bigg]
  \frac{\sigma_z}{z}
  \\ =
  \frac{\sigma_z}{ z }
  + \frac{\sigma_z h \sigma_z}{ z^2 }
  + \frac{\sigma_z h \sigma_z h \sigma_z}{ z^3 }
  + \mathcal{O}(z^{-4})
  \, .
\end{multline}
Similarly, for a general Nambu Green's function $\bG$ with high frequency expansion
$\bG(z) = \sum_{p=1}^\infty \frac{\mathbf{c}_p}{z^p}$,
the inverse is given by
\begin{multline}
  \bG^{-1}(z) =
  \Bigg[ \mathbf{1} + \sigma_z z \sum_{p=2}^\infty \frac{\mathbf{c}_p}{z^p}
    \Bigg]^{-1} \! \! \! \sigma_z z
  \\ =
  \sigma_z z
  - \sigma_z \mathbf{c}_2 \sigma_z
  + \frac{1}{z} \left( - \sigma_z \mathbf{c}_3 \sigma_z + (\sigma_z \mathbf{c}_2)^2 \sigma_z \right)
  \\ + \frac{1}{z^2} \left(
  - \sigma_z \mathbf{c}_4 \sigma_z
  + \sigma_z \mathbf{c}_2 \sigma_z \mathbf{c}_3 \sigma_z
  + \sigma_z \mathbf{c}_3 \sigma_z \mathbf{c}_2 \sigma_z
  - ( \sigma_z \mathbf{c}_2 )^3 \sigma_z
  \right)
  \\ + \mathcal{O}(z^{-3})
  \, ,
\end{multline}
where we have used the fact that $\mathbf{c}_1 = \sigma_z$.
Combining these relations allows us to write down the high-frequency expansion of the self energy, 
$\bS(z) = \sum_{p=0}^\infty \frac{\mathbf{s}_p}{z^p}$, 
in terms of the high-frequency expansion coefficients $\mbf{c}_p$ of the Green's function
\begin{multline}
  \bS(z) = \bG_0^{-1}(z) - \bG^{-1}(z)
  \\ =
  - h
  - \Bigg[ \sum_{q=1}^\infty \left(
    -\sigma_z z \sum_{p=2}^{\infty} \frac{ \mathbf{c}_p}{z^p}
  \right)^q \Bigg] \sigma_z z
  \\ =
  -h
  + \sigma_z \mathbf{c}_2 \sigma_z
  - \frac{1}{z} \left( - \sigma_z \mathbf{c}_3 \sigma_z + [\sigma_z \mathbf{c}_2]^2 \sigma_z \right)
  + \mathcal{O}(z^{-2})
  \, .
\end{multline}
Hence, the lattice Green's function $\bG(\mbf{k}, z)$ with a momentum independent self energy $\bS$,
$\bG^{-1}(\mbf{k}, z) = \sigma_z z - E_{\mbf{k}} - \bS(z)$,
has the tail expansion
\begin{multline}
%  \bG_\mbf{k} =
  \bG(\mbf{k}, z) =
  \frac{\sigma_z}{z}
  + \frac{\sigma_z (E_{\mbf{k}} + \mathbf{s}_0) \sigma_z}{z^2}
  \\ + \frac{1}{z^3} \left(
    \sigma_z \mathbf{s}_1 \sigma_z
    + [\sigma_z ( E_{\mbf{k}} + \mathbf{s}_0 ) ]^2 \sigma_z
  \right)
  + \mathcal{O}(z^{-4})
  \, ,
\end{multline}
where $E_{\mbf{k}} = \mbf{1}(\epsilon_{\mbf{k}} - \mu)$.

% --------------------------------------------------------------------
\section{SFA3 stationary points\\ and order of the superfluid phase transition}
\label{App:Sta_Points}
% --------------------------------------------------------------------

% - - - - - - - - - - - - - - - - - - - - - - - - - - - - - - - - - - 
\begin{figure}
  \includegraphics[scale=1.0]{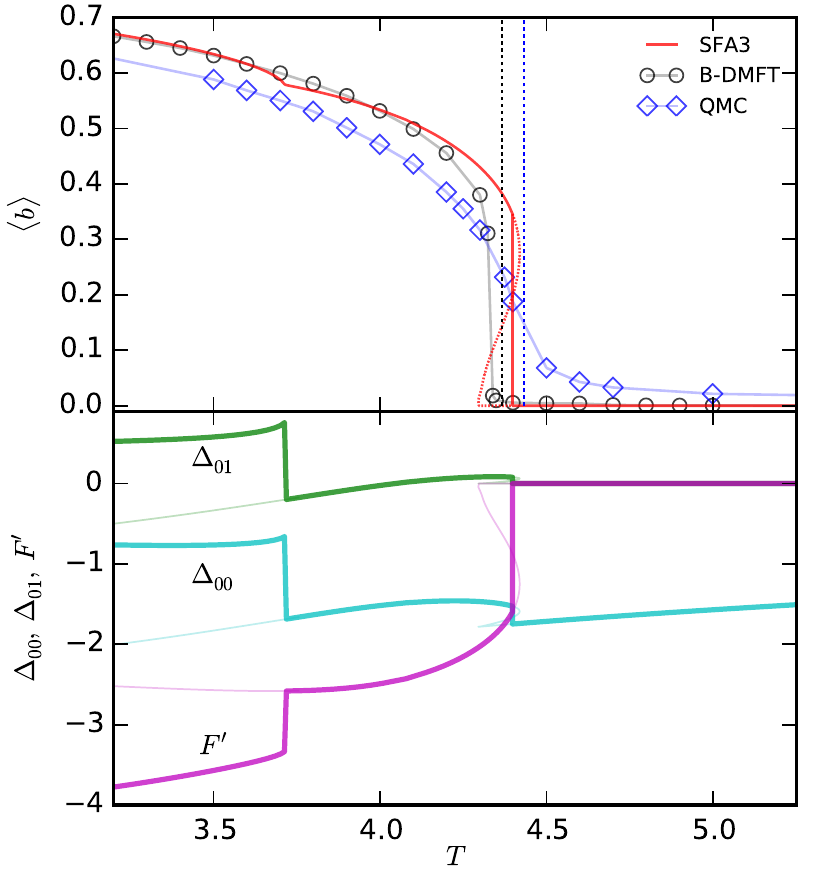}
  \caption{\label{fig:SFTVarParm}(Color online) Superfluid order parameter $\la b \ra$ from SFA3 (red line), B-DMFT (circles), and QMC (diamonds) \cite{Anders:2011uq} (upper panel) and SFA3 variational parameters $\Delta_{00}$, $\Delta_{01}$ and $F$ (lower panel) as a function of temperature $T$ in the vicinity of the superfluid to normal phase transition for the 3d cubic lattice with $U/J=20$ and $\mu/U=0.4$, cf.\ Fig.\ \ref{fig:Thermo}. The location of the B-DMFT and QMC phase transitions are indicated (left and right vertical dotted line respectively). The QMC data were computed for a finite size system with $40^3$ sites, leading to a crossover, while the phase transition line is calculated with finite-size scaling.}
\end{figure}
% - - - - - - - - - - - - - - - - - - - - - - - - - - - - - - - - - - 

In our SFA3 calculations on the Bose-Hubbard model we find several symmetry breaking stationary points with $F'\neq 0$. Most solutions can be discarded by requiring that the single-particle Green's function should fulfill the physical constraints $G_{00}(\mbf{k}, i\omega_0) < 0$ and $\det \bG(\mbf{k},i\omega_0) > 0$ at the zeroth Matsubara frequency $\omega_0 = 0$.

After discarding unphysical stationary points we still find two symmetry broken solutions in the deep superfluid phase. This occurs for temperatures $T/J < 3.7$ for $U/J = 20$ and $\mu / J = 0.4$.
The two solutions are distinguished by the relative phases of the linear symmetry breaking field $F'$ and the anomalous pairing field $\Delta_{01}$ variational parameters. One solution has the fields in-phase, $\arg \Delta_{01} = \arg F'$, and the other solution has the fields in anti-phase, $\arg \Delta_{01} = \arg F' + \pi$, see thin and thick lines respectively for $T < 3.7$ in the lower panel of Fig.\ \ref{fig:SFTVarParm}.

The anti-phase solution disappears through a saddle-node bifurcation \cite{Crawford:1991qy} with one of the unphysical stationary points at $T/J \approx 3.7$, while the in-phase solution prevails up to the normal-phase to superfluid phase transition temperature at $T_c \approx 4.39778$.
However, when comparing free energies it turns out that the anti-phase solution, when present, has the lowest free energy.
Thus within SFA3 when increasing temperature the system jumps from the anti-phase (deep superfluid) to the in-phase (weak) superfluid solution at $T \approx 3.7$. 
The jump between stationary points causes weak discontinuities in observables, e.g., the superfluid order parameter $\phi = \la b \ra$, see the upper panel of Fig.\ \ref{fig:SFTVarParm}.

The appearance of this spurious ``phase-transition'' in the SFA3 calculations might at first be considered a deficiency of the SFT scheme. However, rather than a general SFT issue it is the extremely simplistic SFA3 variational ansatz, with only three parameters that causes this behavior.
In SFA3 the retarded hybridization is reduced to an instantaneous pairing field $\Delta_{01}(\tau) = \delta(\tau) \Delta_{01}$. This one parameter degree-of-freedom is simply not enough to interpolate between the two regimes and instead generates a discontinuity.
As the SFA3 reference system is extended with additional bath sites the SFT calculation is expected to become continuous within the superfluid phase, as is the case for BDMFT.

The upper panel of Fig.\ \ref{fig:SFTVarParm} also shows the detailed behavior of the order parameter $\phi = \la b \ra$ around the normal-phase to superfluid phase transition for SFA3, BDMFT, and QMC.
The superfluid critical temperature within SFA3 is in quantitative agreement with BDMFT and QMC. The phase transition, however, is weakly first order with an accompanying narrow hysteresis region. The thermodynamical ground state solution is accompanied by an unstable superfluid solution that disappears through a saddle-node bifurcation with another stationary point (higher in free energy) that adiabatically connects the superfluid to the normal-phase solution (dotted red line).
The phase transition within SFA3 is located between the BDMFT and QMC results (horizontal dotted lines), and is expected to shift to the BDMFT result as bath sites are added to the SFT reference system.

The existence of three solutions within a hysteresis region is a general feature of first order transitions, not limited to SFA3.
It has also been observed e.g. in DMFT for the paramagnetic Mott to metal transition in the single-band Fermionic-Hubbard model \cite{Strand:2011lr}, and for Gutzwiller variational calculations on two-band generalized Fermi-Hubbard models \cite{Lanata:2012lr}.

% --------------------------------------------------------------------
\section{Comparison with Ref.\ \onlinecite{Arrigoni:2011aa}}
\label{app:Arrigoni}
% --------------------------------------------------------------------

% - - - - - - - - - - - - - - - - - - - - - - - - - - - - - - - - - - 
\begin{figure}
  \includegraphics[scale=1.0]{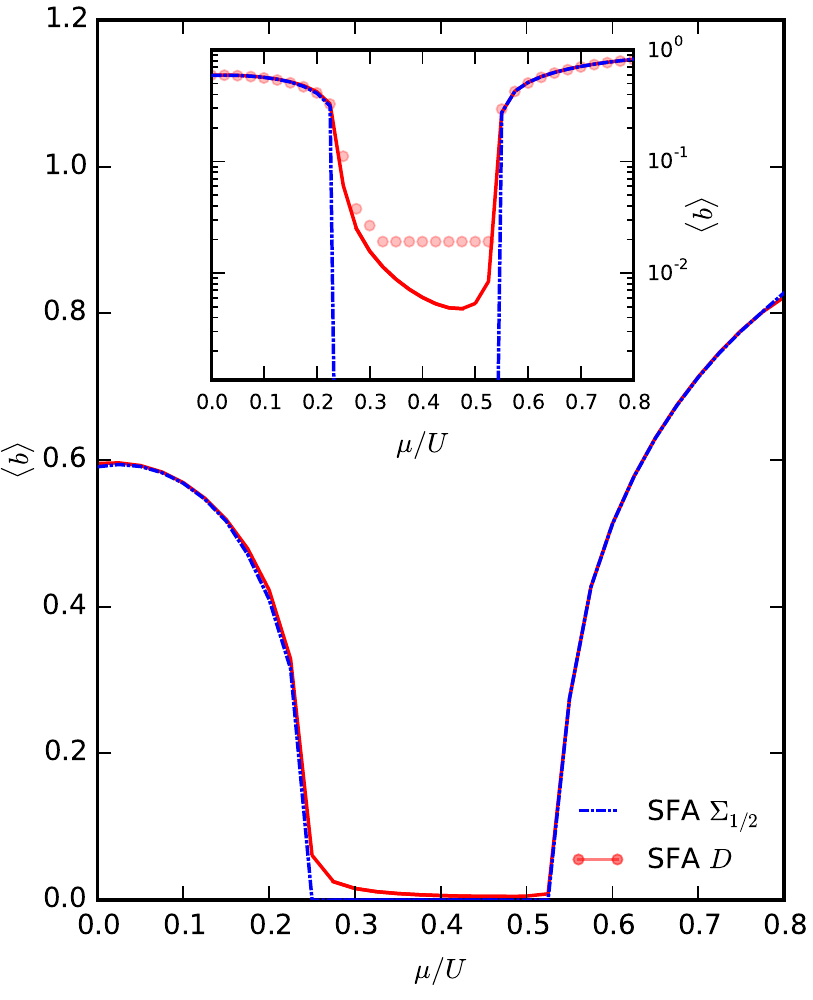}
  \caption{\label{fig:Arrigoni}(Color online) Condensate order-parameter $\langle b \rangle$ of the 2d lattice as a function of chemical potential $\mu$ for $\beta=6.4$ and $U=20$. Both results using the SFT formalism presented here (blue dash-dotted line) and the formalism of Ref.\ \onlinecite{Arrigoni:2011aa} (solid red line) are shown. Inset: Logarithmic vertical axis with original data from Ref.\ \onlinecite{Arrigoni:2011aa} (red markers).}
\end{figure}
% - - - - - - - - - - - - - - - - - - - - - - - - - - - - - - - - - - 

As discussed in the main text self-energy functional theory has already been applied to bosons including $U(1)$ symmetry breaking in Ref.\ \onlinecite{Arrigoni:2011aa}. In this appendix we show how the self-energy functional $\SE$ [Eq.\ (\ref{eq:SESigma})] derived in Section \ref{sec:SE} differs from the result in Ref.\ \onlinecite{Arrigoni:2011aa}.

The self-energy functional $\SED$ of Ref.\ \onlinecite{Arrigoni:2011aa} is built around the Dyson equation ansatz
\begin{align}
  \bG^{-1} \bPhi = \bF - \bD
  \, , \label{eq:DysonD} \\
  \bG^{-1} = \bG_0^{-1} - \bS
  \, , \label{eq:DysonSigmaApp}
\end{align}
where the first-order tensor quantity $\bD$ corresponds to a ``self-energy like'' object for the one-point propagator $\bPhi$. This ansatz differs from the one-point propagator Dyson equation, $\bG_0^{-1} \bPhi = \bF - \bSp$ [Eq.\ (\ref{eq:DysonS12})] obtained from the stationarity of the Baym-Kadanoff functional $\BK$ in Eq.\ (\ref{eq:BK}) and therefore contradicts Refs.\ \cite{De-Dominicis:1964aa, De-Dominicis:1964ab}. The difference being the appearance of the interacting propagator $\bG$ in Eq.\ (\ref{eq:DysonD}) instead of the non-interacting propagator $\bG_0$.

Using the Dyson equation ansatz in Eqs.\ (\ref{eq:DysonD}) and (\ref{eq:DysonSigmaApp}) the authors of Ref.\ \onlinecite{Arrigoni:2011aa} derive the self-energy effective action
\begin{multline}
  \SED[\bD, \bS] =
  \frac{1}{2} (\bF - \bD)^\dagger (\bG_0^{-1} - \bS)^{-1} (\bF - \bD)
  \\
  + \frac{1}{2} \Tr \ln [ -(\bG_0^{-1} - \bS) ]
  + \hFA[\bD, \bS]
  \, ,
  \label{eq:SEArrigoni}
\end{multline}
through a series of transforms of the free energy $\Omega$, where the functional $\hFA \equiv \hFA[\bD, \bS]$ is assumed to be a universal functional in $\bD$ and $\bS$ with variations $\delta_\bD \hFA = \bPhi$ and $\delta_{\bS} \hFA = [ \bG - \bPhi \bPhi^\dagger ] / 2$,
such that the stationarity condition $\delta \SED = 0$ reproduces the Dyson equation ansatz of Eqs.\ (\ref{eq:DysonD}) and (\ref{eq:DysonSigmaApp})
\begin{align}
  \frac{\delta \SED}{\delta \bD} & =
  - \bG (\bF - \bD) + \bPhi = 0
  \, , \label{eq:DysonD1} \\
  \frac{\delta \SED}{\delta \bS} & =
  -\frac{1}{2}(\bG_0^{-1} - \bS)^{-1} + \frac{1}{2} \bG = 0
  \, . \label{eq:DysonSigmaApp1}
\end{align}

We have performed additional calculations employing the self-energy effective action $\SED$ [Eq.\ (\ref{eq:SEArrigoni})] in the construction of the SFT approximation [Eq. (\ref{eq:SFAFunc})] (here denoted by SFA-$\bD$) comparing with the SFT results using the self-energy effective action $\SE$ [Eq.\ (\ref{eq:SESigma})] derived in this work (here denoted by SFA-$\bSp$), see Fig.\ \ref{fig:Arrigoni}. 
The system parameters are chosen to also enable comparison with numerical results published in Ref.\ \onlinecite{Arrigoni:2011aa} (Fig.\ 3b.2), see the inset in Fig.\ \ref{fig:Arrigoni}. The quantitative agreement indicates that our SFA-$\bD$ calculations are consistent with Ref.\ \onlinecite{Arrigoni:2011aa}. 

When sweeping the chemical potential $\mu$ through the unit-filling Mott-lobe at finite temperature and studying the superfluid order parameter $\la b \ra$, we find that SFA-$\bSp$ displays a superfluid to normal-phase transition, while SFA-$\bD$ only yields a crossover (in the range $0.25 < \mu/U < 0.52$). This is the result of the presence of a superfluid solution ($F'\neq 0$) with lower free-energy than the normal solution throughout the entire sweep in $\mu$. In SFA-$\bD$ the free energies of the two solutions therefore never cross, in contrast to SFA-$\bSp$ (see Fig.\ \ref{fig:Omega} a).
The good agreement between the two methods deep in the superfluid phase can be understood from the fact that in this limit $\bG\approx\bG_0$, and therefore Eq.\ (\ref{eq:DysonD}) is essentially equivalent to our symmetry-breaking Dyson equation (\ref{eq:DysonS12}).  However, the absence of a finite temperature superfluid to normal-phase transition in SFA-$\bD$ is unphysical.

\bibliography{/Users/hugstr/Documents/Papers/DMFT_Biblography}

\begin{thebibliography}{81}%
\makeatletter
\providecommand \@ifxundefined [1]{%
 \@ifx{#1\undefined}
}%
\providecommand \@ifnum [1]{%
 \ifnum #1\expandafter \@firstoftwo
 \else \expandafter \@secondoftwo
 \fi
}%
\providecommand \@ifx [1]{%
 \ifx #1\expandafter \@firstoftwo
 \else \expandafter \@secondoftwo
 \fi
}%
\providecommand \natexlab [1]{#1}%
\providecommand \enquote  [1]{``#1''}%
\providecommand \bibnamefont  [1]{#1}%
\providecommand \bibfnamefont [1]{#1}%
\providecommand \citenamefont [1]{#1}%
\providecommand \href@noop [0]{\@secondoftwo}%
\providecommand \href [0]{\begingroup \@sanitize@url \@href}%
\providecommand \@href[1]{\@@startlink{#1}\@@href}%
\providecommand \@@href[1]{\endgroup#1\@@endlink}%
\providecommand \@sanitize@url [0]{\catcode `\\12\catcode `\$12\catcode
  `\&12\catcode `\#12\catcode `\^12\catcode `\_12\catcode `\%12\relax}%
\providecommand \@@startlink[1]{}%
\providecommand \@@endlink[0]{}%
\providecommand \url  [0]{\begingroup\@sanitize@url \@url }%
\providecommand \@url [1]{\endgroup\@href {#1}{\urlprefix }}%
\providecommand \urlprefix  [0]{URL }%
\providecommand \Eprint [0]{\href }%
\providecommand \doibase [0]{http://dx.doi.org/}%
\providecommand \selectlanguage [0]{\@gobble}%
\providecommand \bibinfo  [0]{\@secondoftwo}%
\providecommand \bibfield  [0]{\@secondoftwo}%
\providecommand \translation [1]{[#1]}%
\providecommand \BibitemOpen [0]{}%
\providecommand \bibitemStop [0]{}%
\providecommand \bibitemNoStop [0]{.\EOS\space}%
\providecommand \EOS [0]{\spacefactor3000\relax}%
\providecommand \BibitemShut  [1]{\csname bibitem#1\endcsname}%
\let\auto@bib@innerbib\@empty
%</preamble>
\bibitem [{\citenamefont {Wheatley}(1975)}]{Wheatley:1975aa}%
  \BibitemOpen
  \bibfield  {author} {\bibinfo {author} {\bibfnamefont {J.~C.}\ \bibnamefont
  {Wheatley}},\ }\href {\doibase 10.1103/RevModPhys.47.415} {\bibfield
  {journal} {\bibinfo  {journal} {Rev. Mod. Phys.}\ }\textbf {\bibinfo {volume}
  {47}},\ \bibinfo {pages} {415} (\bibinfo {year} {1975})}\BibitemShut
  {NoStop}%
\bibitem [{\citenamefont {Anderson}(1966)}]{ANDERSON:1966aa}%
  \BibitemOpen
  \bibfield  {author} {\bibinfo {author} {\bibfnamefont {P.~W.}\ \bibnamefont
  {Anderson}},\ }\href {\doibase 10.1103/RevModPhys.38.298} {\bibfield
  {journal} {\bibinfo  {journal} {Rev. Mod. Phys.}\ }\textbf {\bibinfo {volume}
  {38}},\ \bibinfo {pages} {298} (\bibinfo {year} {1966})}\BibitemShut
  {NoStop}%
\bibitem [{\citenamefont {Morsch}\ and\ \citenamefont
  {Oberthaler}(2006)}]{Morsch:2006vn}%
  \BibitemOpen
  \bibfield  {author} {\bibinfo {author} {\bibfnamefont {O.}~\bibnamefont
  {Morsch}}\ and\ \bibinfo {author} {\bibfnamefont {M.}~\bibnamefont
  {Oberthaler}},\ }\href {\doibase 10.1103/RevModPhys.78.179} {\bibfield
  {journal} {\bibinfo  {journal} {Rev. Mod. Phys.}\ }\textbf {\bibinfo {volume}
  {78}},\ \bibinfo {pages} {179} (\bibinfo {year} {2006})}\BibitemShut
  {NoStop}%
\bibitem [{\citenamefont {Bloch}\ \emph {et~al.}(2008)\citenamefont {Bloch},
  \citenamefont {Dalibard},\ and\ \citenamefont {Zwerger}}]{Bloch:2008uq}%
  \BibitemOpen
  \bibfield  {author} {\bibinfo {author} {\bibfnamefont {I.}~\bibnamefont
  {Bloch}}, \bibinfo {author} {\bibfnamefont {J.}~\bibnamefont {Dalibard}}, \
  and\ \bibinfo {author} {\bibfnamefont {W.}~\bibnamefont {Zwerger}},\ }\href
  {\doibase 10.1103/RevModPhys.80.885} {\bibfield  {journal} {\bibinfo
  {journal} {Rev. Mod. Phys.}\ }\textbf {\bibinfo {volume} {80}},\ \bibinfo
  {pages} {885} (\bibinfo {year} {2008})}\BibitemShut {NoStop}%
\bibitem [{\citenamefont {Fisher}\ \emph {et~al.}(1989)\citenamefont {Fisher},
  \citenamefont {Weichman}, \citenamefont {Grinstein},\ and\ \citenamefont
  {Fisher}}]{Fisher:1989kl}%
  \BibitemOpen
  \bibfield  {author} {\bibinfo {author} {\bibfnamefont {M.~P.~A.}\
  \bibnamefont {Fisher}}, \bibinfo {author} {\bibfnamefont {P.~B.}\
  \bibnamefont {Weichman}}, \bibinfo {author} {\bibfnamefont {G.}~\bibnamefont
  {Grinstein}}, \ and\ \bibinfo {author} {\bibfnamefont {D.~S.}\ \bibnamefont
  {Fisher}},\ }\href {\doibase 10.1103/PhysRevB.40.546} {\bibfield  {journal}
  {\bibinfo  {journal} {Phys. Rev. B}\ }\textbf {\bibinfo {volume} {40}},\
  \bibinfo {pages} {546} (\bibinfo {year} {1989})}\BibitemShut {NoStop}%
\bibitem [{\citenamefont {Jaksch}\ \emph {et~al.}(1998)\citenamefont {Jaksch},
  \citenamefont {Bruder}, \citenamefont {Cirac}, \citenamefont {Gardiner},\
  and\ \citenamefont {Zoller}}]{Jaksch:1998vn}%
  \BibitemOpen
  \bibfield  {author} {\bibinfo {author} {\bibfnamefont {D.}~\bibnamefont
  {Jaksch}}, \bibinfo {author} {\bibfnamefont {C.}~\bibnamefont {Bruder}},
  \bibinfo {author} {\bibfnamefont {J.~I.}\ \bibnamefont {Cirac}}, \bibinfo
  {author} {\bibfnamefont {C.~W.}\ \bibnamefont {Gardiner}}, \ and\ \bibinfo
  {author} {\bibfnamefont {P.}~\bibnamefont {Zoller}},\ }\href {\doibase
  10.1103/PhysRevLett.81.3108} {\bibfield  {journal} {\bibinfo  {journal}
  {Phys. Rev. Lett.}\ }\textbf {\bibinfo {volume} {81}},\ \bibinfo {pages}
  {3108} (\bibinfo {year} {1998})}\BibitemShut {NoStop}%
\bibitem [{\citenamefont {Pollet}(2012)}]{Pollet:2012ly}%
  \BibitemOpen
  \bibfield  {author} {\bibinfo {author} {\bibfnamefont {L.}~\bibnamefont
  {Pollet}},\ }\href@noop {} {\bibfield  {journal} {\bibinfo  {journal} {Rep.
  Prog. Phys.}\ }\textbf {\bibinfo {volume} {75}},\ \bibinfo {pages} {094501}
  (\bibinfo {year} {2012})}\BibitemShut {NoStop}%
\bibitem [{\citenamefont {Struck}\ \emph {et~al.}(2012)\citenamefont {Struck},
  \citenamefont {{\"O}lschl{\"a}ger}, \citenamefont {Weinberg}, \citenamefont
  {Hauke}, \citenamefont {Simonet}, \citenamefont {Eckardt}, \citenamefont
  {Lewenstein}, \citenamefont {Sengstock},\ and\ \citenamefont
  {Windpassinger}}]{Struck:2012aa}%
  \BibitemOpen
  \bibfield  {author} {\bibinfo {author} {\bibfnamefont {J.}~\bibnamefont
  {Struck}}, \bibinfo {author} {\bibfnamefont {C.}~\bibnamefont
  {{\"O}lschl{\"a}ger}}, \bibinfo {author} {\bibfnamefont {M.}~\bibnamefont
  {Weinberg}}, \bibinfo {author} {\bibfnamefont {P.}~\bibnamefont {Hauke}},
  \bibinfo {author} {\bibfnamefont {J.}~\bibnamefont {Simonet}}, \bibinfo
  {author} {\bibfnamefont {A.}~\bibnamefont {Eckardt}}, \bibinfo {author}
  {\bibfnamefont {M.}~\bibnamefont {Lewenstein}}, \bibinfo {author}
  {\bibfnamefont {K.}~\bibnamefont {Sengstock}}, \ and\ \bibinfo {author}
  {\bibfnamefont {P.}~\bibnamefont {Windpassinger}},\ }\href
  {http://link.aps.org/doi/10.1103/PhysRevLett.108.225304} {\bibfield
  {journal} {\bibinfo  {journal} {Phys. Rev. Lett.}\ }\textbf {\bibinfo
  {volume} {108}},\ \bibinfo {pages} {225304} (\bibinfo {year}
  {2012})}\BibitemShut {NoStop}%
\bibitem [{\citenamefont {Greschner}\ \emph {et~al.}(2014)\citenamefont
  {Greschner}, \citenamefont {Sun}, \citenamefont {Poletti},\ and\
  \citenamefont {Santos}}]{Greschner:2014aa}%
  \BibitemOpen
  \bibfield  {author} {\bibinfo {author} {\bibfnamefont {S.}~\bibnamefont
  {Greschner}}, \bibinfo {author} {\bibfnamefont {G.}~\bibnamefont {Sun}},
  \bibinfo {author} {\bibfnamefont {D.}~\bibnamefont {Poletti}}, \ and\
  \bibinfo {author} {\bibfnamefont {L.}~\bibnamefont {Santos}},\ }\href
  {http://link.aps.org/doi/10.1103/PhysRevLett.113.215303} {\bibfield
  {journal} {\bibinfo  {journal} {Phys. Rev. Lett.}\ }\textbf {\bibinfo
  {volume} {113}},\ \bibinfo {pages} {215303} (\bibinfo {year}
  {2014})}\BibitemShut {NoStop}%
\bibitem [{\citenamefont {Goldman}\ and\ \citenamefont
  {Dalibard}(2014)}]{Goldman:2014aa}%
  \BibitemOpen
  \bibfield  {author} {\bibinfo {author} {\bibfnamefont {N.}~\bibnamefont
  {Goldman}}\ and\ \bibinfo {author} {\bibfnamefont {J.}~\bibnamefont
  {Dalibard}},\ }\href {http://link.aps.org/doi/10.1103/PhysRevX.4.031027}
  {\bibfield  {journal} {\bibinfo  {journal} {Phys. Rev. X}\ }\textbf {\bibinfo
  {volume} {4}},\ \bibinfo {pages} {031027} (\bibinfo {year}
  {2014})}\BibitemShut {NoStop}%
\bibitem [{\citenamefont {Lin}\ \emph {et~al.}(2011)\citenamefont {Lin},
  \citenamefont {Jim{\'e}nez-Garc{\'\i}a},\ and\ \citenamefont
  {Spielman}}]{Lin:2011aa}%
  \BibitemOpen
  \bibfield  {author} {\bibinfo {author} {\bibfnamefont {Y.-J.}\ \bibnamefont
  {Lin}}, \bibinfo {author} {\bibfnamefont {K.}~\bibnamefont
  {Jim{\'e}nez-Garc{\'\i}a}}, \ and\ \bibinfo {author} {\bibfnamefont {I.~B.}\
  \bibnamefont {Spielman}},\ }\href@noop {} {\bibfield  {journal} {\bibinfo
  {journal} {Nature}\ }\textbf {\bibinfo {volume} {471}},\ \bibinfo {pages}
  {83} (\bibinfo {year} {2011})}\BibitemShut {NoStop}%
\bibitem [{\citenamefont {Struck}\ \emph {et~al.}(2014)\citenamefont {Struck},
  \citenamefont {Simonet},\ and\ \citenamefont {Sengstock}}]{Struck:2014aa}%
  \BibitemOpen
  \bibfield  {author} {\bibinfo {author} {\bibfnamefont {J.}~\bibnamefont
  {Struck}}, \bibinfo {author} {\bibfnamefont {J.}~\bibnamefont {Simonet}}, \
  and\ \bibinfo {author} {\bibfnamefont {K.}~\bibnamefont {Sengstock}},\ }\href
  {http://link.aps.org/doi/10.1103/PhysRevA.90.031601} {\bibfield  {journal}
  {\bibinfo  {journal} {Phys. Rev. A}\ }\textbf {\bibinfo {volume} {90}},\
  \bibinfo {pages} {031601} (\bibinfo {year} {2014})}\BibitemShut {NoStop}%
\bibitem [{\citenamefont {Jim{\'e}nez-Garc{\'\i}a}\ \emph
  {et~al.}(2015)\citenamefont {Jim{\'e}nez-Garc{\'\i}a}, \citenamefont
  {LeBlanc}, \citenamefont {Williams}, \citenamefont {Beeler}, \citenamefont
  {Qu}, \citenamefont {Gong}, \citenamefont {Zhang},\ and\ \citenamefont
  {Spielman}}]{Jimenez-Garcia:2015aa}%
  \BibitemOpen
  \bibfield  {author} {\bibinfo {author} {\bibfnamefont {K.}~\bibnamefont
  {Jim{\'e}nez-Garc{\'\i}a}}, \bibinfo {author} {\bibfnamefont {L.~J.}\
  \bibnamefont {LeBlanc}}, \bibinfo {author} {\bibfnamefont {R.~A.}\
  \bibnamefont {Williams}}, \bibinfo {author} {\bibfnamefont {M.~C.}\
  \bibnamefont {Beeler}}, \bibinfo {author} {\bibfnamefont {C.}~\bibnamefont
  {Qu}}, \bibinfo {author} {\bibfnamefont {M.}~\bibnamefont {Gong}}, \bibinfo
  {author} {\bibfnamefont {C.}~\bibnamefont {Zhang}}, \ and\ \bibinfo {author}
  {\bibfnamefont {I.~B.}\ \bibnamefont {Spielman}},\ }\href
  {http://link.aps.org/doi/10.1103/PhysRevLett.114.125301} {\bibfield
  {journal} {\bibinfo  {journal} {Phys. Rev. Lett.}\ }\textbf {\bibinfo
  {volume} {114}},\ \bibinfo {pages} {125301} (\bibinfo {year}
  {2015})}\BibitemShut {NoStop}%
\bibitem [{\citenamefont {{Baier}}\ \emph {et~al.}(2015)\citenamefont
  {{Baier}}, \citenamefont {{Mark}}, \citenamefont {{Petter}}, \citenamefont
  {{Aikawa}}, \citenamefont {{Chomaz}}, \citenamefont {{Cai}}, \citenamefont
  {{Baranov}}, \citenamefont {{Zoller}},\ and\ \citenamefont
  {{Ferlaino}}}]{Baier:2015aa}%
  \BibitemOpen
  \bibfield  {author} {\bibinfo {author} {\bibfnamefont {S.}~\bibnamefont
  {{Baier}}}, \bibinfo {author} {\bibfnamefont {M.~J.}\ \bibnamefont {{Mark}}},
  \bibinfo {author} {\bibfnamefont {D.}~\bibnamefont {{Petter}}}, \bibinfo
  {author} {\bibfnamefont {K.}~\bibnamefont {{Aikawa}}}, \bibinfo {author}
  {\bibfnamefont {L.}~\bibnamefont {{Chomaz}}}, \bibinfo {author}
  {\bibfnamefont {Z.}~\bibnamefont {{Cai}}}, \bibinfo {author} {\bibfnamefont
  {M.}~\bibnamefont {{Baranov}}}, \bibinfo {author} {\bibfnamefont
  {P.}~\bibnamefont {{Zoller}}}, \ and\ \bibinfo {author} {\bibfnamefont
  {F.}~\bibnamefont {{Ferlaino}}},\ }\href@noop {} {\bibfield  {journal}
  {\bibinfo  {journal} {ArXiv e-prints}\ } (\bibinfo {year} {2015})},\ \Eprint
  {http://arxiv.org/abs/1507.03500} {arXiv:1507.03500 [cond-mat.quant-gas]}
  \BibitemShut {NoStop}%
\bibitem [{\citenamefont {{Zeiher}}\ \emph {et~al.}(2016)\citenamefont
  {{Zeiher}}, \citenamefont {{van Bijnen}}, \citenamefont {{Schau{\ss}}},
  \citenamefont {{Hild}}, \citenamefont {{Choi}}, \citenamefont {{Pohl}},
  \citenamefont {{Bloch}},\ and\ \citenamefont {{Gross}}}]{Zeiher:2016aa}%
  \BibitemOpen
  \bibfield  {author} {\bibinfo {author} {\bibfnamefont {J.}~\bibnamefont
  {{Zeiher}}}, \bibinfo {author} {\bibfnamefont {R.}~\bibnamefont {{van
  Bijnen}}}, \bibinfo {author} {\bibfnamefont {P.}~\bibnamefont
  {{Schau{\ss}}}}, \bibinfo {author} {\bibfnamefont {S.}~\bibnamefont
  {{Hild}}}, \bibinfo {author} {\bibfnamefont {J.-y.}\ \bibnamefont {{Choi}}},
  \bibinfo {author} {\bibfnamefont {T.}~\bibnamefont {{Pohl}}}, \bibinfo
  {author} {\bibfnamefont {I.}~\bibnamefont {{Bloch}}}, \ and\ \bibinfo
  {author} {\bibfnamefont {C.}~\bibnamefont {{Gross}}},\ }\href@noop {}
  {\bibfield  {journal} {\bibinfo  {journal} {ArXiv e-prints}\ } (\bibinfo
  {year} {2016})},\ \Eprint {http://arxiv.org/abs/1602.06313} {arXiv:1602.06313
  [cond-mat.quant-gas]} \BibitemShut {NoStop}%
\bibitem [{\citenamefont {Troyer}\ and\ \citenamefont
  {Wiese}(2005)}]{Troyer:2005aa}%
  \BibitemOpen
  \bibfield  {author} {\bibinfo {author} {\bibfnamefont {M.}~\bibnamefont
  {Troyer}}\ and\ \bibinfo {author} {\bibfnamefont {U.-J.}\ \bibnamefont
  {Wiese}},\ }\href {\doibase 10.1103/PhysRevLett.94.170201} {\bibfield
  {journal} {\bibinfo  {journal} {Phys. Rev. Lett.}\ }\textbf {\bibinfo
  {volume} {94}},\ \bibinfo {pages} {170201} (\bibinfo {year}
  {2005})}\BibitemShut {NoStop}%
\bibitem [{\citenamefont {Potthoff}(2003{\natexlab{a}})}]{Potthoff:2003aa}%
  \BibitemOpen
  \bibfield  {author} {\bibinfo {author} {\bibfnamefont {M.}~\bibnamefont
  {Potthoff}},\ }\href {http://dx.doi.org/10.1140/epjb/e2003-00121-8}
  {\bibfield  {journal} {\bibinfo  {journal} {The European Physical Journal B -
  Condensed Matter and Complex Systems}\ }\textbf {\bibinfo {volume} {32}},\
  \bibinfo {pages} {429} (\bibinfo {year} {2003}{\natexlab{a}})}\BibitemShut
  {NoStop}%
\bibitem [{\citenamefont {Potthoff}(2003{\natexlab{b}})}]{Potthoff:2003ab}%
  \BibitemOpen
  \bibfield  {author} {\bibinfo {author} {\bibfnamefont {M.}~\bibnamefont
  {Potthoff}},\ }\href {http://dx.doi.org/10.1140/epjb/e2003-00352-7}
  {\bibfield  {journal} {\bibinfo  {journal} {The European Physical Journal B -
  Condensed Matter and Complex Systems}\ }\textbf {\bibinfo {volume} {36}},\
  \bibinfo {pages} {335} (\bibinfo {year} {2003}{\natexlab{b}})}\BibitemShut
  {NoStop}%
\bibitem [{\citenamefont {Potthoff}(2006)}]{Potthoff:2006aa}%
  \BibitemOpen
  \bibfield  {author} {\bibinfo {author} {\bibfnamefont {M.}~\bibnamefont
  {Potthoff}},\ }\href@noop {} {\bibfield  {journal} {\bibinfo  {journal}
  {Cond. Mat. Phys.}\ }\textbf {\bibinfo {volume} {9}},\ \bibinfo {pages} {557}
  (\bibinfo {year} {2006})}\BibitemShut {NoStop}%
\bibitem [{\citenamefont {Avella}\ and\ \citenamefont
  {Mancini}(2012)}]{Springer:2012}%
  \BibitemOpen
  \bibinfo {editor} {\bibfnamefont {A.}~\bibnamefont {Avella}}\ and\ \bibinfo
  {editor} {\bibfnamefont {F.}~\bibnamefont {Mancini}},\ eds.,\ \href@noop {}
  {\emph {\bibinfo {title} {Strongly Correlated Systems Theoretical
  Methods}}},\ \bibinfo {series} {Springer Series in Solid-State Sciences},
  Vol.\ \bibinfo {volume} {171}\ (\bibinfo  {publisher} {Springer},\ \bibinfo
  {year} {2012})\BibitemShut {NoStop}%
\bibitem [{\citenamefont {Georges}\ \emph {et~al.}(1996)\citenamefont
  {Georges}, \citenamefont {Kotliar}, \citenamefont {Krauth},\ and\
  \citenamefont {Rozenberg}}]{Georges:1996aa}%
  \BibitemOpen
  \bibfield  {author} {\bibinfo {author} {\bibfnamefont {A.}~\bibnamefont
  {Georges}}, \bibinfo {author} {\bibfnamefont {G.}~\bibnamefont {Kotliar}},
  \bibinfo {author} {\bibfnamefont {W.}~\bibnamefont {Krauth}}, \ and\ \bibinfo
  {author} {\bibfnamefont {M.~J.}\ \bibnamefont {Rozenberg}},\ }\href@noop {}
  {\bibfield  {journal} {\bibinfo  {journal} {Rev. Mod. Phys.}\ }\textbf
  {\bibinfo {volume} {68}},\ \bibinfo {pages} {13} (\bibinfo {year}
  {1996})}\BibitemShut {NoStop}%
\bibitem [{\citenamefont {Kotliar}\ \emph {et~al.}(2006)\citenamefont
  {Kotliar}, \citenamefont {Savrasov}, \citenamefont {Haule}, \citenamefont
  {Oudovenko}, \citenamefont {Parcollet},\ and\ \citenamefont
  {Marianetti}}]{Kotliar:2006aa}%
  \BibitemOpen
  \bibfield  {author} {\bibinfo {author} {\bibfnamefont {G.}~\bibnamefont
  {Kotliar}}, \bibinfo {author} {\bibfnamefont {S.~Y.}\ \bibnamefont
  {Savrasov}}, \bibinfo {author} {\bibfnamefont {K.}~\bibnamefont {Haule}},
  \bibinfo {author} {\bibfnamefont {V.~S.}\ \bibnamefont {Oudovenko}}, \bibinfo
  {author} {\bibfnamefont {O.}~\bibnamefont {Parcollet}}, \ and\ \bibinfo
  {author} {\bibfnamefont {C.~A.}\ \bibnamefont {Marianetti}},\ }\href@noop {}
  {\bibfield  {journal} {\bibinfo  {journal} {Rev. Mod. Phys.}\ }\textbf
  {\bibinfo {volume} {78}},\ \bibinfo {pages} {865} (\bibinfo {year}
  {2006})}\BibitemShut {NoStop}%
\bibitem [{\citenamefont {Potthoff}\ \emph {et~al.}(2003)\citenamefont
  {Potthoff}, \citenamefont {Aichhorn},\ and\ \citenamefont
  {Dahnken}}]{Potthoff:2003ac}%
  \BibitemOpen
  \bibfield  {author} {\bibinfo {author} {\bibfnamefont {M.}~\bibnamefont
  {Potthoff}}, \bibinfo {author} {\bibfnamefont {M.}~\bibnamefont {Aichhorn}},
  \ and\ \bibinfo {author} {\bibfnamefont {C.}~\bibnamefont {Dahnken}},\ }\href
  {\doibase 10.1103/PhysRevLett.91.206402} {\bibfield  {journal} {\bibinfo
  {journal} {Phys. Rev. Lett.}\ }\textbf {\bibinfo {volume} {91}},\ \bibinfo
  {pages} {206402} (\bibinfo {year} {2003})}\BibitemShut {NoStop}%
\bibitem [{\citenamefont {Potthoff}\ and\ \citenamefont
  {Balzer}(2007)}]{Potthoff:2007aa}%
  \BibitemOpen
  \bibfield  {author} {\bibinfo {author} {\bibfnamefont {M.}~\bibnamefont
  {Potthoff}}\ and\ \bibinfo {author} {\bibfnamefont {M.}~\bibnamefont
  {Balzer}},\ }\href {\doibase 10.1103/PhysRevB.75.125112} {\bibfield
  {journal} {\bibinfo  {journal} {Phys. Rev. B}\ }\textbf {\bibinfo {volume}
  {75}},\ \bibinfo {pages} {125112} (\bibinfo {year} {2007})}\BibitemShut
  {NoStop}%
\bibitem [{\citenamefont {Koller}\ and\ \citenamefont
  {Dupuis}(2006)}]{Koller:2006aa}%
  \BibitemOpen
  \bibfield  {author} {\bibinfo {author} {\bibfnamefont {W.}~\bibnamefont
  {Koller}}\ and\ \bibinfo {author} {\bibfnamefont {N.}~\bibnamefont
  {Dupuis}},\ }\href {http://stacks.iop.org/0953-8984/18/i=41/a=019} {\bibfield
   {journal} {\bibinfo  {journal} {Journal of Physics: Condensed Matter}\
  }\textbf {\bibinfo {volume} {18}},\ \bibinfo {pages} {9525} (\bibinfo {year}
  {2006})}\BibitemShut {NoStop}%
\bibitem [{\citenamefont {Arrigoni}\ \emph {et~al.}(2011)\citenamefont
  {Arrigoni}, \citenamefont {Knap},\ and\ \citenamefont {von~der
  Linden}}]{Arrigoni:2011aa}%
  \BibitemOpen
  \bibfield  {author} {\bibinfo {author} {\bibfnamefont {E.}~\bibnamefont
  {Arrigoni}}, \bibinfo {author} {\bibfnamefont {M.}~\bibnamefont {Knap}}, \
  and\ \bibinfo {author} {\bibfnamefont {W.}~\bibnamefont {von~der Linden}},\
  }\href {http://link.aps.org/doi/10.1103/PhysRevB.84.014535} {\bibfield
  {journal} {\bibinfo  {journal} {Physical Review B}\ }\textbf {\bibinfo
  {volume} {84}},\ \bibinfo {pages} {014535} (\bibinfo {year}
  {2011})}\BibitemShut {NoStop}%
\bibitem [{\citenamefont {De~Dominicis}\ and\ \citenamefont
  {Martin}(1964{\natexlab{a}})}]{De-Dominicis:1964aa}%
  \BibitemOpen
  \bibfield  {author} {\bibinfo {author} {\bibfnamefont {C.}~\bibnamefont
  {De~Dominicis}}\ and\ \bibinfo {author} {\bibfnamefont {P.~C.}\ \bibnamefont
  {Martin}},\ }\href
  {http://scitation.aip.org/content/aip/journal/jmp/5/1/10.1063/1.1704062}
  {\bibfield  {journal} {\bibinfo  {journal} {Journal of Mathematical Physics}\
  }\textbf {\bibinfo {volume} {5}},\ \bibinfo {pages} {14} (\bibinfo {year}
  {1964}{\natexlab{a}})}\BibitemShut {NoStop}%
\bibitem [{\citenamefont {De~Dominicis}\ and\ \citenamefont
  {Martin}(1964{\natexlab{b}})}]{De-Dominicis:1964ab}%
  \BibitemOpen
  \bibfield  {author} {\bibinfo {author} {\bibfnamefont {C.}~\bibnamefont
  {De~Dominicis}}\ and\ \bibinfo {author} {\bibfnamefont {P.~C.}\ \bibnamefont
  {Martin}},\ }\href
  {http://scitation.aip.org/content/aip/journal/jmp/5/1/10.1063/1.1704064}
  {\bibfield  {journal} {\bibinfo  {journal} {Journal of Mathematical Physics}\
  }\textbf {\bibinfo {volume} {5}},\ \bibinfo {pages} {31} (\bibinfo {year}
  {1964}{\natexlab{b}})}\BibitemShut {NoStop}%
\bibitem [{\citenamefont {Baym}\ and\ \citenamefont
  {Kadanoff}(1961)}]{Baym:1961tw}%
  \BibitemOpen
  \bibfield  {author} {\bibinfo {author} {\bibfnamefont {G.}~\bibnamefont
  {Baym}}\ and\ \bibinfo {author} {\bibfnamefont {L.~P.}\ \bibnamefont
  {Kadanoff}},\ }\href {\doibase 10.1103/PhysRev.124.287} {\bibfield  {journal}
  {\bibinfo  {journal} {Phys. Rev.}\ }\textbf {\bibinfo {volume} {124}},\
  \bibinfo {pages} {287} (\bibinfo {year} {1961})}\BibitemShut {NoStop}%
\bibitem [{\citenamefont {Baym}(1962)}]{Baym:1962qo}%
  \BibitemOpen
  \bibfield  {author} {\bibinfo {author} {\bibfnamefont {G.}~\bibnamefont
  {Baym}},\ }\href {\doibase 10.1103/PhysRev.127.1391} {\bibfield  {journal}
  {\bibinfo  {journal} {Phys. Rev.}\ }\textbf {\bibinfo {volume} {127}},\
  \bibinfo {pages} {1391} (\bibinfo {year} {1962})}\BibitemShut {NoStop}%
\bibitem [{\citenamefont {Luttinger}\ and\ \citenamefont
  {Ward}(1960)}]{Luttinger:1960aa}%
  \BibitemOpen
  \bibfield  {author} {\bibinfo {author} {\bibfnamefont {J.~M.}\ \bibnamefont
  {Luttinger}}\ and\ \bibinfo {author} {\bibfnamefont {J.~C.}\ \bibnamefont
  {Ward}},\ }\href {http://link.aps.org/abstract/PR/v118/p1417} {\bibfield
  {journal} {\bibinfo  {journal} {Phys. Rev.}\ }\textbf {\bibinfo {volume}
  {118}} (\bibinfo {year} {1960})}\BibitemShut {NoStop}%
\bibitem [{\citenamefont {Byczuk}\ and\ \citenamefont
  {Vollhardt}(2008)}]{Byczuk:2008nx}%
  \BibitemOpen
  \bibfield  {author} {\bibinfo {author} {\bibfnamefont {K.}~\bibnamefont
  {Byczuk}}\ and\ \bibinfo {author} {\bibfnamefont {D.}~\bibnamefont
  {Vollhardt}},\ }\href {\doibase 10.1103/PhysRevB.77.235106} {\bibfield
  {journal} {\bibinfo  {journal} {Phys. Rev. B}\ }\textbf {\bibinfo {volume}
  {77}},\ \bibinfo {pages} {235106} (\bibinfo {year} {2008})}\BibitemShut
  {NoStop}%
\bibitem [{\citenamefont {Hubener}\ \emph {et~al.}(2009)\citenamefont
  {Hubener}, \citenamefont {Snoek},\ and\ \citenamefont
  {Hofstetter}}]{Hubener:2009cr}%
  \BibitemOpen
  \bibfield  {author} {\bibinfo {author} {\bibfnamefont {A.}~\bibnamefont
  {Hubener}}, \bibinfo {author} {\bibfnamefont {M.}~\bibnamefont {Snoek}}, \
  and\ \bibinfo {author} {\bibfnamefont {W.}~\bibnamefont {Hofstetter}},\
  }\href {\doibase 10.1103/PhysRevB.80.245109} {\bibfield  {journal} {\bibinfo
  {journal} {Phys. Rev. B}\ }\textbf {\bibinfo {volume} {80}},\ \bibinfo
  {pages} {245109} (\bibinfo {year} {2009})}\BibitemShut {NoStop}%
\bibitem [{\citenamefont {Hu}\ and\ \citenamefont {Tong}(2009)}]{Hu:2009qf}%
  \BibitemOpen
  \bibfield  {author} {\bibinfo {author} {\bibfnamefont {W.-J.}\ \bibnamefont
  {Hu}}\ and\ \bibinfo {author} {\bibfnamefont {N.-H.}\ \bibnamefont {Tong}},\
  }\href {\doibase 10.1103/PhysRevB.80.245110} {\bibfield  {journal} {\bibinfo
  {journal} {Phys. Rev. B}\ }\textbf {\bibinfo {volume} {80}},\ \bibinfo
  {pages} {245110} (\bibinfo {year} {2009})}\BibitemShut {NoStop}%
\bibitem [{\citenamefont {Anders}\ \emph {et~al.}(2010)\citenamefont {Anders},
  \citenamefont {Gull}, \citenamefont {Pollet}, \citenamefont {Troyer},\ and\
  \citenamefont {Werner}}]{Anders:2010uq}%
  \BibitemOpen
  \bibfield  {author} {\bibinfo {author} {\bibfnamefont {P.}~\bibnamefont
  {Anders}}, \bibinfo {author} {\bibfnamefont {E.}~\bibnamefont {Gull}},
  \bibinfo {author} {\bibfnamefont {L.}~\bibnamefont {Pollet}}, \bibinfo
  {author} {\bibfnamefont {M.}~\bibnamefont {Troyer}}, \ and\ \bibinfo {author}
  {\bibfnamefont {P.}~\bibnamefont {Werner}},\ }\href {\doibase
  10.1103/PhysRevLett.105.096402} {\bibfield  {journal} {\bibinfo  {journal}
  {Phys. Rev. Lett.}\ }\textbf {\bibinfo {volume} {105}},\ \bibinfo {pages}
  {096402} (\bibinfo {year} {2010})}\BibitemShut {NoStop}%
\bibitem [{\citenamefont {{Snoek}}\ and\ \citenamefont
  {{Hofstetter}}(2010)}]{Snoek:2010uq}%
  \BibitemOpen
  \bibfield  {author} {\bibinfo {author} {\bibfnamefont {M.}~\bibnamefont
  {{Snoek}}}\ and\ \bibinfo {author} {\bibfnamefont {W.}~\bibnamefont
  {{Hofstetter}}},\ }\href@noop {} {\bibfield  {journal} {\bibinfo  {journal}
  {ArXiv e-prints}\ } (\bibinfo {year} {2010})},\ \Eprint
  {http://arxiv.org/abs/1007.5223} {arXiv:1007.5223 [cond-mat.quant-gas]}
  \BibitemShut {NoStop}%
\bibitem [{\citenamefont {Anders}\ \emph {et~al.}(2011)\citenamefont {Anders},
  \citenamefont {Gull}, \citenamefont {Pollet}, \citenamefont {Troyer},\ and\
  \citenamefont {Werner}}]{Anders:2011uq}%
  \BibitemOpen
  \bibfield  {author} {\bibinfo {author} {\bibfnamefont {P.}~\bibnamefont
  {Anders}}, \bibinfo {author} {\bibfnamefont {E.}~\bibnamefont {Gull}},
  \bibinfo {author} {\bibfnamefont {L.}~\bibnamefont {Pollet}}, \bibinfo
  {author} {\bibfnamefont {M.}~\bibnamefont {Troyer}}, \ and\ \bibinfo {author}
  {\bibfnamefont {P.}~\bibnamefont {Werner}},\ }\href@noop {} {\bibfield
  {journal} {\bibinfo  {journal} {New J. Phys.}\ }\textbf {\bibinfo {volume}
  {13}},\ \bibinfo {pages} {075013} (\bibinfo {year} {2011})}\BibitemShut
  {NoStop}%
\bibitem [{\citenamefont {Sachdev}(1999)}]{Sachdev:1999fk}%
  \BibitemOpen
  \bibfield  {author} {\bibinfo {author} {\bibfnamefont {S.}~\bibnamefont
  {Sachdev}},\ }\href@noop {} {\emph {\bibinfo {title} {Quantum Phase
  Transitions}}}\ (\bibinfo  {publisher} {Cambridge University Press},\
  \bibinfo {address} {The Edinburgh Building, Cambridge CB2 2RU, UK},\ \bibinfo
  {year} {1999})\BibitemShut {NoStop}%
\bibitem [{\citenamefont {H\"ugel}\ and\ \citenamefont
  {Pollet}(2015)}]{Hugel:2015ab}%
  \BibitemOpen
  \bibfield  {author} {\bibinfo {author} {\bibfnamefont {D.}~\bibnamefont
  {H\"ugel}}\ and\ \bibinfo {author} {\bibfnamefont {L.}~\bibnamefont
  {Pollet}},\ }\href {\doibase 10.1103/PhysRevB.91.224510} {\bibfield
  {journal} {\bibinfo  {journal} {Phys. Rev. B}\ }\textbf {\bibinfo {volume}
  {91}},\ \bibinfo {pages} {224510} (\bibinfo {year} {2015})}\BibitemShut
  {NoStop}%
\bibitem [{\citenamefont {Capogrosso-Sansone}\ \emph
  {et~al.}(2007)\citenamefont {Capogrosso-Sansone}, \citenamefont {Prokof'ev},\
  and\ \citenamefont {Svistunov}}]{Capogrosso-Sansone:2007lh}%
  \BibitemOpen
  \bibfield  {author} {\bibinfo {author} {\bibfnamefont {B.}~\bibnamefont
  {Capogrosso-Sansone}}, \bibinfo {author} {\bibfnamefont {N.~V.}\ \bibnamefont
  {Prokof'ev}}, \ and\ \bibinfo {author} {\bibfnamefont {B.~V.}\ \bibnamefont
  {Svistunov}},\ }\href {\doibase 10.1103/PhysRevB.75.134302} {\bibfield
  {journal} {\bibinfo  {journal} {Phys. Rev. B}\ }\textbf {\bibinfo {volume}
  {75}},\ \bibinfo {pages} {134302} (\bibinfo {year} {2007})}\BibitemShut
  {NoStop}%
\bibitem [{\citenamefont {Hofmann}\ \emph {et~al.}(2013)\citenamefont
  {Hofmann}, \citenamefont {Eckstein}, \citenamefont {Arrigoni},\ and\
  \citenamefont {Potthoff}}]{Hofmann:2013aa}%
  \BibitemOpen
  \bibfield  {author} {\bibinfo {author} {\bibfnamefont {F.}~\bibnamefont
  {Hofmann}}, \bibinfo {author} {\bibfnamefont {M.}~\bibnamefont {Eckstein}},
  \bibinfo {author} {\bibfnamefont {E.}~\bibnamefont {Arrigoni}}, \ and\
  \bibinfo {author} {\bibfnamefont {M.}~\bibnamefont {Potthoff}},\ }\href
  {\doibase 10.1103/PhysRevB.88.165124} {\bibfield  {journal} {\bibinfo
  {journal} {Phys. Rev. B}\ }\textbf {\bibinfo {volume} {88}},\ \bibinfo
  {pages} {165124} (\bibinfo {year} {2013})}\BibitemShut {NoStop}%
\bibitem [{\citenamefont {{Hofmann}}\ \emph {et~al.}(2015)\citenamefont
  {{Hofmann}}, \citenamefont {{Eckstein}},\ and\ \citenamefont
  {{Potthoff}}}]{Hofmann:2015ab}%
  \BibitemOpen
  \bibfield  {author} {\bibinfo {author} {\bibfnamefont {F.}~\bibnamefont
  {{Hofmann}}}, \bibinfo {author} {\bibfnamefont {M.}~\bibnamefont
  {{Eckstein}}}, \ and\ \bibinfo {author} {\bibfnamefont {M.}~\bibnamefont
  {{Potthoff}}},\ }\href@noop {} {\bibfield  {journal} {\bibinfo  {journal}
  {ArXiv e-prints}\ } (\bibinfo {year} {2015})},\ \Eprint
  {http://arxiv.org/abs/1510.05866} {arXiv:1510.05866 [cond-mat.str-el]}
  \BibitemShut {NoStop}%
\bibitem [{\citenamefont {Aoki}\ \emph {et~al.}(2014)\citenamefont {Aoki},
  \citenamefont {Tsuji}, \citenamefont {Eckstein}, \citenamefont {Kollar},
  \citenamefont {Oka},\ and\ \citenamefont {Werner}}]{Aoki:2014kx}%
  \BibitemOpen
  \bibfield  {author} {\bibinfo {author} {\bibfnamefont {H.}~\bibnamefont
  {Aoki}}, \bibinfo {author} {\bibfnamefont {N.}~\bibnamefont {Tsuji}},
  \bibinfo {author} {\bibfnamefont {M.}~\bibnamefont {Eckstein}}, \bibinfo
  {author} {\bibfnamefont {M.}~\bibnamefont {Kollar}}, \bibinfo {author}
  {\bibfnamefont {T.}~\bibnamefont {Oka}}, \ and\ \bibinfo {author}
  {\bibfnamefont {P.}~\bibnamefont {Werner}},\ }\href {\doibase
  10.1103/RevModPhys.86.779} {\bibfield  {journal} {\bibinfo  {journal} {Rev.
  Mod. Phys.}\ }\textbf {\bibinfo {volume} {86}},\ \bibinfo {pages} {779}
  (\bibinfo {year} {2014})}\BibitemShut {NoStop}%
\bibitem [{\citenamefont {Strand}\ \emph
  {et~al.}(2015{\natexlab{a}})\citenamefont {Strand}, \citenamefont
  {Eckstein},\ and\ \citenamefont {Werner}}]{Strand:2015aa}%
  \BibitemOpen
  \bibfield  {author} {\bibinfo {author} {\bibfnamefont {H.~U.~R.}\
  \bibnamefont {Strand}}, \bibinfo {author} {\bibfnamefont {M.}~\bibnamefont
  {Eckstein}}, \ and\ \bibinfo {author} {\bibfnamefont {P.}~\bibnamefont
  {Werner}},\ }\href {http://link.aps.org/doi/10.1103/PhysRevX.5.011038}
  {\bibfield  {journal} {\bibinfo  {journal} {Phys. Rev. X}\ }\textbf {\bibinfo
  {volume} {5}},\ \bibinfo {pages} {011038} (\bibinfo {year}
  {2015}{\natexlab{a}})}\BibitemShut {NoStop}%
\bibitem [{\citenamefont {Kleinert}(1982)}]{Kleinert:1982aa}%
  \BibitemOpen
  \bibfield  {author} {\bibinfo {author} {\bibfnamefont {H.}~\bibnamefont
  {Kleinert}},\ }\href {\doibase 10.1002/prop.19820300402} {\bibfield
  {journal} {\bibinfo  {journal} {Fortschritte der Physik}\ }\textbf {\bibinfo
  {volume} {30}},\ \bibinfo {pages} {187} (\bibinfo {year} {1982})}\BibitemShut
  {NoStop}%
\bibitem [{\citenamefont {Berges}\ \emph {et~al.}(2005)\citenamefont {Berges},
  \citenamefont {Bors{\'a}nyi}, \citenamefont {Reinosa},\ and\ \citenamefont
  {Serreau}}]{Berges:2005aa}%
  \BibitemOpen
  \bibfield  {author} {\bibinfo {author} {\bibfnamefont {J.}~\bibnamefont
  {Berges}}, \bibinfo {author} {\bibfnamefont {S.}~\bibnamefont
  {Bors{\'a}nyi}}, \bibinfo {author} {\bibfnamefont {U.}~\bibnamefont
  {Reinosa}}, \ and\ \bibinfo {author} {\bibfnamefont {J.}~\bibnamefont
  {Serreau}},\ }\href {\doibase http://dx.doi.org/10.1016/j.aop.2005.06.001}
  {\bibfield  {journal} {\bibinfo  {journal} {Annals of Physics}\ }\textbf
  {\bibinfo {volume} {320}},\ \bibinfo {pages} {344} (\bibinfo {year}
  {2005})}\BibitemShut {NoStop}%
\bibitem [{\citenamefont {Cornwall}\ \emph {et~al.}(1974)\citenamefont
  {Cornwall}, \citenamefont {Jackiw},\ and\ \citenamefont
  {Tomboulis}}]{Cornwall:1974aa}%
  \BibitemOpen
  \bibfield  {author} {\bibinfo {author} {\bibfnamefont {J.~M.}\ \bibnamefont
  {Cornwall}}, \bibinfo {author} {\bibfnamefont {R.}~\bibnamefont {Jackiw}}, \
  and\ \bibinfo {author} {\bibfnamefont {E.}~\bibnamefont {Tomboulis}},\
  }\href@noop {} {\bibfield  {journal} {\bibinfo  {journal} {Phys. Rev. D}\
  }\textbf {\bibinfo {volume} {10}},\ \bibinfo {pages} {2428} (\bibinfo {year}
  {1974})}\BibitemShut {NoStop}%
\bibitem [{Note1()}]{Note1}%
  \BibitemOpen
  \bibinfo {note} {Note that the Luttinger-Ward functional $\Phi _{\protect
  \text {\relax \protect \fontsize {5}{6}\protect \selectfont LW}}$ for
  symmetry broken bosons with only a four-particle interaction vertex
  ($\protect \mathaccentV {hat}05E{V}_3 =0$, $\protect \mathaccentV
  {hat}05E{V}_4 \not =0$) still acquires an effective three-particle vertex
  \cite {Kleinert:1982aa} ($\protect \mathaccentV {hat}05E{\nu }_3 = \protect
  \mathaccentV {hat}05E{V}_4 \protect \boldsymbol \Phi $).}\BibitemShut {Stop}%
\bibitem [{\citenamefont {Kleinert}\ and\ \citenamefont
  {Schulte-Frohlinde}(2001)}]{Kleinert:2001fk}%
  \BibitemOpen
  \bibfield  {author} {\bibinfo {author} {\bibfnamefont {H.}~\bibnamefont
  {Kleinert}}\ and\ \bibinfo {author} {\bibfnamefont {V.}~\bibnamefont
  {Schulte-Frohlinde}},\ }\href@noop {} {\emph {\bibinfo {title} {Critical
  properties of Phi$^4$-Theories}}}\ (\bibinfo  {publisher} {World Scientific
  Publishing Co. Pte. Ltd.},\ \bibinfo {address} {Singapore},\ \bibinfo {year}
  {2001})\BibitemShut {NoStop}%
\bibitem [{\citenamefont {Kozik}\ \emph {et~al.}(2015)\citenamefont {Kozik},
  \citenamefont {Ferrero},\ and\ \citenamefont {Georges}}]{Kozik:2015aa}%
  \BibitemOpen
  \bibfield  {author} {\bibinfo {author} {\bibfnamefont {E.}~\bibnamefont
  {Kozik}}, \bibinfo {author} {\bibfnamefont {M.}~\bibnamefont {Ferrero}}, \
  and\ \bibinfo {author} {\bibfnamefont {A.}~\bibnamefont {Georges}},\ }\href
  {\doibase 10.1103/PhysRevLett.114.156402} {\bibfield  {journal} {\bibinfo
  {journal} {Phys. Rev. Lett.}\ }\textbf {\bibinfo {volume} {114}},\ \bibinfo
  {pages} {156402} (\bibinfo {year} {2015})}\BibitemShut {NoStop}%
\bibitem [{\citenamefont {Stan}\ \emph {et~al.}(2015)\citenamefont {Stan},
  \citenamefont {Romaniello}, \citenamefont {Rigamonti}, \citenamefont
  {Reining},\ and\ \citenamefont {Berger}}]{1367-2630-17-9-093045}%
  \BibitemOpen
  \bibfield  {author} {\bibinfo {author} {\bibfnamefont {A.}~\bibnamefont
  {Stan}}, \bibinfo {author} {\bibfnamefont {P.}~\bibnamefont {Romaniello}},
  \bibinfo {author} {\bibfnamefont {S.}~\bibnamefont {Rigamonti}}, \bibinfo
  {author} {\bibfnamefont {L.}~\bibnamefont {Reining}}, \ and\ \bibinfo
  {author} {\bibfnamefont {J.~A.}\ \bibnamefont {Berger}},\ }\href
  {http://stacks.iop.org/1367-2630/17/i=9/a=093045} {\bibfield  {journal}
  {\bibinfo  {journal} {New Journal of Physics}\ }\textbf {\bibinfo {volume}
  {17}},\ \bibinfo {pages} {093045} (\bibinfo {year} {2015})}\BibitemShut
  {NoStop}%
\bibitem [{\citenamefont {Metzner}\ and\ \citenamefont
  {Vollhardt}(1989)}]{Metzner:1989aa}%
  \BibitemOpen
  \bibfield  {author} {\bibinfo {author} {\bibfnamefont {W.}~\bibnamefont
  {Metzner}}\ and\ \bibinfo {author} {\bibfnamefont {D.}~\bibnamefont
  {Vollhardt}},\ }\href@noop {} {\bibfield  {journal} {\bibinfo  {journal}
  {Phys. Rev. Lett.}\ }\textbf {\bibinfo {volume} {62}},\ \bibinfo {pages}
  {324} (\bibinfo {year} {1989})}\BibitemShut {NoStop}%
\bibitem [{\citenamefont {Werner}\ \emph {et~al.}(2006)\citenamefont {Werner},
  \citenamefont {Comanac}, \citenamefont {de' Medici}, \citenamefont {Troyer},\
  and\ \citenamefont {Millis}}]{Werner:2006rt}%
  \BibitemOpen
  \bibfield  {author} {\bibinfo {author} {\bibfnamefont {P.}~\bibnamefont
  {Werner}}, \bibinfo {author} {\bibfnamefont {A.}~\bibnamefont {Comanac}},
  \bibinfo {author} {\bibfnamefont {L.}~\bibnamefont {de' Medici}}, \bibinfo
  {author} {\bibfnamefont {M.}~\bibnamefont {Troyer}}, \ and\ \bibinfo {author}
  {\bibfnamefont {A.~J.}\ \bibnamefont {Millis}},\ }\href {\doibase
  10.1103/PhysRevLett.97.076405} {\bibfield  {journal} {\bibinfo  {journal}
  {Phys. Rev. Lett.}\ }\textbf {\bibinfo {volume} {97}},\ \bibinfo {pages}
  {076405} (\bibinfo {year} {2006})}\BibitemShut {NoStop}%
\bibitem [{\citenamefont {Gull}\ \emph {et~al.}(2011)\citenamefont {Gull},
  \citenamefont {Millis}, \citenamefont {Lichtenstein}, \citenamefont
  {Rubtsov}, \citenamefont {Troyer},\ and\ \citenamefont
  {Werner}}]{Gull:2011lr}%
  \BibitemOpen
  \bibfield  {author} {\bibinfo {author} {\bibfnamefont {E.}~\bibnamefont
  {Gull}}, \bibinfo {author} {\bibfnamefont {A.~J.}\ \bibnamefont {Millis}},
  \bibinfo {author} {\bibfnamefont {A.~I.}\ \bibnamefont {Lichtenstein}},
  \bibinfo {author} {\bibfnamefont {A.~N.}\ \bibnamefont {Rubtsov}}, \bibinfo
  {author} {\bibfnamefont {M.}~\bibnamefont {Troyer}}, \ and\ \bibinfo {author}
  {\bibfnamefont {P.}~\bibnamefont {Werner}},\ }\href {\doibase
  10.1103/RevModPhys.83.349} {\bibfield  {journal} {\bibinfo  {journal} {Rev.
  Mod. Phys.}\ }\textbf {\bibinfo {volume} {83}},\ \bibinfo {pages} {349}
  (\bibinfo {year} {2011})}\BibitemShut {NoStop}%
\bibitem [{\citenamefont {Boyd}\ and\ \citenamefont
  {Vandenberghe}(2004)}]{Boyd:2004fu}%
  \BibitemOpen
  \bibfield  {author} {\bibinfo {author} {\bibfnamefont {S.}~\bibnamefont
  {Boyd}}\ and\ \bibinfo {author} {\bibfnamefont {L.}~\bibnamefont
  {Vandenberghe}},\ }\href@noop {} {\emph {\bibinfo {title} {Convex
  Optimization}}},\ \bibinfo {edition} {7th}\ ed.\ (\bibinfo  {publisher}
  {Cambridge University Press},\ \bibinfo {address} {The Edinburgh Building,
  Cambridge, CB2 8RU, UK},\ \bibinfo {year} {2004})\BibitemShut {NoStop}%
\bibitem [{Note2()}]{Note2}%
  \BibitemOpen
  \bibinfo {note} {The hybrd and hybrj methods of MINPACK as wrapped in SciPy
  \cite {Jones:2001aa}.}\BibitemShut {Stop}%
\bibitem [{\citenamefont {Knap}\ \emph {et~al.}(2010)\citenamefont {Knap},
  \citenamefont {Arrigoni},\ and\ \citenamefont {von~der
  Linden}}]{Knap:2010aa}%
  \BibitemOpen
  \bibfield  {author} {\bibinfo {author} {\bibfnamefont {M.}~\bibnamefont
  {Knap}}, \bibinfo {author} {\bibfnamefont {E.}~\bibnamefont {Arrigoni}}, \
  and\ \bibinfo {author} {\bibfnamefont {W.}~\bibnamefont {von~der Linden}},\
  }\href {http://link.aps.org/doi/10.1103/PhysRevB.81.024301} {\bibfield
  {journal} {\bibinfo  {journal} {Phys. Rev. B}\ }\textbf {\bibinfo {volume}
  {81}},\ \bibinfo {pages} {024301} (\bibinfo {year} {2010})}\BibitemShut
  {NoStop}%
\bibitem [{\citenamefont {Knap}\ \emph {et~al.}(2011)\citenamefont {Knap},
  \citenamefont {Arrigoni},\ and\ \citenamefont {von~der
  Linden}}]{Knap:2011aa}%
  \BibitemOpen
  \bibfield  {author} {\bibinfo {author} {\bibfnamefont {M.}~\bibnamefont
  {Knap}}, \bibinfo {author} {\bibfnamefont {E.}~\bibnamefont {Arrigoni}}, \
  and\ \bibinfo {author} {\bibfnamefont {W.}~\bibnamefont {von~der Linden}},\
  }\href {http://link.aps.org/doi/10.1103/PhysRevB.83.134507} {\bibfield
  {journal} {\bibinfo  {journal} {Physical Review B}\ }\textbf {\bibinfo
  {volume} {83}},\ \bibinfo {pages} {134507} (\bibinfo {year}
  {2011})}\BibitemShut {NoStop}%
\bibitem [{\citenamefont {Huerga}\ \emph {et~al.}(2013)\citenamefont {Huerga},
  \citenamefont {Dukelsky},\ and\ \citenamefont {Scuseria}}]{Huerga:2013aa}%
  \BibitemOpen
  \bibfield  {author} {\bibinfo {author} {\bibfnamefont {D.}~\bibnamefont
  {Huerga}}, \bibinfo {author} {\bibfnamefont {J.}~\bibnamefont {Dukelsky}}, \
  and\ \bibinfo {author} {\bibfnamefont {G.~E.}\ \bibnamefont {Scuseria}},\
  }\href {http://link.aps.org/doi/10.1103/PhysRevLett.111.045701} {\bibfield
  {journal} {\bibinfo  {journal} {Physical Review Letters}\ }\textbf {\bibinfo
  {volume} {111}},\ \bibinfo {pages} {045701} (\bibinfo {year}
  {2013})}\BibitemShut {NoStop}%
\bibitem [{\citenamefont {Ran\ifmmode~\mbox{\c{c}}\else \c{c}\fi{}on}\ and\
  \citenamefont {Dupuis}(2011{\natexlab{a}})}]{Ranifmmode-celse-cfion:2011aa}%
  \BibitemOpen
  \bibfield  {author} {\bibinfo {author} {\bibfnamefont {A.}~\bibnamefont
  {Ran\ifmmode~\mbox{\c{c}}\else \c{c}\fi{}on}}\ and\ \bibinfo {author}
  {\bibfnamefont {N.}~\bibnamefont {Dupuis}},\ }\href {\doibase
  10.1103/PhysRevB.83.172501} {\bibfield  {journal} {\bibinfo  {journal} {Phys.
  Rev. B}\ }\textbf {\bibinfo {volume} {83}},\ \bibinfo {pages} {172501}
  (\bibinfo {year} {2011}{\natexlab{a}})}\BibitemShut {NoStop}%
\bibitem [{\citenamefont {Ran\ifmmode~\mbox{\c{c}}\else \c{c}\fi{}on}\ and\
  \citenamefont {Dupuis}(2011{\natexlab{b}})}]{Ranifmmode-celse-cfion:2011ab}%
  \BibitemOpen
  \bibfield  {author} {\bibinfo {author} {\bibfnamefont {A.}~\bibnamefont
  {Ran\ifmmode~\mbox{\c{c}}\else \c{c}\fi{}on}}\ and\ \bibinfo {author}
  {\bibfnamefont {N.}~\bibnamefont {Dupuis}},\ }\href {\doibase
  10.1103/PhysRevB.84.174513} {\bibfield  {journal} {\bibinfo  {journal} {Phys.
  Rev. B}\ }\textbf {\bibinfo {volume} {84}},\ \bibinfo {pages} {174513}
  (\bibinfo {year} {2011}{\natexlab{b}})}\BibitemShut {NoStop}%
\bibitem [{\citenamefont {Pankov}\ \emph {et~al.}(2002)\citenamefont {Pankov},
  \citenamefont {Kotliar},\ and\ \citenamefont {Motome}}]{Pankov:2002aa}%
  \BibitemOpen
  \bibfield  {author} {\bibinfo {author} {\bibfnamefont {S.}~\bibnamefont
  {Pankov}}, \bibinfo {author} {\bibfnamefont {G.}~\bibnamefont {Kotliar}}, \
  and\ \bibinfo {author} {\bibfnamefont {Y.}~\bibnamefont {Motome}},\ }\href
  {\doibase 10.1103/PhysRevB.66.045117} {\bibfield  {journal} {\bibinfo
  {journal} {Phys. Rev. B}\ }\textbf {\bibinfo {volume} {66}},\ \bibinfo
  {pages} {045117} (\bibinfo {year} {2002})}\BibitemShut {NoStop}%
\bibitem [{\citenamefont {Hugenholtz}\ and\ \citenamefont
  {Pines}(1959)}]{Hugenholtz:1959aa}%
  \BibitemOpen
  \bibfield  {author} {\bibinfo {author} {\bibfnamefont {N.~M.}\ \bibnamefont
  {Hugenholtz}}\ and\ \bibinfo {author} {\bibfnamefont {D.}~\bibnamefont
  {Pines}},\ }\href {\doibase 10.1103/PhysRev.116.489} {\bibfield  {journal}
  {\bibinfo  {journal} {Phys. Rev.}\ }\textbf {\bibinfo {volume} {116}},\
  \bibinfo {pages} {489} (\bibinfo {year} {1959})}\BibitemShut {NoStop}%
\bibitem [{\citenamefont {Rickayzen}(1980)}]{Rickayzen:1980aa}%
  \BibitemOpen
  \bibfield  {author} {\bibinfo {author} {\bibfnamefont {G.}~\bibnamefont
  {Rickayzen}},\ }\href@noop {} {\emph {\bibinfo {title} {Green's functions and
  condensed matter}}}\ (\bibinfo  {publisher} {Academic Press Inc.},\ \bibinfo
  {year} {1980})\BibitemShut {NoStop}%
\bibitem [{Note3()}]{Note3}%
  \BibitemOpen
  \bibinfo {note} {Here specialized for lattices with cosine-dispersion, e.g.,
  hypercubic lattices with nearest-neighbor hopping.}\BibitemShut {Stop}%
\bibitem [{\citenamefont {Jarrell}\ and\ \citenamefont
  {Gubernatis}(1996)}]{Jarrell:1996fj}%
  \BibitemOpen
  \bibfield  {author} {\bibinfo {author} {\bibfnamefont {M.}~\bibnamefont
  {Jarrell}}\ and\ \bibinfo {author} {\bibfnamefont {J.~E.}\ \bibnamefont
  {Gubernatis}},\ }\href@noop {} {\bibfield  {journal} {\bibinfo  {journal}
  {Physics Reports}\ }\textbf {\bibinfo {volume} {269}},\ \bibinfo {pages}
  {133} (\bibinfo {year} {1996})}\BibitemShut {NoStop}%
\bibitem [{\citenamefont {Pippan}\ \emph {et~al.}(2009)\citenamefont {Pippan},
  \citenamefont {Evertz},\ and\ \citenamefont {Hohenadler}}]{Pippan:2009aa}%
  \BibitemOpen
  \bibfield  {author} {\bibinfo {author} {\bibfnamefont {P.}~\bibnamefont
  {Pippan}}, \bibinfo {author} {\bibfnamefont {H.~G.}\ \bibnamefont {Evertz}},
  \ and\ \bibinfo {author} {\bibfnamefont {M.}~\bibnamefont {Hohenadler}},\
  }\href {http://link.aps.org/doi/10.1103/PhysRevA.80.033612} {\bibfield
  {journal} {\bibinfo  {journal} {Phys. Rev. A}\ }\textbf {\bibinfo {volume}
  {80}},\ \bibinfo {pages} {033612} (\bibinfo {year} {2009})}\BibitemShut
  {NoStop}%
\bibitem [{\citenamefont {Anders}\ \emph {et~al.}(2012)\citenamefont {Anders},
  \citenamefont {Werner}, \citenamefont {Troyer}, \citenamefont {Sigrist},\
  and\ \citenamefont {Pollet}}]{Anders:2012kx}%
  \BibitemOpen
  \bibfield  {author} {\bibinfo {author} {\bibfnamefont {P.}~\bibnamefont
  {Anders}}, \bibinfo {author} {\bibfnamefont {P.}~\bibnamefont {Werner}},
  \bibinfo {author} {\bibfnamefont {M.}~\bibnamefont {Troyer}}, \bibinfo
  {author} {\bibfnamefont {M.}~\bibnamefont {Sigrist}}, \ and\ \bibinfo
  {author} {\bibfnamefont {L.}~\bibnamefont {Pollet}},\ }\href {\doibase
  10.1103/PhysRevLett.109.206401} {\bibfield  {journal} {\bibinfo  {journal}
  {Phys. Rev. Lett.}\ }\textbf {\bibinfo {volume} {109}},\ \bibinfo {pages}
  {206401} (\bibinfo {year} {2012})}\BibitemShut {NoStop}%
\bibitem [{\citenamefont {Panas}\ \emph {et~al.}(2015)\citenamefont {Panas},
  \citenamefont {Kauch}, \citenamefont {Kune\ifmmode~\check{s}\else
  \v{s}\fi{}}, \citenamefont {Vollhardt},\ and\ \citenamefont
  {Byczuk}}]{Panas:2015ab}%
  \BibitemOpen
  \bibfield  {author} {\bibinfo {author} {\bibfnamefont {J.}~\bibnamefont
  {Panas}}, \bibinfo {author} {\bibfnamefont {A.}~\bibnamefont {Kauch}},
  \bibinfo {author} {\bibfnamefont {J.}~\bibnamefont
  {Kune\ifmmode~\check{s}\else \v{s}\fi{}}}, \bibinfo {author} {\bibfnamefont
  {D.}~\bibnamefont {Vollhardt}}, \ and\ \bibinfo {author} {\bibfnamefont
  {K.}~\bibnamefont {Byczuk}},\ }\href {\doibase 10.1103/PhysRevB.92.045102}
  {\bibfield  {journal} {\bibinfo  {journal} {Phys. Rev. B}\ }\textbf {\bibinfo
  {volume} {92}},\ \bibinfo {pages} {045102} (\bibinfo {year}
  {2015})}\BibitemShut {NoStop}%
\bibitem [{\citenamefont {Strand}\ \emph
  {et~al.}(2015{\natexlab{b}})\citenamefont {Strand}, \citenamefont
  {Eckstein},\ and\ \citenamefont {Werner}}]{Strand:2015ac}%
  \BibitemOpen
  \bibfield  {author} {\bibinfo {author} {\bibfnamefont {H.~U.~R.}\
  \bibnamefont {Strand}}, \bibinfo {author} {\bibfnamefont {M.}~\bibnamefont
  {Eckstein}}, \ and\ \bibinfo {author} {\bibfnamefont {P.}~\bibnamefont
  {Werner}},\ }\href {\doibase 10.1103/PhysRevA.92.063602} {\bibfield
  {journal} {\bibinfo  {journal} {Phys. Rev. A}\ }\textbf {\bibinfo {volume}
  {92}},\ \bibinfo {pages} {063602} (\bibinfo {year}
  {2015}{\natexlab{b}})}\BibitemShut {NoStop}%
\bibitem [{\citenamefont {Ayral}\ and\ \citenamefont
  {Parcollet}(2015)}]{Ayral:2015ab}%
  \BibitemOpen
  \bibfield  {author} {\bibinfo {author} {\bibfnamefont {T.}~\bibnamefont
  {Ayral}}\ and\ \bibinfo {author} {\bibfnamefont {O.}~\bibnamefont
  {Parcollet}},\ }\href {\doibase 10.1103/PhysRevB.92.115109} {\bibfield
  {journal} {\bibinfo  {journal} {Phys. Rev. B}\ }\textbf {\bibinfo {volume}
  {92}},\ \bibinfo {pages} {115109} (\bibinfo {year} {2015})}\BibitemShut
  {NoStop}%
\bibitem [{\citenamefont {Fetter}\ and\ \citenamefont
  {Walecka}(2003)}]{Fetter:2003aa}%
  \BibitemOpen
  \bibfield  {author} {\bibinfo {author} {\bibfnamefont {A.~L.}\ \bibnamefont
  {Fetter}}\ and\ \bibinfo {author} {\bibfnamefont {J.~D.}\ \bibnamefont
  {Walecka}},\ }\href@noop {} {\emph {\bibinfo {title} {Quantum Theory of
  Many-Particle Systems}}}\ (\bibinfo  {publisher} {Dover Publications Inc.},\
  \bibinfo {address} {31 East 2nd Street, Mineloa, New York 11501},\ \bibinfo
  {year} {2003})\BibitemShut {NoStop}%
\bibitem [{\citenamefont {Negele}\ and\ \citenamefont
  {Orland}(1998)}]{Negele:1998aa}%
  \BibitemOpen
  \bibfield  {author} {\bibinfo {author} {\bibfnamefont {J.~W.}\ \bibnamefont
  {Negele}}\ and\ \bibinfo {author} {\bibfnamefont {H.}~\bibnamefont
  {Orland}},\ }\href@noop {} {\emph {\bibinfo {title} {Quantum Many-Particle
  Systems}}}\ (\bibinfo  {publisher} {Westview Press},\ \bibinfo {year}
  {1998})\BibitemShut {NoStop}%
\bibitem [{\citenamefont {Mahan}(2000)}]{Mahan:2000tg}%
  \BibitemOpen
  \bibfield  {author} {\bibinfo {author} {\bibfnamefont {G.~D.}\ \bibnamefont
  {Mahan}},\ }\href@noop {} {\emph {\bibinfo {title} {Many-Particle
  Physics}}},\ \bibinfo {edition} {3rd}\ ed.\ (\bibinfo  {publisher} {Kluwer
  Academic/Plenum Publishers},\ \bibinfo {address} {233 Spring Street, New
  York, New York 10013},\ \bibinfo {year} {2000})\BibitemShut {NoStop}%
\bibitem [{\citenamefont {Matsubara}(1955)}]{Matsubara:1955aa}%
  \BibitemOpen
  \bibfield  {author} {\bibinfo {author} {\bibfnamefont {T.}~\bibnamefont
  {Matsubara}},\ }\href {\doibase 10.1143/PTP.14.351} {\bibfield  {journal}
  {\bibinfo  {journal} {Prog. Theor. Phys.}\ }\textbf {\bibinfo {volume}
  {14}},\ \bibinfo {pages} {351} (\bibinfo {year} {1955})}\BibitemShut
  {NoStop}%
\bibitem [{\citenamefont {Altland}\ and\ \citenamefont
  {Simons}(2010)}]{Atland:2006nx}%
  \BibitemOpen
  \bibfield  {author} {\bibinfo {author} {\bibfnamefont {A.}~\bibnamefont
  {Altland}}\ and\ \bibinfo {author} {\bibfnamefont {B.}~\bibnamefont
  {Simons}},\ }\href@noop {} {\emph {\bibinfo {title} {Condensed Matter Field
  Theory}}},\ \bibinfo {edition} {2nd}\ ed.\ (\bibinfo  {publisher} {Cambridge
  University Press},\ \bibinfo {address} {The Edinburgh Building, Cambridge CB2
  8RU, UK},\ \bibinfo {year} {2010})\BibitemShut {NoStop}%
\bibitem [{\citenamefont {Kleinert}(2009)}]{Kleinert:2009aa}%
  \BibitemOpen
  \bibfield  {author} {\bibinfo {author} {\bibfnamefont {H.}~\bibnamefont
  {Kleinert}},\ }\href@noop {} {\emph {\bibinfo {title} {Path Integrals in
  Quantum Mechanics, Statistics, Polymer Physics, and Financial Markets}}},\
  \bibinfo {edition} {5th}\ ed.,\ 5 Toh Tuck Link, Singapore 596224\ (\bibinfo
  {publisher} {World Scientific Publishing Co. Pte. Ltd.},\ \bibinfo {year}
  {2009})\BibitemShut {NoStop}%
\bibitem [{\citenamefont {Crawford}(1991)}]{Crawford:1991qy}%
  \BibitemOpen
  \bibfield  {author} {\bibinfo {author} {\bibfnamefont {J.~D.}\ \bibnamefont
  {Crawford}},\ }\href {\doibase 10.1103/RevModPhys.63.991} {\bibfield
  {journal} {\bibinfo  {journal} {Rev. Mod. Phys.}\ }\textbf {\bibinfo {volume}
  {63}},\ \bibinfo {pages} {991} (\bibinfo {year} {1991})}\BibitemShut
  {NoStop}%
\bibitem [{\citenamefont {Strand}\ \emph {et~al.}(2011)\citenamefont {Strand},
  \citenamefont {Sabashvili}, \citenamefont {Granath}, \citenamefont
  {Hellsing},\ and\ \citenamefont {\"Ostlund}}]{Strand:2011lr}%
  \BibitemOpen
  \bibfield  {author} {\bibinfo {author} {\bibfnamefont {H.~U.~R.}\
  \bibnamefont {Strand}}, \bibinfo {author} {\bibfnamefont {A.}~\bibnamefont
  {Sabashvili}}, \bibinfo {author} {\bibfnamefont {M.}~\bibnamefont {Granath}},
  \bibinfo {author} {\bibfnamefont {B.}~\bibnamefont {Hellsing}}, \ and\
  \bibinfo {author} {\bibfnamefont {S.}~\bibnamefont {\"Ostlund}},\ }\href
  {\doibase 10.1103/PhysRevB.83.205136} {\bibfield  {journal} {\bibinfo
  {journal} {Phys. Rev. B}\ }\textbf {\bibinfo {volume} {83}},\ \bibinfo
  {pages} {205136} (\bibinfo {year} {2011})}\BibitemShut {NoStop}%
\bibitem [{\citenamefont {Lanat\`a}\ \emph {et~al.}(2012)\citenamefont
  {Lanat\`a}, \citenamefont {Strand}, \citenamefont {Dai},\ and\ \citenamefont
  {Hellsing}}]{Lanata:2012lr}%
  \BibitemOpen
  \bibfield  {author} {\bibinfo {author} {\bibfnamefont {N.}~\bibnamefont
  {Lanat\`a}}, \bibinfo {author} {\bibfnamefont {H.~U.~R.}\ \bibnamefont
  {Strand}}, \bibinfo {author} {\bibfnamefont {X.}~\bibnamefont {Dai}}, \ and\
  \bibinfo {author} {\bibfnamefont {B.}~\bibnamefont {Hellsing}},\ }\href
  {\doibase 10.1103/PhysRevB.85.035133} {\bibfield  {journal} {\bibinfo
  {journal} {Phys. Rev. B}\ }\textbf {\bibinfo {volume} {85}},\ \bibinfo
  {pages} {035133} (\bibinfo {year} {2012})}\BibitemShut {NoStop}%
\bibitem [{\citenamefont {Jones}\ \emph {et~al.}(01  )\citenamefont {Jones},
  \citenamefont {Oliphant}, \citenamefont {Peterson} \emph
  {et~al.}}]{Jones:2001aa}%
  \BibitemOpen
  \bibfield  {author} {\bibinfo {author} {\bibfnamefont {E.}~\bibnamefont
  {Jones}}, \bibinfo {author} {\bibfnamefont {T.}~\bibnamefont {Oliphant}},
  \bibinfo {author} {\bibfnamefont {P.}~\bibnamefont {Peterson}},  \emph
  {et~al.},\ }\href {http://www.scipy.org/} {\enquote {\bibinfo {title}
  {{Scipy}: Open source scientific tools for {Python}},}\ } (\bibinfo {year}
  {2001--})\BibitemShut {NoStop}%
\end{thebibliography}%
% --------------------------------------------------------------------

% --------------------------------------------------------------------
\end{document}